\documentclass[12pt]{article}
\usepackage{epsfig, epsfig, graphics, lscape, amsmath, amsbsy, amstext, natbib}

\renewcommand{\baselinestretch}{1.5}

\begin{document}
\newcommand{\ve}[1]{{\mbox{\boldmath ${#1}$}}}
\renewcommand{\baselinestretch}{1.2}
%\lhead[\fancyplain{} \leftmark]{}
%\chead[]{}
%\rhead[]{\fancyplain{}\rightmark}
%\cfoot{}
%\headrulewidth=0pt
\markright{
%\hbox{\footnotesize\rm Statistica Sinica
%{\footnotesize\bf ??}(200?), 000-000}\hfill
}

\begin{titlepage}
\title{
{\Large  \bf Spatially Varying Coefficient Model   for Neuroimaging Data with Jump Discontinuities  }
 \author{
 Hongtu Zhu$^\star$,
 Department  of Biostatistics   \\
 and Biomedical Research Imaging Center \\
 University of North Carolina at Chapel Hill\\
 Chapel Hill, NC 27599, USA\\
 Jianqing Fan$^\dag$,
 \\Department of Oper Res and Fin. Eng \\
 Princeton University, 
 Princeton, NJ 08540\\
 and
 Linglong Kong$^\ddag$
 \\
 Department of Mathematical and Statistical Sciences \\ 
  University of Alberta, 
  Edmonton, AB Canada T6G 2G1 
 } 
}
\date{}
\maketitle 
\thispagestyle{empty}

%\begin{frontmatter}
%\title{ Spatial Varying Coefficient Model for Neuroimaging Data}
%\runtitle{ Spatial  Varying Coefficient Model}
%\thankstext{T1}{Footnote to the title with the `thankstext' command.}

%\begin{aug}
%\author{\snm{Hongtu Zhu}\thanksref{t1}\ead[label=e1]{hzhu@bios.unc.edu}},
%\author{\snm{Jianqing Fan}\thanksref{t2}\ead[label=e2]{jqfan@Princeton.EDU}}

%\thankstext{t1}{The research of Zhu  was supported by NSF grant BCS-08-26844 and NIH
%grants  RR025747-01,  P01CA142538-01,   MH086633, EB005149-01 and  AG033387.}
%\thankstext{t2}{Fan's research was supported by Supported by NSF Grant DMS-07-14554 and NIH Grant R01-GM072611..
% The content is
%solely the responsibility of the authors and does not necessarily
%represent the official views of the NSF or the NIH.}
%\runauthor{H. Zhu and  J. Fan}

%\affiliation{University of North Carolina and Princeton  University}

%\address{
%\printead{e1}\\
%\phantom{E-mail:\ }\printead*{e3}}

%\address{
%\printead{e2}}
%\end{aug}

\begin{abstract}
\noindent
Motivated by recent work on studying massive  imaging data in  various
neuroimaging  studies,  we propose  a novel spatially varying coefficient
model (SVCM)  to capture the  varying association between
  imaging measures  in a three-dimensional  (3D)
volume (or 2D surface) with a set of covariates.
Two stylized features of  neuorimaging data are the presence of
multiple piecewise  smooth  regions   with unknown edges and jumps and substantial spatial correlations.
 To specifically account for these two features, SVCM includes
  a measurement model with multiple varying coefficient functions,
    a jumping surface  model    for each varying coefficient function, and  a functional principal component model.
We develop a three-stage estimation procedure to simultaneously estimate the varying coefficient functions and  the spatial correlations.
The estimation procedure includes a  fast multiscale adaptive
estimation and testing procedure
 to  independently estimate each   varying  coefficient function, while preserving its edges among different piecewise-smooth regions.
  We   systematically  investigate the asymptotic properties (e.g.,   consistency and asymptotic normality) of the  multiscale adaptive parameter
estimates.    We also establish the uniform
convergence rate of the estimated spatial covariance function
and its associated eigenvalues and eigenfunctions.     Our Monte Carlo simulation and real data analysis have confirmed
  the excellent performance of SVCM.
\end{abstract}

%\begin{keyword}[class=AMS]
%\kwd[Primary ]{62G05}
%\kwd{62G08}
%\kwd[; secondary ]{62G20}
%\end{keyword}

 \noindent {\it Key Words}:
Asymptotic normality; Functional principal component analysis;  Jumping surface model;  Kernel;
 Spatial varying coefficient model;   Wald test.

 \end{titlepage}

\section{Introduction}

% See for example  among many others.

The aims of this paper are to develop a spatially varying coefficient model (SVCM)
to  delineate   association between massive imaging data and  a set of covariates of interest, such
as age, and to characterize the spatial
variability of the imaging data.  Examples of 
such imaging  data  include T1 weighted magnetic resonance imaging (MRI),
 functional MRI,   and diffusion tensor imaging,    among many others    \citep{Friston2007, Thompson2002, Mori2002, Lazar2008}. 
 In  neuroimaging studies,
 following spatial normalization,
  imaging
  data   usually consists of data points  from  different subjects (or scans)  at 
 a large number of locations (called voxels) in  a common 3D
 volume (without loss of generality), which is called  a {\it template}.    
We assume that all imaging data have been registered to a template throughout the paper. 

To analyze such massive imaging data, researchers face at least two main challenges.
 The first one is to characterize   varying  association between   imaging data and   covariates, while
preserving important  features, such as edges and jumps,  and  the  shape and  spatial  extent
  of    effect images.
Due to the physical and biological reasons, imaging data are  usually expected to contain spatially contiguous  regions or effect regions  with relatively sharp edges
\citep{Chumbley2009, Chan2005, Tabelow2008, Tabelow2008b}.  For instance, 
normal
  brain tissue can   generally be classified into three broad tissue types including white matter, gray matter,  and cerebrospinal fluid.
These three tissues can be roughly separated  by using  MRI due to their imaging intensity differences and relatively intensity homogeneity within each tissue.
The second  challenge is  to characterize  spatial correlations among a large number of voxels, usually in the tens thousands to millions, for imaging data.
 Such spatial correlation
structure and variability are important for achieving better prediction accuracy, for increasing the sensitivity of signal detection,  and for characterizing the random variability of imaging data across subjects 
\citep{Cressie2011, Spence2007}.  

% Similar tissue segmentation can be observed in diffusion weighted  image (Fig. 1 (c)), while one may observe more subtle  structures in white matter based on diffusion tensor image (Fig. 1 ({\bf d})).
%However,
%   different distant brain regions may activate  under either the same   cognitive task  for different subjects (Fig. 1 (f) and (h)) or  different cognitive tasks for the same subject
%    (Fig. 1 (e) and (g)) in
%functional MRI.

  There are two major statistical  methods including  voxel-wise   methods    and multiscale adaptive   methods
  for addressing the first challenge.
 Conventional  voxel-wise approaches  involve  in  Gaussian smoothing imaging data,  independently fitting a
statistical model
 to  imaging data  at each   voxel,   and   generating   statistical maps of test
statistics and  $p$-values  \citep{Lazar2008,  Worsley2004}.
As shown in    \cite{Chumbley2009} and \cite{Li2011},   voxel-wise methods  are generally not optimal in power since it ignores 
 the spatial information of imaging data. Moreover,  the 
  use of Gaussian smoothing  can
blur the image data near the edges of the
spatially contiguous regions and  thus  introduce substantial bias in statistical results \citep{YueLoh2010}.

%The first  major methods   explicitly incorporate
%the spatial non-independence in the $p$-value or test  statistic images. These methods
%usually combine strength from both
 % the magnitude of testing statistic and    cluster extent, where the cluster is defined as the number of
 %contiguous significant voxels above a specific threshold \citep{Worsley2004}.

There is a great interest in the development of  multiscale adaptive   methods to adaptively smooth  neuroimaging data, which is often characterized by a high noise level and a low 
signal-to-noise ratio 
  \citep{Tabelow2008, Tabelow2008b,Polzehl2010,Li2011, Qiu2005, Qiu2007}.
Such multiscale adaptive   methods   not only increase signal-to-noise ratio,  
but also preserve  important    features (e.g., edge)   of imaging data. 
 For instance, 
 in   \cite{Polzehl2000, Polzehl2006},   a novel   propagation-separation  approach was developed
  to  adaptively and spatially smooth  a single image without   explicitly detecting  edges. Recently,
  there are a few attempts to extend those adaptive smoothing methods to smoothing multiple images from a single subject 
   \citep{Tabelow2008, Tabelow2008b, Polzehl2010}.
In \cite{Li2011},    a multiscale adaptive
regression model, which integrates     the propagation-separation approach  and
voxel-wise approach,  was developed  for a large class of  parametric models.

There are      two major statistical  models, including Markov random fields  and low rank models, for  addressing the second challenge.
The Markov random field  models
  explicitly
  use  the Markov property  of  an undirected graph
to
characterize spatial dependence
 among spatially connected
voxels
      \citep{Besag1986, LiMRF2009}.
However, it can be
restrictive to assume a specific type of spatial correlation  structure,
such as   Markov random fields,  for very large spatial data sets besides its computational complexity \citep{Cressie2011}.
 In spatial statistics,  low rank models,    also called spatial random effects models,  use a linear combination of `known' spatial basis functions to approximate spatial dependence structure  in
 a single spatial map \citep{Cressie2011}.  The low rank models have a close connection with  the functional principal component analysis  model for characterizing spatial correlation structure in
multiple images, in which   spatial basis functions are directly estimated  \citep{Zipunnikov2011, Ramsay2005, MR2278365}.

%\begin{figure}[!ht]
%\vspace*{12pt}
%\centerline{\includegraphics[bb=60mm 90mm 177mm 194mm, width=.9\textwidth]{fig1.png}}
%\centerline{\includegraphics[width=.9\textwidth]{fig1.png}}
%\caption{ Representative functional neuroimaging data: (a)  the right internal
%capsule tract, (b) fractional anisotropy (FA) values measured at 75 grid points from 40
%randomly selected infants, (c) 3D plot of FA functional curves from 40 randomly
%selected infants, ({\bf d}) the Bonferroni corrected $p$ values of the $t$ statistics from a selected slice and a selected voxel, (e) and (f) the
%estimated hemodynamic response functions (HRF) corresponding to two stimulus categories  from 14 subjects.
% } \label{tract3fig0}
%\end{figure}

The goal of this article is to develop   SVCM and its estimation procedure  to simultaneously address the two challenges discussed above.
SVCM  has three   features: piecewise smooth, spatially  correlated, and
spatially adaptive, while its estimation procedure is  fast, accurate and individually updated.
Major contributions of the paper are as follows.
\begin{itemize}
\item{}  Compared with the existing multiscale adaptive   methods, SVCM  first integrates
 a jumping surface model   to delineate the piecewise  smooth feature of  raw and  effect images and
 the functional principal component model to explicitly  incorporate the spatial
   correlation  structure of   raw  imaging data.
\item{}
   A comprehensive three-stage estimation  procedure
    is developed to  adaptively and spatially  improve estimation accuracy and capture spatial correlations.
   \item{}   Compared with the existing methods, we use a fast and accurate estimation method to independently smooth each of effect images, while consistently estimating their standard deviation images.
\item{}
  We  systematically establish
consistency and asymptotic distribution of the adaptive parameter estimators  under two different scenarios including  piecewise-smooth and piecewise-constant varying coefficient functions.
In particular, we   introduce several adaptive boundary conditions to delineate the relationship  between the amount of jumps and the sample size.
Our conditions and theoretical results   differ substantially from  those for the propagation-separation type methods \citep{Polzehl2000, Polzehl2006, Li2011}.
\end{itemize}

The rest of this paper is organized as follows. In Section 2, we
describe SVCM and its three-stage estimation procedure and establish  the
theoretical properties. In Section 3, we present
a set  of   simulation studies with the known ground truth  to
examine the finite sample performance of the three-stage  estimation procedure for SVCM.
In Section 4,  we apply    the proposed methods in a  real imaging dataset on  attention deficit hyperactivity disorder (ADHD).
In Section 5, we conclude the paper with some discussions.
Technical conditions 
are given in Section 6. Proofs and additional results are given in a supplementary document.

\section{Spatial Varying Coefficient Model with Jumping Discontinuities}

\subsection{Model Setup}

We consider  imaging  measurements  in a   template    
and clinical variables (e.g., age, gender, and height) from $n$ subjects.
  Let $\mathcal{D}$  represent a 3D  volume  and 
  ${\bf d}$ and ${\bf d}_0$, respectively,  denote 
   a point and   the center of a voxel in $\mathcal{D}$. 
  Let 
  $\mathcal D_0$ be the union of all centers ${\bf d}_0$ in $\mathcal D$  and  $N_D$ equal  the number of
voxels in $\mathcal{D}_0$.
Without loss of generality, ${\mathcal D}$ is assumed to be a compact set in $R^3$.
  For the $i$-th 
subject, we observe  an $m\times 1$ vector of imaging measures $y_{i}({\bf d}_0)$ at  ${\bf d}_0\in {\mathcal D}_0$, which leads to  an $m N_D\times 1$ vector of measurements across ${\mathcal D}_0,$
denoted by ${\bf Y}_{i, {\mathcal D}_0}=\{y_{i}({\bf d}_0): {\bf d}_0\in
\mathcal{D}_0\}$.  For notational simplicity, we   set $m=1$ and consider a 3D volume throughout the paper.
%  In neuroimaging studies,  imaging measurements can include   the shape
%representation of the surfaces of cortical or   subcortical
%structures,
%fMRI signals,
 % diffusion tensors, and so on \citep{Thompson2002,  Li2011, YueLoh2010}.

The proposed {\it spatial varying coefficient model} (SVCM) consists of  three components: a measurement model,  a jumping surface model, and a functional component analysis model.
The measurement model characterizes the association between imaging measures and covariates  and is given by
\begin{equation}
\label{SVCMeq1}
 {y}_{i}({\bf d})= {\bf
x}_i^T{\ve \beta}({\bf d})+\eta_{i}({\bf d})+\epsilon_{i}({\bf d})~~~ \mbox{for all}~~i=1, \ldots, n~~\mbox{and}~~{\bf d}\in \mathcal D, 
\end{equation}
where ${\bf x}_i=(x_{i1}, \ldots, x_{ip})^T$ is a $p\times 1$ vector of  covariates,  ${\ve \beta}({\bf d})=(\beta_{1}({\bf d}), \ldots, \beta_{p}({\bf d}))^T$ is a $p\times 1$
vector of  coefficient functions of $d$,  $\eta_{i}({\bf d})$ characterizes individual image variations from ${\bf x}_i^T{\ve\beta}({\bf d})$,
and $\epsilon_{i}({\bf d})$ are measurement errors. 
%Although we only observe  ${y}_{i}({\bf d})$ at ${\bf d}={\bf d}_0\in \mathcal D_0$, model (\ref{SVCMeq1}) is assumed for all 
%${\bf d}\in\mathcal D$.    
Moreover,  $\{\eta_{i}({\bf d}):
{\bf d}\in {\mathcal D}\}$ is a stochastic process indexed by
${\bf d}\in {\mathcal D}$ that captures the within-image
dependence.
We assume  that they are mutually  independent and ${\eta}_i({\bf d})$
and ${\epsilon}_i({\bf d})$ are independent and identical copies of
SP$({\bf 0}, \Sigma_{\eta})$ and SP$({\bf 0}, \Sigma_\epsilon)$,
respectively, where SP$(\mu, \Sigma)$ denotes  a stochastic process
vector with  mean function $\mu({\bf d})$ and covariance function
$\Sigma({\bf d}, {\bf d}')$.  Moreover, $\epsilon_i({\bf d})$ and $\epsilon_i({\bf d}')$ are
 independent for ${\bf d}\not = {\bf d}'$ and  thus  $\Sigma_\epsilon
 ({\bf d}, {\bf d}') =0$ for ${\bf d} \not = {\bf d}'$.  Therefore, the
covariance function of  $\{{\bf y}_i({\bf d}): {\bf d}\in \mathcal D\}$, conditioned on ${\bf x}_i$,
is given by
\begin{equation} \label{SVCMeq2}
\Sigma_y({\bf d}, {\bf d}')=\mbox{Cov}({\bf y}_i({\bf d}), {\bf y}_i({\bf d}'))=
\Sigma_{\eta}({\bf d}, {\bf d}')+\Sigma_\epsilon({\bf d}, {\bf d}){\bf 1}({\bf d}={\bf d}').
\end{equation}

%Model (\ref{Tract3eq1}) can be regarded as a  multivariate varying coefficient model. T
%his model has been widely studied and developed  for longitudinal, time series, and functional data.
%See for example  \cite{MR1742497, MR1769751, MR1959093, MR2425354, MR2504204, MR1666699, Ramsay2005, MR2087972, MR1888349, Zhang2007, MR2523900} among many others. Most references consider the scenario that both the number of data points  and the location points can be different for different individuals.  In our setting, however, we assume the common design across all subjects.  This design has been widely used in
% modern imaging techniques    \cite{Fass2008,  Buzsaki2006, Friston2009,   Zhu:2010}.
%For instance, for     most fMRI studies,   the same imaging acquisition sequence with
%t

\begin{figure}
%%\vspace*{0pt}
%%\begin{center}
%%\begin{tabular}{@{}c@{} @{}c@{} }
%% {\includegraphics[width=0.5\hsize]{JSM_disjointFig11.pdf}}&
%% {\includegraphics[width=0.5\hsize]{JSM_betaFig21.pdf}}\\
%% \\
%%(a) & (b) \\
%% {\includegraphics[width=0.5\hsize]{JSM_betaNoiseFig4.pdf}}&
%%  {\includegraphics[width=0.5\hsize]{Theorey1.pdf}}
%% \\
%%(c) & ({\bf d})
%%\end{tabular}
%%\end{center}
%%\vspace*{0pt}
%%\vspace*{0pt}
\begin{center}
\begin{tabular}{@{}c@{} @{}c@{} }
 {\includegraphics[width=0.45\hsize]{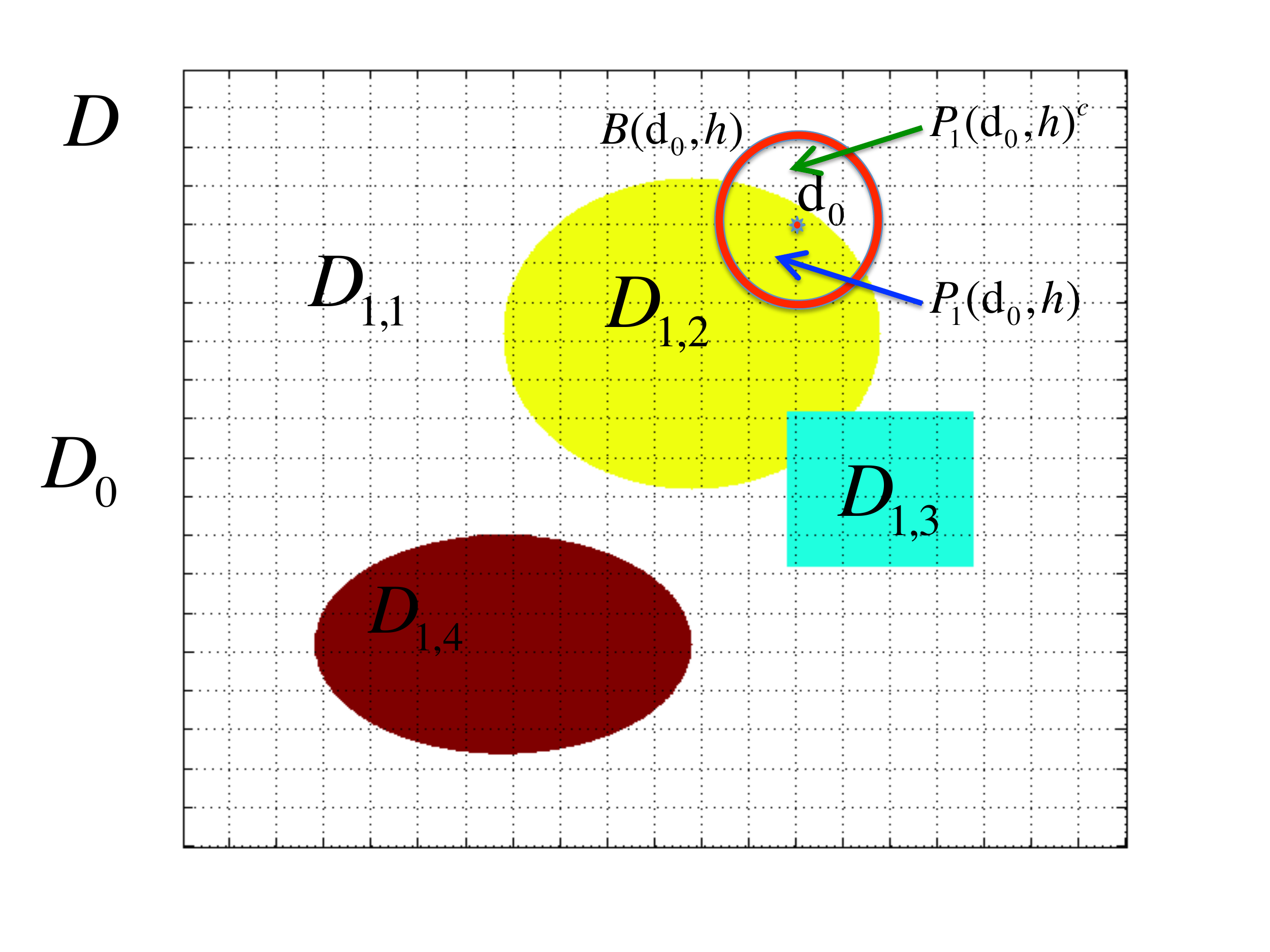}}&
 {\includegraphics[width=0.45\hsize]{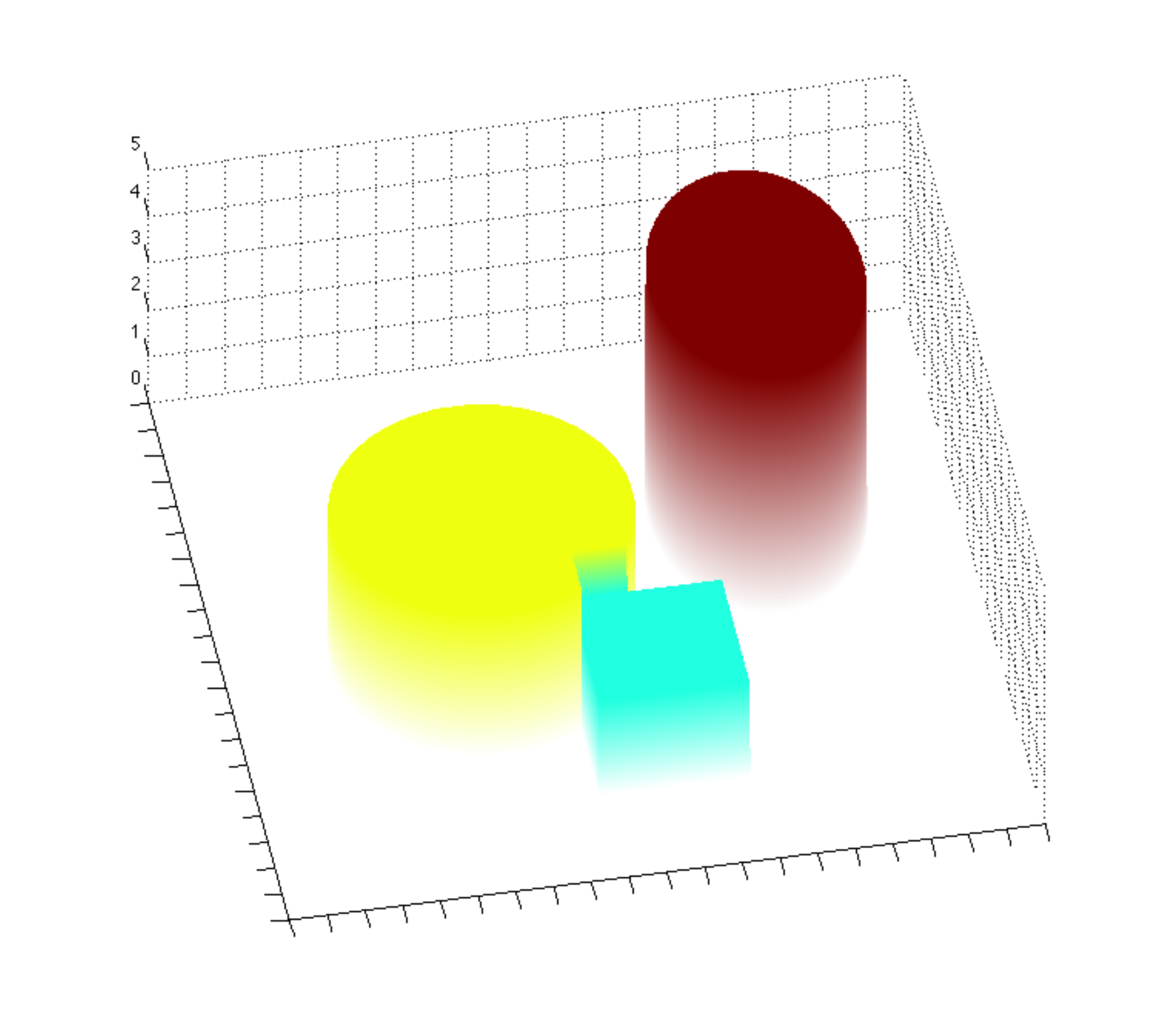}}\\
 \\
(a) & (b) \\
 {\includegraphics[width=0.45\hsize]{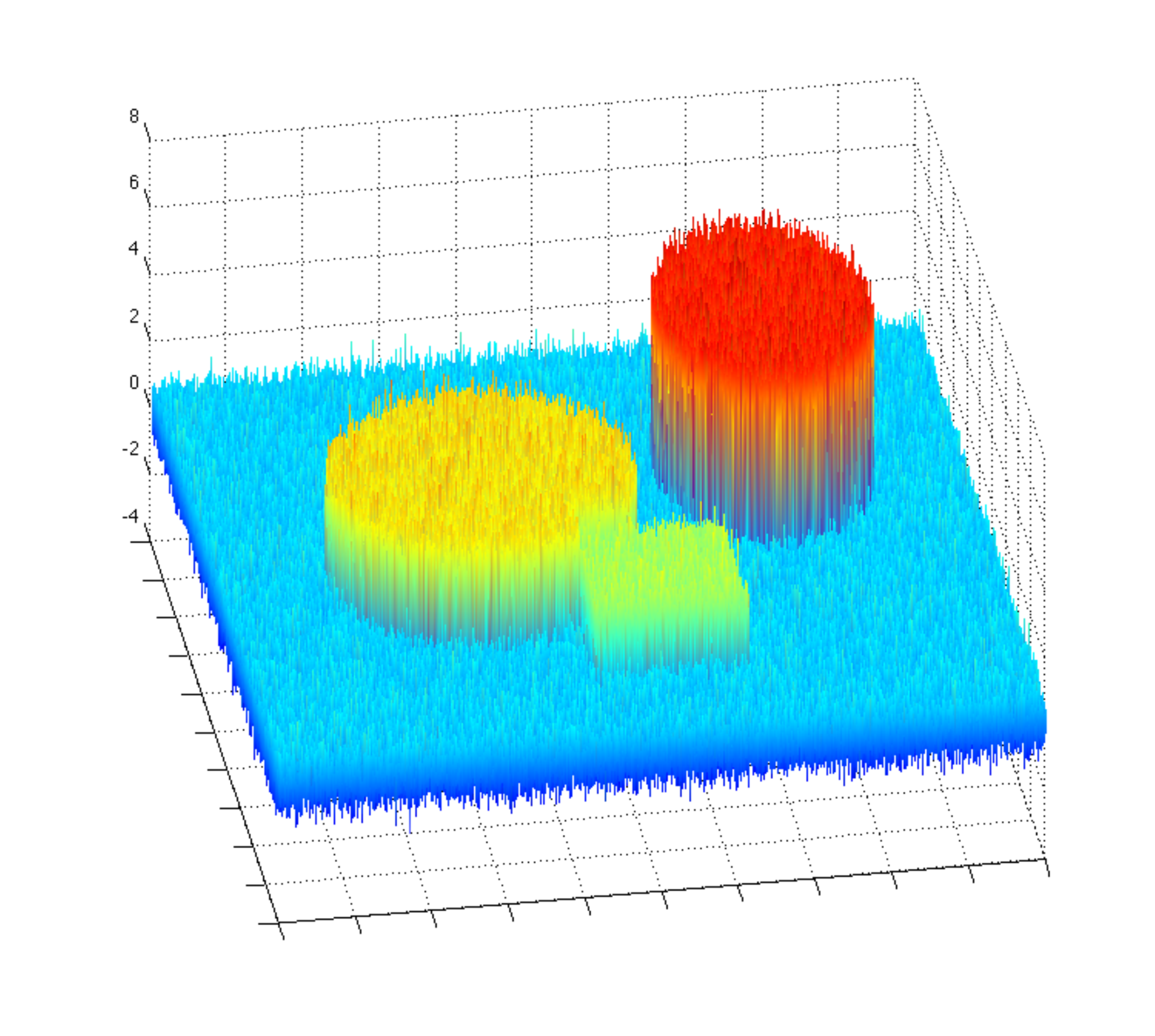}}&
  {\includegraphics[width=0.45\hsize]{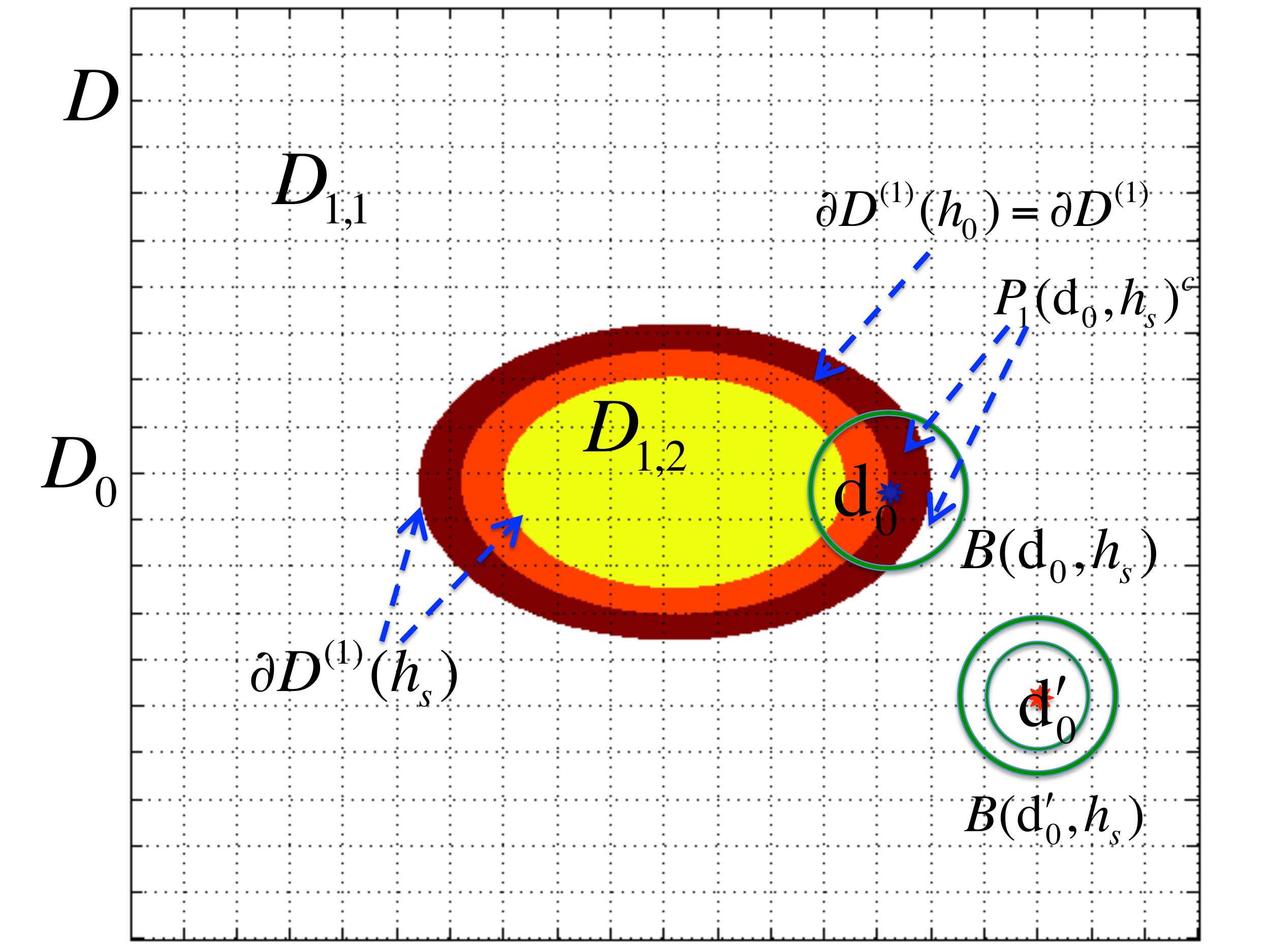}}
 \\
(c) & ({\bf d})
\end{tabular}
\end{center}
\vspace*{0pt}
\caption{Illustration of a jumping surface model for $\beta_1({\bf d})$ and boundary sets over a two-dimensional region $D$:     (a)  $\mathcal D$, $\mathcal D_0$, a  disjoint partition of $\mathcal D$ as the union of four disjoint regions with white, yellow,   blue green, and red representing $\mathcal D_{1, 1},$ $\mathcal D_{1, 2},$
$\mathcal D_{1, 3},$ and $\mathcal D_{1, 4},$ 
a representative voxel ${\bf d}_0\in \mathcal D_0$, an open ball of ${\bf d}_0$, $B({\bf d}_0, h)$, a maximal path-connected set $P_1({\bf d}_0, h)$, and $P_1({\bf d}_0, h)^c$;  (b)  three-dimensional shaded surface of  true $\{\beta_1({\bf d}): {\bf d}\in {\mathcal D}\}$ map;  (c)  three-dimensional shaded surface of  estimated $\{\hat\beta_1({\bf d}_0): {\bf d}_0\in {\mathcal D}_0\}$ map; and ({  d})     $\mathcal D$, $\mathcal D_0$, a  disjoint partition of $\mathcal D={\mathcal D}_{1, 1}\cup {\mathcal D}_{1, 2}$,   $\partial D^{(1)}(h_0)\subset \partial D^{(1)}(h_s)$,
two representative voxels ${\bf d}_0$ and ${\bf d}_0'$ in  $\mathcal D_0$, two open balls of ${\bf d}_0'\in {\mathcal D}_{1, 1}$,  an open ball of ${\bf d}_0\in \partial D^{(1)}(h_s)\cap \mathcal D_0$, $B({\bf d}_0, h_s)$,  and $P_1({\bf d}_0, h_s)^c$. }
\label{tract3newfig2}
\end{figure}

The second component of the SVCM is a jumping surface model for  each of  $\{\beta_j({\bf d}): {\bf d}\in {\mathcal D}\}_{j\leq p}$.   Imaging data $\{y_i({\bf d}_0): {\bf d}_0\in {\mathcal D}_0\}$
  can usually be regarded as a noisy version of a piecewise-smooth function of ${\bf d}\in {\mathcal D}$ with jumps or edges.  In many neuroimaging data,  those jumps or edges often reflect the   functional and/or structural changes, such as white matter and gray matter,  across the brain.
Therefore,  the varying  function  $\{\beta_j({\bf d}): {\bf d}\in {\mathcal D}\}$
in model  (\ref{SVCMeq1})
may
inherit the piecewise-smooth feature from imaging data for $j=1, \ldots, p$, but allows to have different jumps and edges. 
%Furthermore, it is more reasonable to assume that different $\{\beta_j({\bf d}): {\bf d}\in {\mathcal D}\}$ images have different jumps or edges, since
%  different covariates  may play different roles in characterizing the piecewise-smooth pattern in imaging data.
%
%
%
%Specifically,  the jumping surface model for  $\{\beta_j({\bf d}): {\bf d}\in\mathcal D\}$ makes three assumptions including disjoint partition, piecewise smoothness, and local patch as follows.
Specially, we make the following assumptions.
\begin{itemize}
\item{(i)} (Disjoint Partition) There is a finite and
disjoint partition   $\{{\mathcal D}_{j,l}: l=1, \cdots, L_j\}$ of
${\mathcal D}$ such that each ${\mathcal D}_{j, l}$ is a connected region
of ${\mathcal D}$ and its interior, denoted by ${\mathcal D}_{j,l}^o$,
is nonempty,
%${\mathcal D}_{j, l}\cap {\mathcal D}_{j, l'}=\emptyset$ if
%$l\not=l'$ and $ {\mathcal D}=\cup_{l=1}^{L_j} {\mathcal D}_{j, l}$,
where $L_j$ is a fixed, but unknown integer. See Figure  \ref{tract3newfig2}  (a),  (b), and (d) for an illustration.
\item{(ii)} (Piecewise Smoothness)
$\beta_j({\bf d})$ is a smooth function of ${\bf d}$ within each ${\mathcal D}_{j, l}^o$ for $l=1, \ldots, L_j$, but $\beta_j({\bf d})$ is discontinuous on  $
\partial {\mathcal
D}^{(j)}=
{\mathcal
D}\setminus[\cup_{l=1}^{L_j} {\mathcal D}_{j, l}^o]$, which is the union of
the boundaries of all ${\mathcal D}_{j,l}$. 
See Figure  \ref{tract3newfig2}  (b) for an illustration.
%Thus,  $\cup_{l=1}^{L_j} {\mathcal D}_{j, l}^o$, is the union of
%continuity regions of $\beta_j({\bf d})$ as a function of $d$, whereas the
%boundary of ${\mathcal D}\setminus[\cup_{l=1}^{L_j} {\mathcal D}_{j, l}^o]$
%is the jumping surface of $\beta_j({\bf d})$. Note that $\beta_j({\bf d})$ in distant regions
%${\mathcal D}_{j, l}$ are allowed to have similar  signals. 

\item{(iii)} (Local Patch)
For any ${\bf d}_0\in {\mathcal D}_0$ and $h>0$, let $B({\bf d}_0, h)$ be 
an open ball of ${\bf d}_0$ with radius $h$ 
%It is assumed that $B({\bf d}_0, h)=P_j({\bf d}_0, h)\cup P_j({\bf d}_0, h)^c$ for any $h>0$, where $P_j({\bf d}_0, h)^c=B({\bf d}_0, h)\setminus P_j({\bf d}_0, h)$  and 
%${\bf d}_0$ belongs to 
%   a maximal path-connected    set  $P_j({\bf d}_0, h)$,  in which    $\beta_j({\bf d})$ is a smooth function of ${\bf d}$, while  $P_j({\bf d}_0, h)$ contains an open set.
and  $P_j({\bf d}_0, h)$  
   a maximal path-connected    set  in $B({\bf d}_0, h)$,  in which    $\beta_j({\bf d})$ is a smooth function of ${\bf d}$. Assume that  $P_j({\bf d}_0, h)$, which will be called a {\sl local patch}, contains an open set.
See Figure  \ref{tract3newfig2}   for a graphical illustration.
\end{itemize}

 The  jumping surface model can be regarded as a generalization of  various models  for delineating  changes at unknown location (or time). See, for example,  \cite{Khodadadi2004} for  
 an annotated bibliography of change point problem and regression.
 The disjoint partition and piecewise smoothness assumptions characterize the shape and  smoothness of $\beta_j({\bf d})$ in $ {\mathcal D}$, whereas
the local patch  assumption primarily characterizes the local shape of $\beta_j({\bf d})$ at each voxel ${\bf d}_0\in {\mathcal D}_0$ across different scales (or radii).  
%The set $P_j({\bf d}_0, h)$ is called the {\it local
%patch} of voxel ${\bf d}_0$ at the scale $h$.
For ${\bf d}_0\in [\cup_{l=1}^{L_j} {\mathcal D}_{j, l}^o]\cap {\mathcal D}_0$, there exists a radius $h({\bf d}_0)$ such that $B({\bf d}_0, h({\bf d}_0))\subset \cup_{l=1}^{L_j} {\mathcal D}_{j, l}^o$. In this case, for 
$h\leq h({\bf d}_0)$, we have 
$P_j({\bf d}_0, h)=B({\bf d}_0, h)$ and $P_j({\bf d}_0, h)^c=\emptyset$, whereas $P_j({\bf d}_0, h)^c$ may not equal the empty set for large $h$ since $B({\bf d}_0, h)$ may cross  different ${\mathcal D}_{j,l}^o$s.
For  ${\bf d}_0\in \partial {\mathcal
D}^{(j)}\cap {\mathcal D}_0$,   $P_j({\bf d}_0, h)^c\not=\emptyset$ for all $h>0$.  Since $P_j({\bf d}_0, h)$ contains an open set for any $h>0$, it eliminates the case of ${\bf d}_0$ being an isolated point. See Figure  \ref{tract3newfig2}  (a) and (d) for an illustration.
%Note that   the partition  of ${\mathcal D}$ and their associated boundaries    are not known  a priori, which are  difficult to infer. 

The last component of the SVCM is a functional principal component analysis model for  $\eta_i({\bf d})$.
%Specifically, it is assumed  that
%$\eta_{i}({\bf d})$ is square-integrable and admits  the Karhunen-Loeve expansion given below.
Let $\lambda_{1}\geq \lambda_{2}\geq \ldots \geq 0$ be
ordered values of the eigenvalues of the linear operator determined by
$\Sigma_{\eta}$  with $\sum_{l=1}^\infty\lambda_{l}<\infty$
and the $\psi_{l}({\bf d})$s' be the corresponding orthonormal
eigenfunctions  (or principal components)
 \citep{LiHsing2010,  MR2278365}.  Then, $\Sigma_\eta$  admits the spectral decomposition: 
\begin{equation}\label{SVCMeq3}
\Sigma_{\eta}({\bf d}, {\bf d}')=\sum_{l=1}^\infty \lambda_{l} \psi_{l}({\bf d})\psi_{l}({\bf d}').
\end{equation}
The eigenfunctions $\psi_{l}({\bf d})$ form an orthonormal basis on the space of square-integrable functions on ${\mathcal D}$, and
$\eta_{i}({\bf d})$  admits the Karhunen-Loeve expansion as follows:
\begin{equation}  \label{SVCMeq4}
\eta_{i}({\bf d})=\sum_{l=1}^\infty \xi_{i,l}\psi_{l}({\bf d}),
\end{equation}
 where
$\xi_{i,l}=\int_{s\in {\mathcal D}} \eta_{i}(s)\psi_{l}(s)d{\mathcal V}(s)$ is referred
to as the $l$-th functional principal component score  of the
$i$th subject, in which $d{\mathcal V}(s)$ denotes the Lebesgue measure.
 The $\xi_{i,l}$ are uncorrelated random variables with $E(\xi_{i,l})=0$ and $E(\xi_{i,l}\xi_{i,k})=\lambda_{l}{\bf 1}(l=k)$.
If $\lambda_{l}\approx 0$ for $l\geq L_S+1$, then model (\ref{SVCMeq1}) can be approximated by
\begin{equation}
\label{SVCMeq5}
 {y}_{i}({\bf d}) \approx {\bf
x}_i^T{\ve \beta}({\bf d})+\sum_{l=1}^{L_S} \xi_{i,l}\psi_{l}({\bf d})+\epsilon_{i}({\bf d}).
\end{equation}
In (\ref{SVCMeq5}), since $\xi_{i, l}$ are random variables and $\psi_l({\bf d})$ are `unknown'  but fixed  basis functions,
 it can be regarded as a {\it varying coefficient spatial mixed effects model}. Therefore, model (\ref{SVCMeq5}) is  a mixed effects representation of model (\ref{SVCMeq1}).

Model (\ref{SVCMeq5}) differs  significantly  from  other models in the existing literature.
Most   varying coefficient models assume some degrees of  smoothness on   varying coefficient functions, while they do not model
  the within-curve
dependence    \citep{WuChiang1998}.   See \citet{MR2425354} for a
comprehensive   review of  varying coefficient models.
Most spatial mixed effects models in spatial statistics  assume that  spatial basis functions are known and  regression coefficients do not vary across ${\bf d}$
    \citep{Cressie2011}.
Most  functional principal component analysis models focus on  characterizing spatial correlation among
multiple observed functions when ${\mathcal D}\in R^1$ \citep{Zipunnikov2011, Ramsay2005, MR2278365}.
%Due to these major differences, we are facing many challenges in accurately estimating piecewisely smoothed  $\ve\beta({\bf d})$ and the eigenvalues and eigenfunctions of $\Sigma_\eta({\bf d}, {\bf d}')$ for model (\ref{SVCMeq1}), which will be solved below.

\subsection{Three-stage Estimation Procedure}

%To estimate the  coefficient functions  in ${\bf B}({\bf d})$, we develop an adaptive  {\it local polynomial kernel
%smoothing } technique  \cite{Fan1996, Wand1995, Wu2006, Ramsay2005, Welsh2006, Zhang2007}.

\begin{figure}[ht!]
\vspace*{0pt}
\begin{center}
\includegraphics[width=.95\textwidth]{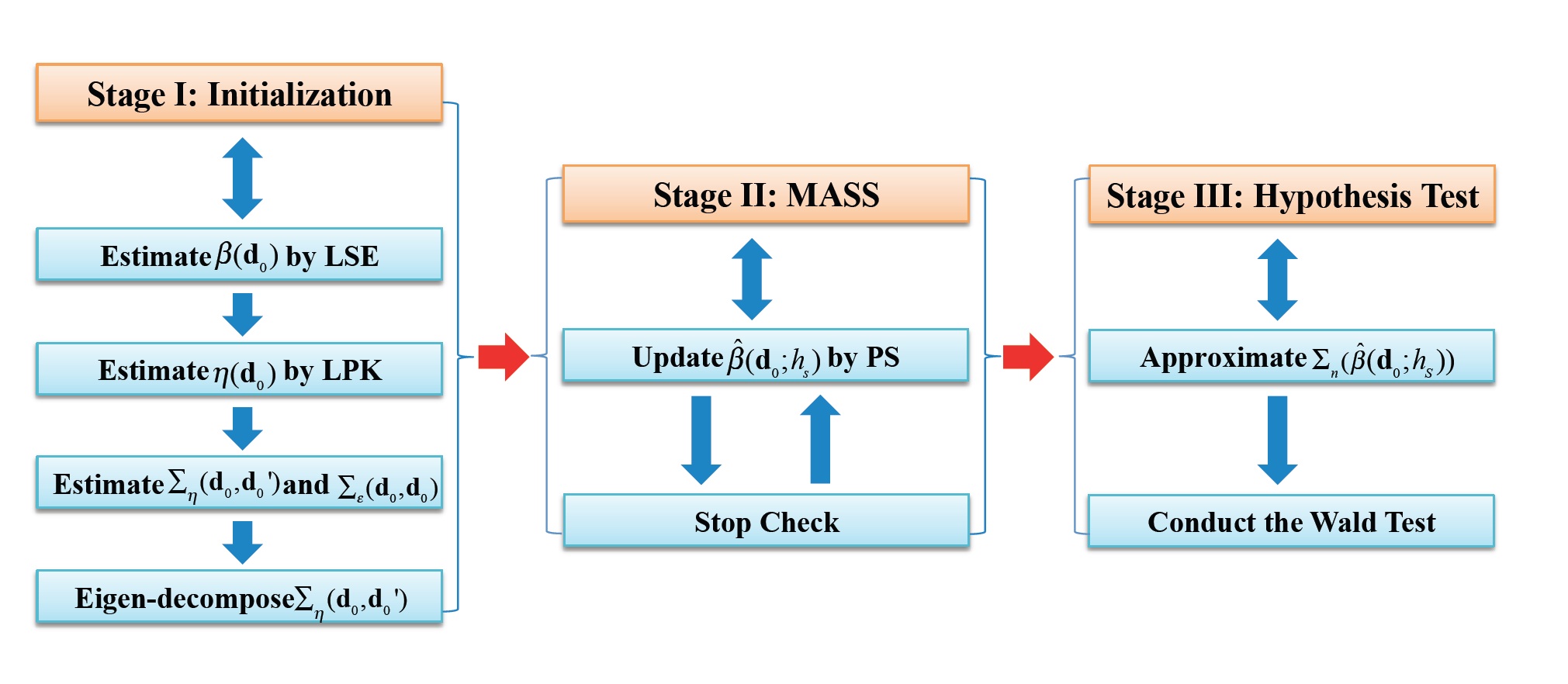}
\end{center}
\vspace*{0pt}
\caption{A schematic overview   of the three stages of SVCM: Stage (I) is the initialization step, Stage (II) is the Multiscale Adaptive and Sequential Smoothing (MASS) method, and Stage (III) is the hypothesis test. }
\label{figpara}
\end{figure}

  We develop a three-stage estimation
procedure as follows. See Figure \ref{figpara} for a schematic overview of SVCM. 
%Specifically,
%the  three stages include Stage (I): an initial estimation; Stage (II): a multiscale adaptive and sequential smoothing method;  and Stage (III): a test procedure.
%The key ideas of each stage are given as follows.
\begin{itemize}
\item{Stage (I)}:  Calculate the
 least squares estimate of ${\ve\beta}({\bf d}_0)$, denoted by $\hat{\ve\beta}({\bf d}_0)$,  across all voxels in ${\mathcal D}_0$, and 
estimate  $\{\Sigma_\epsilon({\bf d}_0, {\bf d}_0):  {\bf d}_0\in \mathcal D_0\}$,    $\{\Sigma_\eta({\bf d}, {\bf d}'): ({\bf d}, {\bf d}')\in {\mathcal D}^2\}$ and  its eigenvalues and eigenfunctions.
\item{Stage (II)}:  Use the propagation-seperation   method to   adaptively and spatially  smooth  each component of   $\hat{\ve\beta}({\bf d}_0)$ across all ${\bf d}_0\in {\mathcal D}_0$.
%Build   a sequence of nested spheres with
%increasing bandwidths $h_1<\cdots<h_S=r_0$ ranging from the
%smallest bandwidth $h_1$ to the largest bandwidth $h_S=r_0$ at each ${\bf d}\in
%{\mathcal D}$.
%Then we obtain an estimate of $\beta({\bf d})$, denoted by $\hat \beta(d; h_l)=(\hat\beta_1(d; h_l), \ldots, \hat\beta_p(d; h_l))^T$;
\item{Stage (III)}:  Approximate  the asymptotic covariance matrix of the final estimate of $\ve\beta({\bf d}_0)$ and
calculate test statistics  across all voxels ${\bf d}_0\in {\mathcal D}_0$.
\end{itemize}
This is more refined idea than the two-stage procedure proposed in \cite{fanzhang99, fanzhang02}.

  \subsubsection{Stage (I)}

Stage (I) consists of four steps.

Step (I.1) is to   calculate
the least squares estimate of ${\ve\beta}({\bf d}_0)$, which equals
 $\hat{\ve\beta}({\bf d}_0)=\Omega_{X, n}^{-1}\sum_{i=1}^n {\bf x}_iy_i({\bf d}_0)$ across
all voxels ${\bf d}_0\in {\mathcal D}_0$, where    $\Omega_{X, n}=\sum_{i=1}^n {\bf x}_i^{\otimes 2}$, in which
   ${\bf a}^{\otimes 2}={\bf a}{\bf a}^T$ for any vector
${\bf a}$. See Figure  \ref{tract3newfig2}  (c)  for a graphical illustration of $\{ \hat{\ve\beta}({\bf d}_0): {\bf d}_0\in \mathcal D_0\}$.

Step (I.2)  is to estimate $\eta_i({\bf d})$ for all ${\bf d}\in {\mathcal D}$. We  employ the   local linear regression technique to estimate
all individual functions  $\eta_{i}({\bf d})$.
% \cite{Fan1996, Wand1995, Wu2006, Ramsay2005, Welsh2006, Zhang2007}.
Let $\partial_d{\eta}_{i}({\bf d})=\partial\eta_i({\bf d})/\partial {\bf d}$,   $C_{i}({\bf d})=({\eta}_{i}({\bf d}),
h\partial_d{\eta}_{i}({\bf d})^T)^T$, and     ${\bf z}_{h}({\bf d}_m-{\bf d})=(1, (d_{m, 1}-d_1)/h, (d_{m, 2}-d_2)/h, (d_{m, 3}-d_3)/h)^T$, where ${\bf d}=(d_{1}, d_{ 2}, d_{ 3})^T$ and ${\bf d}_m=(d_{m,1}, d_{m, 2}, d_{m, 3})^T\in {\mathcal D}_0$.  We use  Taylor series   expansion to expand ${\eta}_{i}({\bf d}_m)$
at ${\bf d}$ leading to   $$
\eta_{i}({\bf d}_m)=C_{i}({\bf d})^T{\bf z}_{h}({\bf d}_m-{\bf d}).
$$    We
develop an algorithm to estimate $C_{i}({\bf d})$ as follows.
 Let   $K_{loc}(\cdot)$ be a univariate kernel function and
$K_h({\bf d}_m-{\bf d})=h^{-3}   \prod_{k=1}^3K_{loc}((d_{m,k}-d_k)/h)$ be the rescaled kernel function with
a bandwidth $h$. For each  $i$, we estimate $C_{i}({\bf d})$ by minimizing the
weighted least squares function given by
$$
\hat C_i({\bf d})=\mbox{argmin}_{C_i({\bf d})} \sum_{{\bf d}_m \in {\mathcal D}_0} \{r_i({\bf d}_m)-C_{i}({\bf d})^T{\bf z}_{h}({\bf d}_m-{\bf d})\}^2K_{h}({\bf d}_m-{\bf d}).
$$
where $r_i({\bf d}_m)=
{y}_{i}({\bf d}_m)-{\bf x}_i^T\hat {\ve\beta}({\bf d}_m)$. It can be shown that  
\begin{equation}  \label{SVCMeq7n1}
\hat C_{i}({\bf d}) = \{\sum_{{\bf d}_m\in {\mathcal D}_0 }K_{h}({\bf d}_m-{\bf d}){\bf
z}_{h}({\bf d}_m-{\bf d})^{\otimes 2}\}^{-1} \sum_{{\bf d}_m\in {\mathcal D}_0 } K_{h}({\bf d}_m-{\bf d}){\bf z}_{h}({\bf d}_m-{\bf d})r_i({\bf d}_m), 
\end{equation}
Let 
 $\hat R_i=(r_i({\bf d}_0): {\bf d}_0\in {\mathcal D}_0)$ be an $N_D\times 1$ vector of estimated residuals and notice that $\hat {\ve\eta}_i(\bf d)$ is the first component of $C_{i}({\bf d})$. Then,   we have   
\begin{equation}
\hat {\ve\eta}_i=(\hat \eta_{i}({\bf d}_0): {\bf d}_0\in {\mathcal D}_0)=S_i\hat R_i~~\mbox{and}~~  
\hat \eta_i({\bf d})=(1, 0, 0, 0)\hat C_i({\bf d}),   \label{SVCMeq8}
\end{equation}
where  $S_i$ is an  $N_D\times N_D$  smoothing matrix \citep{Fan1996}.
We  pool the data from all $n$ subjects and select the
optimal bandwidth $h$, denoted by $\tilde h$, by
minimizing the generalized cross-validation (GCV) score given by
\begin{equation} \label{SVCMeq9}
\mbox{GCV}(h)=\sum_{i=1}^n \frac{\hat R_{i}^T(I_{D}-S_{i})^T(I_{D}-S_{i})\hat R_{i}
}{[1-N_D^{-1}\mbox{tr}(S_{i})]^2},
\end{equation}
where $I_D$ is an $N_D\times N_D$ identity matrix.
Based on  $\tilde h$, we can use (\ref{SVCMeq8}) to
estimate $\eta_{i}({\bf d})$  for all $i$.

Step (I.3)  is to estimate $\Sigma_\eta({\bf d}, {\bf d}')$ and $\Sigma_\epsilon({\bf d}_0, {\bf d}_0)$.
Let   $\hat
{\epsilon}_i({\bf d}_0)= {y}_{i}({\bf d}_0)- {\bf
x}_i^T\hat {\ve\beta}({\bf d}_0)-\hat{\eta}_{i}({\bf d}_0) $ be estimated residuals for $i=1,
\ldots, n$ and ${\bf d}_0\in {\mathcal D}_0$. We estimate
$\Sigma_\epsilon({\bf d}_0, {\bf d}_0)$ by
\begin{equation} \label{SVCMeq11}
\hat\Sigma_\epsilon({\bf d}_0, {\bf d}_0)=
n^{-1}\sum_{i=1}^n
\hat{\epsilon}_{i}({\bf d}_0)^{
2}
\end{equation}
and
$\Sigma_\eta({\bf d}, {\bf d}')$ by  the sample covariance matrix:
\begin{equation}\label{SVCMeq10}
\hat\Sigma_\eta({\bf d}, {\bf d}')=(n-p)^{-1}\sum_{i=1}^n \hat {\eta}_{i}({\bf d})
 \hat {\eta}_{i}({\bf d}').
\end{equation}
% When $N_D$ is   large, it can be numerically  infeasible to  calculate $\hat \Sigma_\eta({\bf d}, {\bf d}')$ for all pairs of ${\bf d}, {\bf d}'\in {\mathcal D}$ and thus  we take an alternative approach given  in Step (I.4) below.

 Step (I.4) is to estimate   the eigenvalue-eigenfunction pairs of $\Sigma_\eta$ by using the singular value decomposition.
  Let  ${\bf V}=[\hat {\ve\eta}_{1}, \cdots, \hat{\ve\eta}_{n}]$ be an $N_ D\times n$ matrix.
Since   $n$ is much smaller than $N_D$,
we can easily calculate the eigenvalue-eigenvector pairs of the $n\times n$ matrix  ${\bf V}^T{\bf V}$, denoted by $\{(\hat\lambda_i, \hat{\ve \xi}_i): i=1, \cdots, n\}$.
It can be shown that $\{(\hat\lambda_i, {\bf V}\hat{\ve \xi}_i): i=1, \cdots, n\}$  are the eigenvalue-eigenvector pairs of the $N_D\times N_D$ matrix ${\bf VV}^T$.
In applications, one usually considers  large  $\hat\lambda_l$ values, while dropping  small $\hat\lambda_l$s.
It is common to choose
  a value of ${L_S}$  so that the cumulative eigenvalue  $\sum_{l=1}^{L_S}\hat\lambda_l/\sum_{l=1}^n\hat\lambda_l$ is above a prefixed threshold, say 80\% \citep{Zipunnikov2011, LiHsing2010, MR2278365}.
 Furthermore,   the
$l$th SPCA scores   can be computed
using
\begin{equation} \label{SVCMeq17}
\hat\xi_{i,l}=\sum_{m=1}^{N_D}\hat\eta_{i}({\bf d}_m)\hat\psi_{
l}({\bf d}_m){\mathcal V}({\bf d}_{m})
\end{equation}
for $l=1, \ldots, {L_S}$,
 where ${\mathcal V}({\bf d}_{m})$ is the volume of voxel ${\bf d}_m$.

  \subsubsection{Stage (II)}
Stage (II) is  a multiscale adaptive and sequential smoothing (MASS) method.
 The key idea of MASS is to use the propagation-separation  method \citep{Polzehl2000, Polzehl2006} to individually   smooth
each    least squares estimate image $\{\hat\beta_j({\bf d}_0): {\bf d}_0\in {\mathcal D}_0\}$ for $j=1, \ldots, p$.
  MASS starts with building   a sequence of nested spheres with
increasing bandwidths $0=h_0<h_1<\cdots<h_S=r_0$ ranging from the
smallest bandwidth $h_1$ to the largest bandwidth $h_S=r_0$ for each ${\bf d}_0\in
{\mathcal D}_0$. At bandwidth $h_1$, based on the information contained in
$\{\hat{\ve\beta}({\bf d}_0): {\bf d}_0\in {\mathcal D}_0\}$, we sequentially  calculate adaptive weights $\omega_j({\bf d}_0, {\bf d}_0'; h_1)$ between voxels ${\bf d}_0$ and ${\bf d}_0'$, which depends on the distance $\|{\bf d}_0 - {\bf d}_0\|$ and spacial similarity $|\hat{\beta}_j ({\bf d}_0) - \hat{\beta}_j ({\bf d}_0)|$,
and  update  $\hat{\beta}_j({\bf d}_0; h_1)$   for all ${\bf d}_0\in {\mathcal D}_0$ for $j=1, \cdots, p$.
At   bandwidth $h_2$, we repeat the same process using $\{\hat{\ve\beta}({\bf d}_0; h_1): {\bf d}_0\in {\mathcal D}_0\}$ to compute spatial similarities.
  In this way,
 we can sequentially determine  $\omega_j({\bf d}_0, {\bf d}_0'; h_s)$ and  $\hat{\beta}_j({\bf d}_0; h_s)$ for each component of ${\ve\beta}({\bf d}_0)$
as the bandwidth ranges  from $h_1$ to
$h_S=r_0$.
% A path diagram  of MASS is given
%below:
%$$
%\begin{array}{c}
%\hat \beta_1({\bf d}_0) \\
% \vdots \\
% \hat\beta_p({\bf d}_0)
% \end{array}  \left. \begin{array}{c}
%\rightarrow \\
% \vdots \\
%\rightarrow
% \end{array}\right.
%  \begin{array}{c}
%  \omega_1({\bf d}_0, {\bf d}_0'; h_1) \\
% \vdots \\
%  \omega_p({\bf d}_0, {\bf d}_0'; h_1)
% \end{array}
%  \left. \begin{array}{c}
%\rightarrow \\
% \vdots \\
%\rightarrow
% \end{array}\right.
% \begin{array}{c}
%\hat \beta_1({\bf d}_0; h_1) \\
% \vdots \\
% \hat\beta_p({\bf d}_0; h_1)
% \end{array}
%\left. \begin{array}{c}
%\cdots   \\
% \vdots \\
%\cdots
% \end{array}\right.
%       \begin{array}{c}
% \omega_1({\bf d}_0, {\bf d}_0'; h_S) \\
% \vdots \\
% \omega_p({\bf d}_0, {\bf d}_0'; h_S)
% \end{array}
% \left. \begin{array}{c}
%\rightarrow \\
% \vdots \\
%\rightarrow
% \end{array}\right.
% \begin{array}{c}
%\hat \beta_1({\bf d}_0; h_S) \\
% \vdots \\
% \hat\beta_p({\bf d}_0; h_S)
% \end{array}
%$$
Moreover, as shown below,
we have found a simple way of calculating the standard deviation of $\hat{\beta}_j({\bf d}_0; h_s)$.

%  MASS integrates the key features  of   the original propagation-separation method developed for a standard nonparametric model  of  a single image
%  \citep{Polzehl2006, Polzehl2000}.
% These two existing methods   use the same weight for all components of
%  $\ve\beta({\bf d})$ and thus they can only  update the whole $\ve\beta({\bf d})$ together.   It is difficult to generalize them to independently update each component of $\ve\beta({\bf d})$,
%  since calculating   standard deviations of the adaptive estimates  of $\ve\beta({\bf d})$ can be very challenging under this scenario.  In contrast,
%   MASS sequently
%  determines weights  between  any pair of voxels  solely for the $j-$th component of $\ve\beta({\bf d})$
% and only updates  $\hat{\beta}_j({\bf d}_0; h_s)$ without changing the other components of $\ve\beta({\bf d})$ for  each $j=1, \cdots, p$.

%At each iteration, the computation  involved  for  MASS is minor since  it  the
%$p$ times  as that for the voxel-wise approach. Thus, this
%multiscale adaptive method provides an efficient /Users/hzhu/Documents/my papers/volumeStatistics/Theorey_update/2012_04_08_SimulationWriting/SVCM-1.pdfmethod for  flexibly
%exploring the neighboring areas of each voxel for each component of ${\ve\beta}({\bf d})$. Since MANE
%sequentially includes more data at each iteration, it will
%adaptively increase the statistical  efficiency in estimating
 % ${\beta}_j({\bf d})$ in a homogenous region and decrease  the variation
%of the weights $\omega_j({\bf d}_0, {\bf d}_0'; h)$ for $j=1, \cdots, p$.

MASS consists of three steps including (II.1) an initialization step, (II.2) a sequentially  adaptive estimation step,  and (II.3) a stop checking step,
each of which involves in  the specification of several parameters.  Since  propagation-separation and  the choice of their associated parameters  have been discussed in details in
\cite{Polzehl2010} and \cite{Li2011}, we briefly mention them here for the completeness.  In the initialization step (II.1), 
we take a geometric series $\{
h_{s}=c_h^s: s=1, \ldots, S\}$ of radii with $ h_{0}=0$, where
$c_h > 1$, say $c_h=1.10$.
 We suggest relatively small  $c_h$
to prevent incorporating too many neighboring voxels.
   % We then set $s=1$  and $  h_{1}=c_h$.

In the sequentially   adaptive estimation step (II.2), starting from $s=1$  and $  h_{1}=c_h$, at step $s$,
we compute spatial adaptive locally weighted average 
estimate  $\hat {\beta}_j({\bf d}_0; h_s)$ based on $\{\hat\beta_{j}({\bf d}_0): {\bf d}_0\in {\mathcal D}_0\}$ and $\{\hat\beta_{j}({\bf d}_0;  h_{s-1}): {\bf d}\in {\mathcal D}_0\}$, where
$\hat\beta_j({\bf d}_0;  h_0)=\hat\beta_j({\bf d}_0)$. 
 Specifically,  for each $j$,
  we construct
   a weighted quadratic function
 \begin{equation}  \label{SVCMeq11new}
   \ell_n({\beta}_j({\bf d}_0); h_s)=\sum_{ {\bf d}_m\in
   B({\bf d}_0, h_s)\cap {\mathcal D}_0}\{\hat \beta_{j}({\bf d}_m)-\beta_j({\bf d}_0)\}^2 \omega_j({\bf d}_0, {\bf d}_m; h_s),
 \end{equation}
 where $\omega_j({\bf d}_0, {\bf d}_m; h_s)$, which will be defined below,  characterizes the similarity between
 $\hat\beta_{j}({\bf d}_m; h_{s-1})$ and $\hat\beta_j({\bf d}_0;  h_{s-1})$.
We then  calculate
\begin{equation}\label{SVCMeq12}
\hat\beta_j({\bf d}_0; h_s)
 =\mbox{argmin}_{\beta_j({\bf d}_0)} \ell_n({\beta}_j({\bf d}_0); h_s)
={\sum_{{\bf d}_m\in B({\bf d}_0, h_s)\cap {\mathcal D}_0}\tilde \omega_j({\bf d}_0, {\bf d}_m; h_s) \hat\beta_j({\bf d}_m)},
\end{equation}
where
$\tilde\omega_j({\bf d}_0, {\bf d}_m; h_s)=\omega_j({\bf d}_0, {\bf d}_m; h_s)/\sum_{{\bf d}_{m'}\in B({\bf d}_0, h_s)\cap {\mathcal D}_0}\omega_j({\bf d}_0, {\bf d}_{m'}; h_s)$. 
%a weighted sum of the raw estimates $\{\hat\beta_j({\bf d}_m): {\bf d}_m\in B({\bf d}_0, h_s)\}$.
%Equation (\ref{SVCMeq12}) is equivalent to using the original propagation-separation to smooth  each
%$\{\hat\beta_j({\bf d}): {\bf d}\in {\mathcal D}_0\}$ image.  We will show below that (\ref{SVCMeq12}) leads to   theoretical simplicity and numerical  stablity.

Let
$\Sigma_n(\hat{\beta}_j({\bf d}_0; h_{s}))$ be the asymptotic variance of $\hat\beta_j({\bf d}_0; h_{s})$.
For  ${\beta}_j({\bf d}_0)$,
we compute the similarity between voxels ${\bf d}_0$ and ${\bf d}_0'$, denoted by
  $D_{\beta_j}({\bf d}_0, {\bf d}_0'; h_{s-1})$, and
   the adaptive weight $\omega_j({\bf d}_0, {\bf d}_0'; h_s)$, which are, respectively, defined as
 \begin{eqnarray}\label{SVCMeq13}
D_{\beta_j}({\bf d}_0, {\bf d}_0'; h_{s-1})&=& \{\hat{\beta}_j({\bf d}_0;   h_{s-1})  -
\hat{\beta}_j({\bf d}_0';   h_{s-1})\}^2/\Sigma_n(\hat{\beta}_j({\bf d}_0;
h_{s-1})),\\
  \omega_j({\bf d}_0, {\bf d}_0'; h_s)&=&K_{loc}(||{\bf d}_0-{\bf d}_0'||_2/h_{s})K_{st}(D_{{\beta}_j}({\bf d}_0, {\bf d}_0'; h_{s-1})/C_{n}), \nonumber
\end{eqnarray}
 where   $K_{st}(u)$  is   a nonnegative kernel function  with compact support,
$C_{n}$ is a tuning parameter depending on $n$, and
$||\cdot||_2$ denotes the Euclidean norm of a vector.

%The choice of the two kernel functions $K_{st}(u)$ and $K_{loc}(u)$ is critical for the propagation-separation  method.
 The weights $K_{loc}(||{\bf d}_0-{\bf d}_0'||_2/h_{s})$ give
less weight to the voxel ${\bf d}_0'$ that is far
from the voxel ${\bf d}_0$.
The weights $K_{st}(u)$   downweight  the voxels ${\bf d}_0'$ with large  $D_{{\beta}_j}({\bf d}_0, {\bf d}_0'; h_{s-1})$, which indicates a large difference between  $\hat{\beta}_j({\bf d}_0'; h_{s-1})$ and
$\hat{\beta}_j({\bf d}_0;  h_{s-1})$.
%Therefore, $ \omega_j({\bf d}_0, {\bf d}_0'; h_s)$  primarily depends on the $j-$th component of $\ve\beta({\bf d})$ across ${\bf d}\in \mathcal D_0$.
In practice, we set   $K_{loc}(u)=(1-u)_{+}$.
Although
different choices of
   $K_{st}(\cdot)$ have been suggested in
   the
 propagation-separation method
   \citep{Polzehl2000, Polzehl2006, Polzehl2010, Li2011}, we have tested these   kernel functions  and found that $K_{st}(u)=\exp(-u)$ performs reasonably well. Another good choice of $K_{st}(u)$ is
$\min(1, 2(1-u))_+$.
Moreover, theoretically, as shown in \cite{Scott1992} and \cite{MR1212173}, they have examined the efficiency of different kernels for weighted least squares estimators, but
extending their results to the  propagation-separation method  needs some further investigation.

 The scale $C_n$ is used to penalize the similarity between any two voxels ${\bf d}_0$ and ${\bf d}_0'$ in a similar manner to bandwidth, and an appropriate choice of  $C_n$  is crucial for the behavior of the propagation-separation  method.    
  As discussed in \citep{Polzehl2000, Polzehl2006},  a propagation condition independent of the observations at hand can be used to specify $C_n$.  
  The basic idea of the propagation condition is that the impact of the statistical penalty in $K_{st}(D_{{\beta}_j}({\bf d}_0, {\bf d}_0'; h_{s-1})/C_{n})$ should be negligible under a homogeneous model $\beta_j({\bf d})\equiv \mbox{constant}$ yielding almost free smoothing within homogeneous regions.  However, we take an alternative approach to choose $C_n$ here. Specifically,  a good choice of  $C_n$ should balance between the sensitivity and specificity of MASS.  Theoretically, as shown in Section 2.3,  $C_n$ should satisfy $C_n/n=o(1)$ and $C_n^{-1}\log(N_D)=o(1)$. 
%When imaging measures in  voxels $d$ and $d'$ are  similar to each other, which leads to a small value of $D({\bf d}_0, {\bf d}_0'; h_s)$,    a small
%  $C_n$ can lead to a relative large $D({\bf d}_0, {\bf d}_0'; h_s)/C_n$ and  thus it
%    may decrease    the specificity of MASS in combining such similar voxels
%    in the interior
%  of each $\mathcal D_{j, l}^o$.
%However, if   a small similarity exists between the voxels $d$ and $d'$,  a large $C_n$  leads to small $D({\bf d}_0, {\bf d}_0'; h_s)/C_n$ and thus it  may decrease
%the sensitivity of MASS in separating these voxels.
We choose $C_n=n^{0.4}\chi_1^2(0.8)$ based on our experiments, where  $\chi_1^2(a)$ is the upper
$a$-percentile of the $\chi_1^2$-distribution.

% We calculate a weighted least squares  estimate of ${A}_j({\bf d})$ as
%follows.
%      Let   $K(\cdot)$ be a kernel function, such as the
%Gaussian   and  uniform kernels \cite{Fan1996, Wand1995}.
%For   a fixed
%bandwidth $h$ and each $k$,

%where   $K_{h_{j}}(\cdot)=K(\cdot/h_{j})/h_{j}$ is a rescaled kernel function.

%Let us now introduce some matrix operators.
%Let ${\bf a}^{\otimes 2}={\bf a}{\bf a}^T$ for any vector ${\bf a}$
%and $C\otimes D$ be the Kronecker product of   two matrices $C$ and
%$D$. For an $M_1\times M_2$ matrix $C=(c_{j,l})$, denote
%$\mbox{vec}(C)=(c_{1,1}, \ldots, c_{M_1, 1},  \ldots, c_{1, M_2},
%\ldots, c_{M_1, M_2})^T$. Let  $\hat {A}_j({\bf d})$ be the minimizer of
%(\ref{SVCMeq5}). Then, $ \hat {A}_j({\bf d}_0; h_s) $ is given by
%the weighted least squares estimate of ${A}_j({\bf d})$.
%We can show that
% $\mbox{vec}(\hat {A}_j({\bf d}))=(\hat b_{j1}({\bf d}), \hat{\dot b}_{j1}({\bf d}), \ldots,
% \hat  b_{jp}({\bf d}), \hat {\dot b}_{jp}({\bf d}))^T$
%is given by
%\begin{equation}
%\label{SVCMeq6}
%\Sigma_j({\bf d}_0; h_{s})^{-1}
%\sum_{i=1}^n\sum_{{\bf d}_m\in B(d,
%h_s)}  \tilde\omega_j(d, {\bf d}_m; h_s)
%[{x}_{ij}{\bf z}_{h_{s}}({\bf d}_m-{\bf d})] [y_{i}({\bf d}_m)-\sum_{k\not=j} x_{ik}\hat \beta_k({\bf d}_m, h_{s-1})],
%\end{equation}
%where $\Sigma_j({\bf d}_0; h_{s})=\sum_{i=1}^n\sum_{{\bf d}_m\in B({\bf d}_0, h_s)}
%\tilde\omega_j(d, {\bf d}_m; h_s)[{x}_{ij}^2  {\bf
%z}_{h_{s}}({\bf d}_m-{\bf d})^{\otimes 2}]$. Thus, we have
 %\begin{equation}
 %\label{Tract3eq8}
%\hat {\beta}_j({\bf d}_0; h_s)=  (1, 0)    \hat{ A}_j({\bf d}_0; h_s).
%\end{equation}

We now calculate
$\Sigma_n(\hat{\beta}_j
({\bf d}_0; h_{s}))$. 
  By treating the weights  $ \tilde\omega_j({\bf d}_0, {\bf d}_m; h_s)$ as `fixed' constants, we can approximate $\Sigma_n(\hat\beta_j({\bf d}_0; h_s))$ by  
\begin{equation}  \label{SVCMeq14}
 \sum_{{\bf d}_m, {\bf d}_{m'}\in B({\bf d}_0, h_s)\cap \mathcal D_0} \tilde\omega_j({\bf d}_0, {\bf d}_m; h_s)\tilde\omega_j({\bf d}_0, {\bf d}_{m'}; h_s) \mbox{Cov}(\hat\beta_j({\bf d}_m), \hat\beta_j({\bf d}_{m'})),
\end{equation}
where $\mbox{Cov}(\hat\beta_j({\bf d}_m), \hat\beta_j({\bf d}_{m'}))$ can be estimated by
\begin{equation} \label{SVCMeq15}
{\bf e}_{j, p}^T\Omega_{X, n}^{-1}{\bf e}_{j, p}\{\hat\Sigma_\eta({\bf d}_m, {\bf d}_{m'})+\hat \Sigma_\epsilon({\bf d}_m, {\bf d}_m){\bf 1}({\bf d}_m={\bf d}_{m'})\},
\end{equation}
in which ${\bf e}_{j, p}$ is a $p\times 1$ vector with the $j$-th element 1 and others $0$. %Then, we can substitute the estimates of $\Sigma_\eta({\bf d}, {\bf d}')$ and $\Sigma_\epsilon(d, d)$ in (\ref{SVCMeq10}) and
%(\ref{SVCMeq11}) into (\ref{SVCMeq14}) and (\ref{SVCMeq15}) to approximate $\Sigma_n(\hat\beta_j({\bf d}_0; h_s))$.
We will examine the consistency of   approximation   
(\ref{SVCMeq14}) later.

In the stop checking step (II.3),
after the first iteration,  we start to calculate a stopping criterion based on a normalized distance between
$\hat{\beta}_j({\bf d}_0)$ and  $ \hat{\beta}_j({\bf d}_0; h_{s})$  given by
 \begin{equation}\label{SVCMeq16}
D(\hat{\beta}_j({\bf d}_0),  \hat{\beta}_j({\bf d}_0; h_{s}))=\{\hat{\beta}_j({\bf d}_0)  - \hat{\beta}_j({\bf d}_0;   h_{s})\}^2
/\Sigma_n(\hat{\beta}_j({\bf d}_0)).
\end{equation}
Then, we  check whether
$\hat{ \beta}_j({\bf d}_0; h_s)$ is in a  confidence ellipsoid of $\hat{\beta}_j({\bf d}_0)$ given by $\{{\beta}_j({\bf d}_0):
D(\hat{\beta}_j({\bf d}_0),  {\beta}_j({\bf d}_0))\leq  C_s\}$, where $C_s$ %is a varying threshold.
%To prevent a large  $D(\hat{\beta}_j({\bf d}_0),  \hat{\beta}_j({\bf d}_0; h_{s}))$,  we set    
is taken as $C_s=\chi_1^2({0.80/s})$ in our implementation.
  If  $D(\hat{\beta}_j({\bf d}_0), \hat {\beta}_j({\bf d}_0; h_{s}))$ is
greater than  $ C_s$,   then  we   set $\hat{\beta}_j({\bf d}_0,
h_S)=\hat{\beta}_j({\bf d}_0, h_{s-1})$
 and $s=S$ for the $j$-th component and voxel ${\bf d}_0$.   If $s=S$ for  all components in all voxels, we stop. If
   $ D(\hat{\beta}_j({\bf d}_0), \hat{\beta}_j({\bf d}_0;
h_{s}))\leq   C_s $, then we    set $
h_{s+1}=c_h h_{s}$, increase $s$ by 1 and continue with the   step (II.1).
It should be noted that different components of $\hat{\ve\beta}({\bf d}_0; h)$ may stop at different bandwidths.

We usually set the maximal step $S$   to be relatively small, say between 10 and 20,  and
thus each $B({\bf d}_0, h_{S})$ only contains
a relatively small number of voxels.
As $S$   increases,
the number of neighboring voxels in $B({\bf d}_0, h_S)$ increases exponentially. It increases  the chance of oversmoothing
$\beta_j({\bf d}_0)$ when ${\bf d}_0$ is near the edge of distinct regions.
% and the parameters change slowly with other
%locations.
Moreover, in order to prevent oversmoothing
$\beta_j({\bf d}_0)$,  we compare $\hat {\beta}_j({\bf d}_0; h_{s})$ with
the least squares estimate $\hat{\beta}_j({\bf d}_0)$ and
 gradually decrease $C_s$ with the number of iteration.

   \subsubsection{Stage (III)}

Based on $\hat{\ve\beta}({\bf d}_0; h_S)$,  we can further construct test statistics to examine
scientific  questions associated with ${\ve\beta}({\bf d}_0)$.
For instance, such questions may compare brain
structure across different groups (normal controls versus patients) or
detect  change in brain structure across time.
 These questions  can be formulated as the
  linear hypotheses about ${\ve\beta}({\bf d}_0)$ given by
\begin{equation} \label{SVCMeq18}
 H_{0}({\bf d}_0): R_1{\ve\beta}({\bf d}_0)= {\bf b}_0~~~ \mbox{vs.}~~~ H_{1}({\bf d}_0): R_1{\ve\beta}({\bf d}_0)\not= {\bf b}_0,
\end{equation}
 where  $R_1$ is an $r\times k$ matrix of full row rank
 and  ${\bf b}_0$ is an $r\times 1$ specified vector.
 We 
 %test the null hypothesis $H_{0}({\bf d}_0): R_1{\ve\beta}({\bf d}_0)={\bf b}_0$  using 
use the Wald test statistic
\begin{equation}\label{SVCMeq19}
W_\beta({\bf d}_0; h)=\{R_1\hat{\ve\beta}({\bf d}_0; h_S)-{\bf b}_0\}^T
\{R_1\Sigma_n(\hat{\ve\beta}({\bf d}_0; h_S))R_1^T\}^{-1}
 \{R_1\hat{\ve\beta}({\bf d}_0; h_S)
-{\bf b}_0\}
\end{equation}
for problem (\ref{SVCMeq18}), 
 where $\Sigma_n(\hat{\ve\beta}({\bf d}_0; h_S))$ is the covariance matrix of $\hat{\ve\beta}({\bf d}_0; h_S)$.
 
  We propose an approximation of $\Sigma_n(\hat{\ve\beta}({\bf d}_0; h_S))$. According to (\ref{SVCMeq12}), we know that
  $$
  \hat{\ve\beta}({\bf d}_0; h_S)=\sum_{{\bf d}_m\in B({\bf d}_0, h_S)}\tilde {\ve\omega}({\bf d}_0, {\bf d}_m; h_S)\circ \hat{\ve\beta}({\bf d}_m)
  $$
where $\ve a\circ \ve b$ denotes   the Hadamard product  of matrices $\ve a$ and $\ve b$ and $\tilde {\ve\omega}({\bf d}_0, {\bf d}_m; h)$ is a $p\times 1$ vector determined by the weights
$\tilde \omega_j({\bf d}_0, {\bf d}_m; h)$ in Stage II.
Let $J_p$ be the $p^2\times p$ selection matrix \citep{Liu1999}.
Therefore,  $\Sigma_n(\hat{\ve\beta}({\bf d}_0; h_S))$ can be approximated by
\begin{eqnarray*}
&&\sum_{{\bf d}_m, {\bf d}_m'\in B({\bf d}_0, h_S)}\mbox{Cov}(\tilde {\ve\omega}({\bf d}_0, {\bf d}_m; h_S)\circ \hat{\ve\beta}({\bf d}_m), \tilde{\ve \omega}({\bf d}_0, {\bf d}_m'; h_S)\circ \hat{\ve\beta}({\bf d}_m')) \\
&\approx&\sum_{{\bf d}_m, {\bf d}_m'\in B({\bf d}_0, h_S)}\hat\Sigma_y({\bf d}_m, {\bf d}_m')
 J_p^T\{[\tilde {\ve\omega}({\bf d}_0, {\bf d}_m; h_S)\tilde{\ve \omega}({\bf d}_0, {\bf d}_m'; h_S)^T]\otimes \Omega_{X, n}^{-1}\}J_p.
 \end{eqnarray*}

\subsection{Theoretical Results}

We systematically investigate the asymptotic properties of all estimators obtained from the three-stage estimation procedure.
Throughout the paper, we only consider  a finite number of
iterations and bounded $r_0$ for     MASS, since  a
  brain volume is always bounded.
Without otherwise stated, we assume  that $o_p(1)$ and $O_p(1)$  hold  uniformly across all ${\bf d}$ in either  ${\mathcal D}$ or ${\mathcal D}_0$ throughout the paper.
Moreover,  the sample size $n$ and the number of voxels $N_D$ are allowed to diverge to infinity.
  We state the
following theorems, whose detailed assumptions and proofs can be
found in  Section 6 and a supplementary document.

Let $\ve\beta_*({\bf d}_0)=(\beta_{1*}({\bf d}_0), \ldots, \beta_{p*}({\bf d}_0))^T$ be the true value of $\ve\beta({\bf d}_0)$ at voxel ${\bf d}_0$.
  We first establish  the uniform convergence rate of    $\{\hat{\ve\beta}({\bf d}_0): {\bf d}_0\in {\mathcal D}_0\}$.

\noindent {\bf Theorem 1}. {\em Under assumptions (C1)-(C4)
in Section 6, as $n\rightarrow\infty$,  we have
\begin{itemize}
 \item{(i)} $\sqrt{n}[\hat{\ve\beta}({\bf d}_0)-{\ve\beta}_*({\bf d}_0)]\rightarrow^L N({\ve 0}, \Omega_X^{-1} \Sigma_y({\bf d}_0, {\bf d}_0)  ) $ for any ${\bf d}_0\in {\mathcal D}_0$,
 where $\rightarrow^L$ denotes convergence in distribution;
\item{(ii)}  $
 \sup_{{\bf d}_0\in {\mathcal D}_0} ||\hat{\ve\beta}({\bf d}_0)-{\ve\beta}_*({\bf d}_0)||_2=O_p( \sqrt{n^{-1}\log (1+N_D)})
 %=o_p(1).
 $
 \end{itemize}
}

%According to the best of our knowledge,  Theorem 1 (i) is the first result on

{\sc Remark 1}.   Theorem 1 (i) just restates a standard asymptotic normality of the least squares estimate of $\ve\beta({\bf d}_0)$ at any given voxel ${\bf d}_0\in {\mathcal D}_0$.
Theorem 1 (ii) states that the maximum of $||\hat{\ve\beta}({\bf d}_0)-{\ve\beta}_*({\bf d}_0)||_2$ across all ${\bf d}_0\in {\mathcal D}_0$  is at the order of $\sqrt{n^{-1}\log (1+N_D)}$. If $\log (1+N_D)$ is relatively small compared with $n$,  then the estimation errors converge uniformly to zero in probability. In practice, $N_D$ is determined by imaging resolution and its value
 can be much larger than the sample size. For instance, in most applications, $N_D$ can be as large as $100^3$ and $\log(1+N_D)$ is around 15.
 In a study with several hundreds subjects, $n^{-1}\log (1+N_D)$ can be relatively small.
%Moreover, Theorem 1 allows  both $n$ and $N_D$  diverging to infinity.

We next study  the uniform convergence
rate of $\hat\Sigma_\eta$ and its associated eigenvalues and
eigenfunctions. We also establish the uniform convergence of $\hat\Sigma_\epsilon({\bf d}_0, {\bf d}_0)$.

\noindent {\bf Theorem 2}. {\it
 Under assumptions (C1)-(C8)
in Section 6,  we have the following results:
  \begin{eqnarray*}
&(i)&
\sup_{({\bf d}, {\bf d}')\in {\mathcal D}^2}|\hat\Sigma_\eta({\bf d}, {\bf d}')-\Sigma_\eta({\bf d}, {\bf d}')|=
o_p(1); \\
&(ii) & \int_{\mathcal D} [\hat\psi_{l}({\bf d})-\psi_{l}({\bf d})]^2d{\mathcal V}({\bf d})=o_p(1)~ \mbox{and}~
   |\hat\lambda_{l}-\lambda_{l}|= o_p(1)~~\mbox{ for}~~  l=1, \ldots, E;~ \\
    &(iii)&
\sup_{{\bf d}_0\in {\mathcal D}_0}|\hat\Sigma_\epsilon({\bf d}_0, {\bf d}_0)-\Sigma_\epsilon({\bf d}_0, {\bf d}_0)|=
o_p(1);
\end{eqnarray*}
 where    $E$ will be  described in assumption (C8)  and $\hat\psi_{l}({\bf d})$ is the estimated eigenvector, computed from $\hat \psi_{l} =\mbox{\bf V} \boldsymbol{\xi}_l$.

}

{\sc Remark 2}.
 Theorem 2 (i) and (ii) characterize  the uniform weak  convergence
  of $\hat\Sigma_\eta(\cdot, \cdot)$ and the convergence of    $\hat\psi_{l}(\cdot)$ and
$\hat\lambda_{l}$. These results can be regarded as an extension
of  Theorems 3.3-3.6 in \citet{LiHsing2010}, which
established the uniform strong convergence rates of these estimates
 under a simple model. Specifically,  in  \citet{LiHsing2010}, they considered $y_i({\bf d})= \mu({\bf d})+\eta_i({\bf d})+\epsilon_i({\bf d})$ and assumed that $\mu({\bf d})$ is  twice differentiable.
  Another key difference is that in \citet{LiHsing2010}, they
employed all cross products
  $y_{i}({\bf d})y_{i}({\bf d}')$ for ${\bf d}\not = {\bf d}'$ and then used the local polynomial kernel to
estimate $\Sigma_\eta({\bf d}, {\bf d}')$.  In contrast, our approach is computationally  simple and $\hat\Sigma_{\eta}({\bf d}, {\bf d}')$ is positive definite. Theorem 2 (iii) characterizes   the uniform weak  convergence
  of $\hat\Sigma_\epsilon({\bf d}_0, {\bf d}_0)$ across all voxels ${\bf d}_0\in\mathcal D_0$.

To investigate the asymptotic properties of $\hat{\beta}_j({\bf d}_0; h_s)$, we need to  characterize points close to  and far from the boundary set $\partial {\mathcal D}^{(j)}$.
For a given bandwidth $h_s$, we first  define $h_s$-boundary sets:
\begin{equation}
\partial {\mathcal
D}^{(j)}(h_s)=\{ {\bf d}\in {\mathcal D}:  B({\bf d},  h_s)\cap   \partial {\mathcal
D}^{(j)}\not =\emptyset\}~~\mbox{and}~~
\partial {\mathcal
D}^{(j)}_0(h_s)=\partial {\mathcal
D}^{(j)}(h_s)\cap {\mathcal  D}_0.  \label{ThemEq1}
\end{equation}
Thus,  $\partial {\mathcal
D}^{(j)}(h_s)$ can be regarded as a band with radius $h_s$ covering the boundary set $\partial {\mathcal
D}^{(j)}$, while $\partial {\mathcal
D}^{(j)}_0(h_s)$ contains all grid points within such band.
  It is easy to show that for a sequence of bandwidths $h_0=0<h_1<\cdots<h_S$, we have
\begin{equation}
\partial {\mathcal
D}^{(j)}(h_0)=\partial {\mathcal
D}^{(j)}\subset    \cdots \subset \partial {\mathcal
D}^{(j)}(h_S) ~~~\mbox{and}~~~
\partial {\mathcal
D}^{(j)}_0(h_0)\subset    \cdots \subset \partial {\mathcal
D}^{(j)}_0(h_S).  \label{ThemEq2}
 \end{equation}
 Therefore, for a fixed bandwidth $h_s$, any point ${\bf d}_0\in\mathcal D_0$ belongs to either $ {\mathcal D}\setminus \partial {\mathcal
D}^{(j)}(h_s)$ or $\partial {\mathcal
D}^{(j)}(h_s)$. For each ${\bf d}_0\in {\mathcal D}\setminus \partial {\mathcal
D}^{(j)}(h_s)$,  there exists one and only one ${\mathcal D}_{j, l}$ such that
\begin{equation}
B({\bf d}_0,  h_0)\subset \cdots\subset B({\bf d}_0, h_s)\subset {\mathcal D}_{j, l}^o.
\label{ThemEq3}
\end{equation}
See Figure \ref{tract3newfig2} (d) for an illustration.

 We first investigate the asymptotic behavior of $\hat\beta_j({\bf d}_0; h_s)$ when $\beta_{j*}({\bf d})$ is piecewise constant.
 That is, $\beta_{j*}({\bf d})$ is a constant in
  ${\mathcal D}_{j, l}^o$   and for any ${\bf d}'\in \partial {\mathcal D}^{(j)}$,  there exists a ${\bf d}\in \cup_{l=1}^{L_j} {\mathcal D}_{j, l}^o$ such that
 $\beta_{j*}({\bf d})=\beta_{j*}({\bf d}')$.
Let $  \tilde \beta_{j*}({\bf d}_0; h_s)= \sum_{{\bf d}_m\in B({\bf d}_0, h_s)\cap {\mathcal D}_0}\tilde \omega_j({\bf d}_0, {\bf d}_m; h_s) \beta_{j*}({\bf d}_m)$ be the pseudo-true value of $\beta_j({\bf d}_0)$ at scale $h_s$ in voxel ${\bf d}_0$.
For all ${\bf d}_0\in {\mathcal D}\setminus \partial {\mathcal
D}^{(j)}(h_S)$,  we have $  \tilde \beta_{j*}({\bf d}_0; h_s)=\beta_{j*}({\bf d}_0)$ for all $s\leq S$ due to (\ref{ThemEq3}).  In contrast,  for ${\bf d}_0\in \partial {\mathcal
D}^{(j)}(h_S)$, $\tilde \beta_{j*}({\bf d}_0; h_s)$ may vary from $h_0$ to $h_S$. 
%We define  the smallest  absolute value  of   all possible  jumps at scale $h_s$ as 
% $${\ve u}^{(j)}(h_s)= \min\{ |\Delta_{j*}({\bf d}_0, {\bf d}'_0)|:  \Delta_{j*}({\bf d}_0, {\bf d}'_0)\not=0, ~~ {\bf d}_0'\in B({\bf d}_0, h_s)\cap {\mathcal D}_0,~\mbox{and}~{\bf d}_0\in \mathcal D_0\},  $$ 
%where $\Delta_{j*}({\bf d}_0, {\bf d}_0')=\beta_{j*}({\bf d}_0)-\beta_{j*}({\bf d}'_0)$.  
In this case, we are able to   establish  several important theoretical results to
characterize the asymptotic behavior of $\hat{\ve\beta}({\bf d}_0; h_s)$ even when $h_S$ does not converge to zero.
We need additional notation as follows:
   \begin{eqnarray}
  &&
 \hat\Delta_j({\bf d}_0)=\hat\beta_j({\bf d}_0)-\beta_{j*}({\bf d}_0) ~~~\mbox{and}~~~\Delta_{j*}({\bf d}_0, {\bf d}_0')=\beta_{j*}({\bf d}_0)-\beta_{j*}({\bf d}_0'), \nonumber \\  
&&  \omega_{j}^{(0)}({\bf d}_0, {\bf d}_0'; h_s)= K_{loc}(||{\bf d}_0-{\bf d}_0'||_2/h_{s})K_{st}(0) {\bf 1}(\beta_{j*}({\bf d}_0)=\beta_{j*}({\bf d}_0')),   \label{Th3Eq1}\\
&& \tilde\omega_{j}^{(0)}({\bf d}_0, {\bf d}_0'; h_s)= {\omega_{j}^{(0)}({\bf d}_0, {\bf d}_0'; h_s)}/{\sum_{{\bf d}_m\in B({\bf d}_0, h_s)\cap {\mathcal D}_0} \omega_{j}^{(0)}({\bf d}_0, {\bf d}_m; h_s)},
 \nonumber  \\
% && \hat\Sigma^{(k)}(\sqrt{n}\hat\beta_{j}({\bf d}_0; h_s))= {\bf e}_{j, p}^T\Omega_{X, n}^{-1}{\bf e}_{j, p}\sum_{{\bf d}_m, {\bf d}_m'\in B({\bf d}_0, h_s)\cap {\mathcal D}_0} \tilde\omega_j^{(k)}(d, {\bf d}_m; h_s) \tilde\omega_j^{(k)}(d, {\bf d}_m'; h_s)\hat\Sigma_y({\bf d}_m, {\bf d}_m'), \nonumber  \\
  && \Sigma_j^{(0)}({\bf d}_0; h_s)= {\bf e}_{j, p}^T\Omega_{X}^{-1}{\bf e}_{j, p}\sum_{{\bf d}_m, {\bf d}_m'\in B({\bf d}_0, h_s)\cap {\mathcal D}_0}\tilde \omega_{j}^{(0)}({\bf d}_0, {\bf d}_m; h_s) \tilde\omega_{j}^{(0)}({\bf d}_0, {\bf d}_m'; h_s)\Sigma_y({\bf d}_m, {\bf d}_m'). \nonumber
  \end{eqnarray}

\noindent   {\bf Theorem 3.}      {\it Under assumptions (C1)-(C10) in Section 6 for piecewise constant $\{\beta_{j*}({\bf d}): {\bf d}\in {\mathcal D}\}$,
 we have    the following results for all $0\leq s\leq S$:

(i) $\sup_{{\bf d}_0\in {\mathcal D}_0}|\tilde \beta_{j*}({\bf d}_0; h_s)- \beta_{j*}({\bf d}_0)|=o_p(\sqrt{\log(1+N_D)/n})$;

(ii) $\hat{\beta}_j({\bf d}_0; h_s)-\beta_{j*}({\bf d}_0)= \sum_{{\bf d}_m\in B({\bf d}_0, h_s)\cap \mathcal D_0}   \tilde \omega_{j}^{(0)}({\bf d}_0, {\bf d}_m; h_s) \hat\Delta_j({\bf d}_m)  [1+o_p(1)]$;

 (iii)
$\sup_{{\bf d}_0\in {\mathcal D}_0}|\hat\Sigma(\sqrt{n}\tilde\beta_{j*}({\bf d}_0; h_s))-\Sigma_{j}^{(0)}({\bf d}_0; h_s)|=o_p(1); $

(iv) $\sqrt{n}[\hat{\beta}_j({\bf d}_0; h_s)-\beta_{j*}({\bf d}_0)]$ converges in distribution to a normal distribution with mean zero and variance $\Sigma_{j}^{(0)}({\bf d}_0; h_s)$ as $n\rightarrow\infty$.

 }

{\sc Remark 3}. 
     Theorem 3 shows that MASS has several important features for a piecewise constant   function $\beta_{j*}({\bf d})$. For instance, 
  Theorem 3 (i) quantifies the maximum absolute difference (or bias) between the true value
 ${\beta}_{j*}({\bf d}_0)$ and the pseudo true value  $\tilde \beta_{j*}({\bf d}_0; h_s)$ across all ${\bf d}_0\in {\mathcal D}_0$ for any $s$. 
  Since $\tilde \beta_{j*}({\bf d}_0; h_s)- \beta_{j*}({\bf d}_0)=0$ for
 ${\bf d}_0\in  {\mathcal D}\setminus \partial {\mathcal
D}^{(j)}(h_s)$,  this result delineates the potential bias     for voxels  ${\bf d}_0$ in $ \partial {\mathcal
D}^{(j)}(h_s)$.
% Theorem 3 (ii) indicates the asymptotic equivalence between $\hat{\beta}_j({\bf d}_0; h_s)-\beta_{j*}({\bf d})$ and $\sum_{{\bf d}_m\in B({\bf d}_0, h_s)}   \tilde \omega_{j}^{(0)}(d, {\bf d}_m; h_s) \hat\Delta_j({\bf d}_m)$.
%Theorem 3 (iii) ensures that
%$\hat\Sigma(\sqrt{n}\tilde\beta_{j*}({\bf d}_0; h_s))$ is a uniform consistent estimator of $\Sigma_{j}({\bf d}_0; h_s)$  across ${\bf d}\in {\mathcal D}_0$.
 Theorem 3 (iv) ensures that $\sqrt{n}[\hat{\beta}_j({\bf d}_0; h_s)-\beta_{j*}({\bf d}_0)]$ is   asymptotically normally
 distributed. Moreover, as shown in the supplementary document, 
 $\Sigma_{j}^{(0)}({\bf d}_0; h_s)$   is smaller than 
the asymptotic variance of the raw estimate $\hat{\beta}_j({\bf d}_0)$.    
As a result, MASS  increases statistical power of testing $H_0({\bf d}_0)$.

  We now consider  a much complex  scenario  when $\beta_{j*}({\bf d})$ is piecewise smooth.
In this case,   $  \tilde \beta_{j*}({\bf d}_0; h_s)$ may vary from $h_0$ to $h_S$ for all voxels ${\bf d}_0\in   {\mathcal
D}_0$ regardless whether ${\bf d}_0$ belongs to  $\partial {\mathcal D}^{(j)}(h_s)$ or not.
We can   establish important theoretical results to
characterize the asymptotic behavior of $\hat{\ve\beta}({\bf d}_0; h_s)$ only when $h_s=O(\sqrt{\log(1+N_D)/n})=o(1)$ holds.
We need some additional notation as follows:
   \begin{eqnarray}
 &&  \omega_{j}^{(1)}({\bf d}_0, {\bf d}_0'; h_s)= K_{loc}(||{\bf d}_0-{\bf d}_0'||_2/h_{s})K_{st}(0) {\bf 1}(|\beta_{j*}({\bf d}_0)-\beta_{j*}({\bf d}_0')|\leq O(h_s)),   \\
&& \tilde\omega_{j}^{(1)}({\bf d}_0, {\bf d}_0'; h_s)= {\omega_{j}^{(1)}({\bf d}_0, {\bf d}_0'; h_s)}/{\sum_{{\bf d}_m\in B({\bf d}_0, h_s)\cap {\mathcal D}_0} \omega_{j}^{(1)}({\bf d}_0, {\bf d}_m; h_s)},
 \nonumber  \\
% && \hat\Sigma^{(k)}(\sqrt{n}\hat\beta_{j}({\bf d}_0; h_s))= {\bf e}_{j, p}^T\Omega_{X, n}^{-1}{\bf e}_{j, p}\sum_{{\bf d}_m, {\bf d}_m'\in B({\bf d}_0, h_s)\cap {\mathcal D}_0} \tilde\omega_j^{(k)}(d, {\bf d}_m; h_s) \tilde\omega_j^{(k)}(d, {\bf d}_m'; h_s)\hat\Sigma_y({\bf d}_m, {\bf d}_m'), \nonumber  \\
  && \Sigma_j^{(1)}({\bf d}_0; h_s)= {\bf e}_{j, p}^T\Omega_{X}^{-1}{\bf e}_{j, p}\sum_{{\bf d}_m, {\bf d}_m'\in B({\bf d}_0, h_s)\cap {\mathcal D}_0}\tilde \omega_{j}^{(1)}({\bf d}_0, {\bf d}_m; h_s) \tilde\omega_{j}^{(1)}({\bf d}_0, {\bf d}_m'; h_s)\Sigma_y({\bf d}_m, {\bf d}_m'). \nonumber
  \end{eqnarray}

\noindent   {\bf Theorem 4.}      {\it
  Suppose assumptions (C1)-(C9) and (C11) in Section 6 hold for piecewise continuous $\{\beta_{j*}({\bf d}): {\bf d}\in {\mathcal D}\}$. For all $0\leq s\leq S$,  we have    the following results:

(i) $\sup_{{\bf d}_0\in {\mathcal D}_0}|\tilde \beta_{j*}({\bf d}_0; h_s)- \beta_{j*}({\bf d}_0)|=O_p(h_s)$;

(ii) $\hat{\beta}_j({\bf d}_0; h_s)-\tilde\beta_{j*}({\bf d}_0; h_s)= \sum_{{\bf d}_m\in B({\bf d}_0, h_s)\cap \mathcal D_0}   \tilde \omega_{j}^{(1)}({\bf d}_0, {\bf d}_m; h_s) \hat\Delta_j({\bf d}_m) [1+o_p(1)]$;

 (iii)
$\sup_{{\bf d}_0\in {\mathcal D}_0}|\hat\Sigma(\sqrt{n}\tilde\beta_{j*}({\bf d}_0; h_s))-\Sigma_{j}^{(1)}({\bf d}_0; h_s)|=o_p(1). $

(iv) $\sqrt{n}[\hat{\beta}_j({\bf d}_0; h_s)-\tilde\beta_{j*}({\bf d}_0; h_s)]$ converges in distribution to a normal distribution with mean zero and variance $\Sigma_{j}^{(1)}({\bf d}_0; h_s)$ as $n\rightarrow\infty$.

 }

{\sc Remark 4}.     Theorem 4 characterizes several key features of  MASS   for a piecewise continuous function $\beta_{j*}({\bf d})$. 
    These results  differ  significantly from those for the piecewise constant case, but under weaker assumptions. 
For instance,  Theorem 4 (i) quantifies the  bias of the pseudo true value  $\tilde \beta_{j*}({\bf d}_0; h_s)$ relative to  the true value
${\beta}_{j*}({\bf d}_0)$ across all ${\bf d}_0\in {\mathcal D}_0$ for a fixed $s$.  Even for  voxels inside  the smooth areas of $\beta_{j*}({\bf d})$, the bias
 $O_p(h_s)$ is still much higher than the standard bias at the rate of $h_s^2$ due to the presence of
$K_{st}(D_{{\beta}_j}({\bf d}_0, {\bf d}_0'; h_{s-1})/C_{n})$
 \citep{Fan1996, Wand1995}.  If we set $K_{st}(u)={\bf 1}(u\in [0, 1])$ and $\beta_{j*}({\bf d})$ is twice differentiable,  then
 the  bias of    $\tilde \beta_{j*}({\bf d}_0; h_s)$ relative to ${\beta}_{j*}({\bf d}_0)$  may be reduced to $O_p(h_s^2)$.
%  Theorem 4 (ii) establishes the asymptotic equivalence between $\hat{\beta}_j({\bf d}_0; h_s)-\tilde \beta_{j*}({\bf d}_0; h_s)$ and $\sum_{{\bf d}_m\in B({\bf d}_0, h_s)}   \tilde \omega_{j}^{(1)}(d, {\bf d}_m; h_s) \hat\Delta_j({\bf d}_m)$.
%Theorem 4 (iii) ensures that
%$\hat\Sigma(\sqrt{n}\tilde\beta_{j*}({\bf d}_0; h_s))$ is a uniform consistent estimator of $\Sigma_{j}^{(1)}({\bf d}_0; h_s)$  across ${\bf d}\in {\mathcal D}_0$.
Theorem 4 (iv) ensures that $\sqrt{n}[\hat{\beta}_j({\bf d}_0; h_s)-\tilde\beta_{j*}({\bf d}_0; h_s)]$ is   asymptotically normally
distributed. Moreover,   as shown in the supplementary document, 
 $\Sigma_{j}^{(1)}({\bf d}_0; h_s)$   is smaller than 
the asymptotic variance of the raw estimate $\hat{\beta}_j({\bf d}_0)$, and thus     
  MASS  can increase  statistical power in testing $H_0({\bf d}_0)$ even for the piecewise continuous case. 

\section{Simulation Studies}

In this section, we conducted a set of Monte Carlo simulations to compare MASS with voxel-wise methods   from three different aspects.
Firstly, we examine  the finite sample performance  of
  $\hat{\ve\beta}({\bf d}_0; h_s)$ at different signal-to-noise ratios.
Secondly, we examine   the accuracy of  the estimated eigenfunctions of $\Sigma_\eta({\bf d}, {\bf d}')$. Thirdly, we   assess both Type I and II error rates of
the Wald test statistic. 
For the sake of space, we only present some  selected results below and put additional simulation results in the supplementary document.

\begin{figure}[ht!]
\vspace*{0pt}
\begin{center}
\includegraphics[width=.95\textwidth]{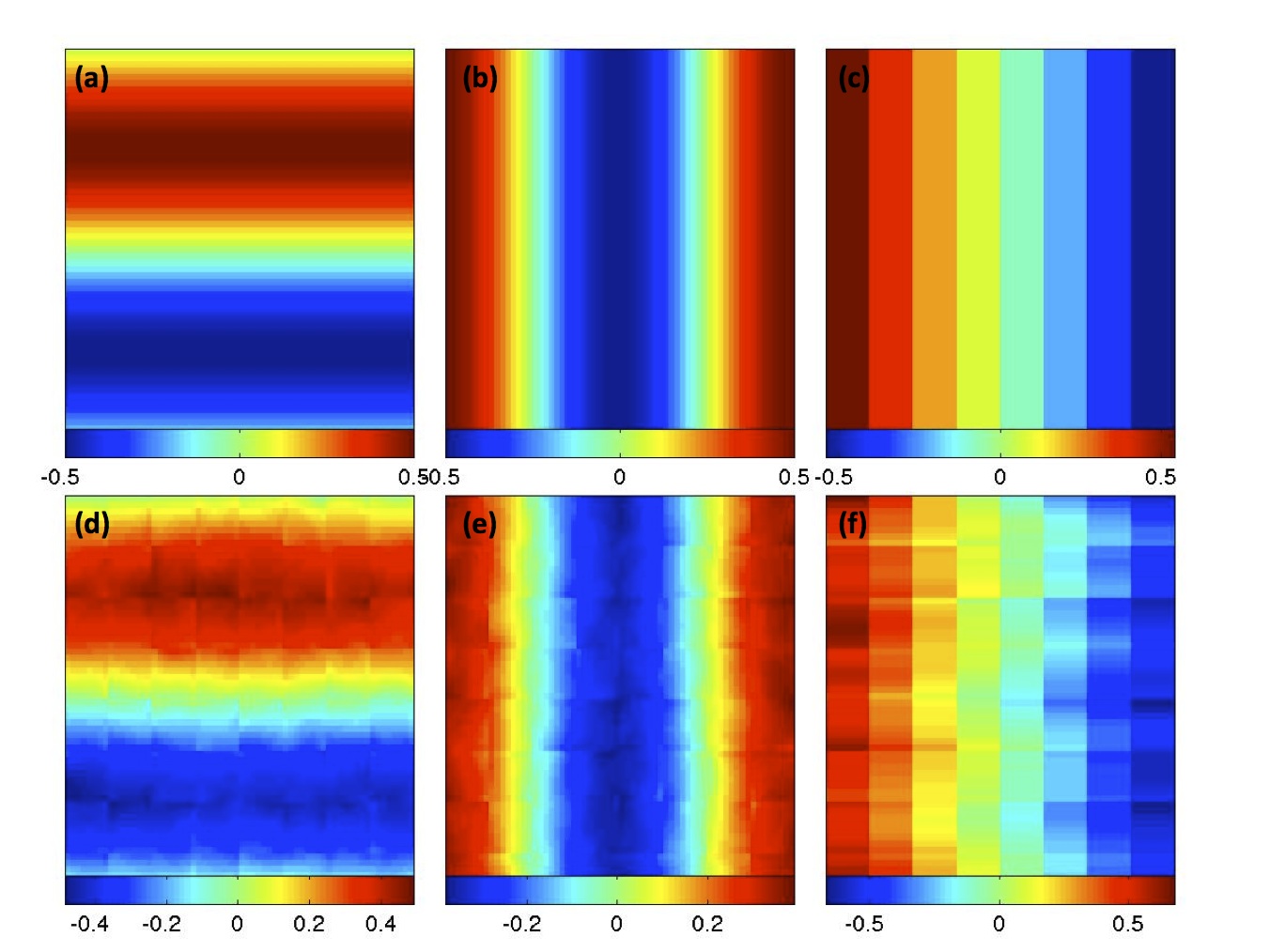}
\end{center}
\vspace*{0pt}
\caption{Simulation results:  a selected slice of (a)  true $\psi_1({\bf d})$; (b) true $\psi_2({\bf d})$; (c) true $\psi_3({\bf d})$;
({d}) $\hat\psi_1({\bf d})$; (e) $\hat\psi_2({\bf d})$; and (f) $\hat\psi_3({\bf d})$. }
\label{fig2}
\end{figure}

We simulated data at all 32,768 voxels   on the $64 \times 64 \times 8$ phantom image for $n=60$ (or $80$) subjects. At each ${\bf d}_0=(d_{0,1}, d_{0,2}, d_{0,3})^T$ in ${\mathcal
D}_0$,  $Y_i({\bf d}_0)$ was simulated according to
\begin{equation}
y_i({\bf d}_0) = {\bf x}_i^T{\ve\beta}({\bf d}_0) + \eta_i({\bf d}_0)+\epsilon_i({\bf d}_0)~~~\mbox{for}~~i=1, \ldots, n,
\end{equation}
where   ${\bf x}_i=(x_{i1},x_{i2},x_{i3})^T$,  ${\ve\beta}({\bf d}_0)=(\beta_1({\bf d}_0), \beta_2({\bf d}_0), \beta_3({\bf d}_0))^T$,  and $\epsilon({\bf d}_0)\sim N(0,1)$ or $\chi(3)^2-3$, in which  $\chi^2(3)-3$  is a very skewed distribution.
Furthermore, we  set $\eta_i({\bf d}_0)= \sum_{l=1}^3 \xi_{il}\psi_l({\bf d}_0)$, where $\xi_{il}$ are independently generated according to
 $\xi_{i1}\sim N(0,0.6),$  $\xi_{i2}\sim N(0,0.3),$  and $\xi_{i3}\sim N(0,0.1),$
$
\psi_1({\bf d}_0)=0.5\sin(2\pi d_{0, 1}/64),$ $ \psi_2({\bf d}_0)=0.5\cos(2\pi d_{0, 2}/64),$ and  $\psi_3({\bf d}_0)=\sqrt{1/2.625}(9/8-d_{0,3}/4).
$
The first eigenfunction $\psi_1({\bf d}_0)$ changes only along $d_{0,1}$ direction, while it keeps constant in the other two directions. The other two eigenfunctions, $\psi_2({\bf d}_0)$ and $\psi_3({\bf d}_0)$, were chosen in a similar way (Figure ~\ref{fig2}).
We set    $x_{i1}=1$ and generated $x_{i2}$   independently   from a Bernoulli distribution with success rate $0.5$  and $x_{i3}$   independently   from the uniform distribution on $[1,2]$.
The covariates $x_{i2}$ and $x_{i3}$ were chosen to represent group identity and scaled age, respectively.

\begin{figure}[ht!]
\vspace*{0pt}
\begin{center}
\includegraphics[width=.95\textwidth]{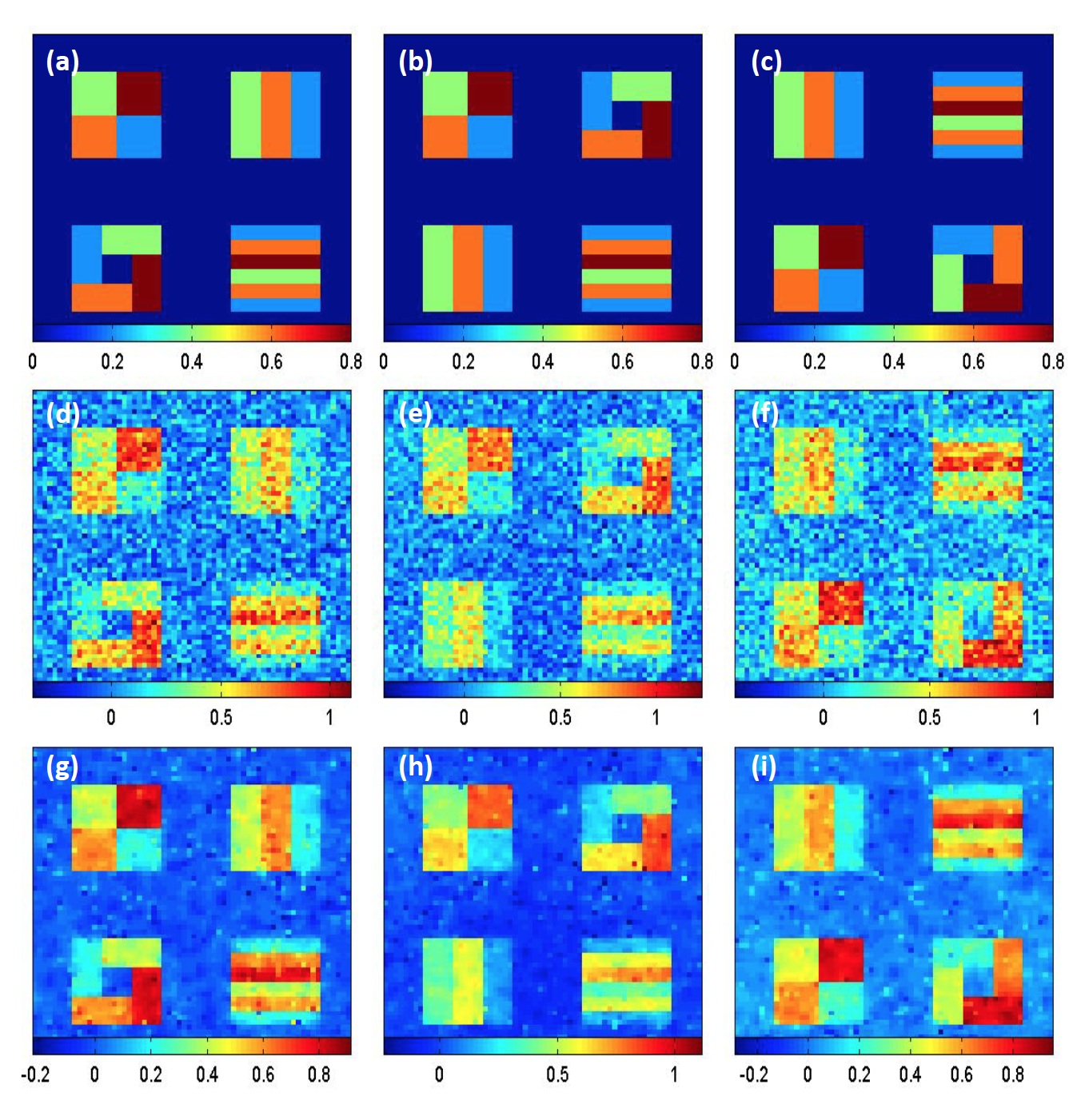}
\end{center}
\vspace*{0pt}
\caption{Simulation results:  a selected slice of  (a) true $\beta_1({\bf d})$; (b) true $\beta_2({\bf d})$; (c) true $\beta_3({\bf d})$;
 (d)  $\hat\beta_1({\bf d}_0)$; (e) $\hat\beta_2({\bf d}_0)$; (f) $\hat\beta_3({\bf d}_0)$;
  (g)  $\hat\beta_1({\bf d}_0;  h_{10})$;  (h) $\hat\beta_2({\bf d}_0; h_{10})$;  and (i) $\hat\beta_3({\bf d}_0; h_{10})$.}
\label{fig3}
\end{figure}

We chose  different pattens for different $\beta_j({\bf d})$ images  in order to examine the finite sample
 performance of our estimation method under different scenarios.
 We set all the $8$ slices along the coronal axis to be identical for each of $\beta_j({\bf d})$ images.
As shown in Figure ~\ref{fig3},   each slice of the three different $\beta_j({\bf d})$ images has  four different blocks and   5 different regions of interest (ROIs)  with varying patterns and shape.
 The  true values of $\beta_j({\bf d})$ were varied from  $0$ to $0.8$, respectively, and were displayed
for all ROIs with  navy blue, blue, green, orange and brown colors representing $0,0.2,0.4,0.6,$ and $0.8$, respectively.

We fitted  the SVCM model (\ref{SVCMeq1}) with the same set of covariates to  a simulated data set, and then  applied the  three-stage estimation procedure described in Section 2.2 to calculate adaptive parameter
estimates across all pixels at 11 different scales.
In MASS, we set  $h_s=1.1^s$ for $s=0, \ldots, S=10$.
 Figure ~\ref{fig3} shows some selected slices of  $\hat{\ve\beta}({\bf d}_0; h_s)$ at  $s=0$ (middle panels) and $s=10$ (lower panels).
 Inspecting  Figure ~\ref{fig3} reveals that all $\hat{\beta}_j({\bf d}_0; h_{10})$ outperform   their corresponding $\hat{\beta}_j({\bf d}_0)$ in terms of variance and detected ROI patterns.
Following the method described in Section 2.2, we estimated $\eta_i({\bf d})$ based on the residuals $y_i({\bf d}_0)-{\bf x}_i^T\hat{\ve \beta}({\bf d}_0)$ by using the local linear smoothing method  and then calculate   $\hat{\eta}_i({\bf d})$.  Figure ~\ref{fig2} shows some selected slices of the  first three estimated eigenfunctions. Inspecting  Figure ~\ref{fig2} reveals that  $\hat\eta_i({\bf d})$ are relatively close to the true eigenfunctions and can capture the  main feature in the true eigenfunctions, which vary in one direction and are constant in the other two directions. However, we do observe some minor block effects, which may be caused by using  the block smoothing method to estimate $\eta_i({\bf d})$.

 Furthermore, for $\hat{\ve\beta}({\bf d}_0; h_s)$, we calculated the bias, the empirical standard error (RMS), the mean of the estimated standard errors (SD), and  the ratio of RMS over SD (RE)  at each voxel of the five ROIs based on the results obtained from the 200 simulated data sets.
 For the sake of space, we only presented some selected results based on $\hat\beta_3({\bf d}_0)$ and $\hat\beta_3({\bf d}_0; h_{10})$  obtained from $N(0, 1)$ distributed  data with $n=60$ in   Table 1.
 The biases are slightly increased from $h_0$  to $h_{10}$ (Table 1), whereas  RMS and SD at $h_5$ and $h_{10}$ are much smaller than those at $h_0$ (Table 1). In addition, the RMS and its corresponding SD are relatively close to each other
at all scales for both the normal and Chi-square distributed data (Table 1).
Moreover,  SDs in these voxels of  ROIs with $\beta_3({\bf d}_0)>0$ are  larger than SDs in those voxels of
 ROI with $\beta_3({\bf d}_0)=0$, since  the interior of ROI with $\beta_3({\bf d}_0)=0$ contains more pixels (Figure ~\ref{fig3} (c)). Moreover,   the SDs at steps $h_0$ and $h_{10}$ show clear spatial patterns caused by spatial correlations. The RMSs also show some evidence of spatial patterns.  The biases, SDs, and RMSs of $\beta_3({\bf d}_0)$ are smaller in  the normal distributed data than in the chi-square distributed data (Table 1), because the signal-to-noise ratios (SNRs)  in the normal distributed data are bigger than those  SNRs in the chi-square distributed data.
Increasing sample size and signal-to-noise ratio decreases the bias, RMS and SD of parameter estimates (Table 1).

To assess both Type I and II error rates at the voxel level, we tested the hypotheses $ H_{0}({\bf d}_0):~\beta_j({\bf d}_0)=0$ versus $H_{1}({\bf d}_0): \beta_j({\bf d}_0)\neq 0$ for $j=1,2,3$ across all ${\bf d}_0\in \mathcal{D}_0$. We applied the same MASS procedure at  scales $h_0$ and $h_{10}$.  The $-\log_{10}(p)$ values  on some selected slices  are shown in the supplementary document.   The $200$ replications were used to calculate the  estimates (ES) and standard errors (SE) of rejection rates at  $\alpha =5\%$ significance level. Due to space limit, we only report the results of testing $\beta_2({\bf d}_0)=0$. The other two tests have similar results and are omitted here. For $ W_\beta({\bf d}_0; h)$, the Type I rejection rates in ROI with $\beta_2({\bf d}_0)=0$ are relatively accurate for all scenarios, while the statistical power for rejecting the null hypothesis in  ROIs with $\beta_2({\bf d}_0)\not=0$   significantly increases with  radius $h_s$  and signal-to-noise ratio (Table 2).
As expected, increasing   $n$ improves  the statistical power for detecting  $\beta_2({\bf d}_0)\not=0$.

\section{Real Data Analysis}

 We applied  SVCM to the Attention Deficit Hyperactivity Disorder (ADHD) data from the New York University (NYU)  site as a part of  the ADHD-200 Sample Initiative  \\  (\textsf{http://fcon$\underline{\mbox{ }}$1000.projects.nitrc.org/indi/adhd200/}).
  ADHD-200 Global Competition is a grassroots initiative event  to accelerate the scientific community's understanding of the neural basis of ADHD through the implementation of open data-sharing and discovery-based science.  Attention deficit hyperactivity disorder (ADHD) is one of the most common childhood disorders and can continue through adolescence and adulthood \citep{Polanczyk2007}. Symptoms include difficulty staying focused and paying attention, difficulty controlling behavior, and hyperactivity (over-activity).    It affects about 3 to 5 percent of children globally and diagnosed in about 2 to 16 percent of school aged children \citep{Polanczyk2007}.  ADHD has three subtypes, namely, predominantly hyperactive-impulsive type, predominantly inattentive type, and combined type.

  The NYU data set consists of $174$ subjects  (99 Normal Controls (NC) and 75  ADHD subjects with  combined hyperactive-impulsive). Among them, there are   $112$ males whose  mean age is $11.4$ years with standard deviation $7.4$ years and $62$ females whose mean   age is $11.9$ years with standard deviation $10$ years.
Resting-state functional MRIs and T1-weighted MRIs  were acquired for each subject.     We only use the T1-weighted MRIs here.
We processed
 the T1-weighted MRIs    by  using a standard image processing pipeline   detailed in the supplementary document.  
 Such pipeline 
  consists of  AC (anterior commissure) and -PC (posterior commissure) correction,   bias field correction,  skull-stripping, intensity inhomogeneity correction,   cerebellum removal,  segmentation, and nonlinear registration.
We  segmented each brain into three different tissues including grey matter (GM), white matter (WM),   and cerebrospinal fluid (CSF).
We  used the  RAVENS maps to quantify the local volumetric group differences  for the whole brain and each of the segmented tissue type (GM, WM,  and CSF) respectively, using the deformation field  that we  obtained during registration  \citep{Davatzikos2001}. RAVENS methodology is based on a volume-preserving spatial transformation, which ensures that no volumetric information is lost during the process of spatial normalization, since this process changes an individual�s brain morphology to conform it to the morphology of the Jacob template \citep{Kabani1998}.

\begin{figure}[ht!]
\vspace*{0pt}
\begin{center}
\includegraphics[width=.95\textwidth,height=.56\textheight]{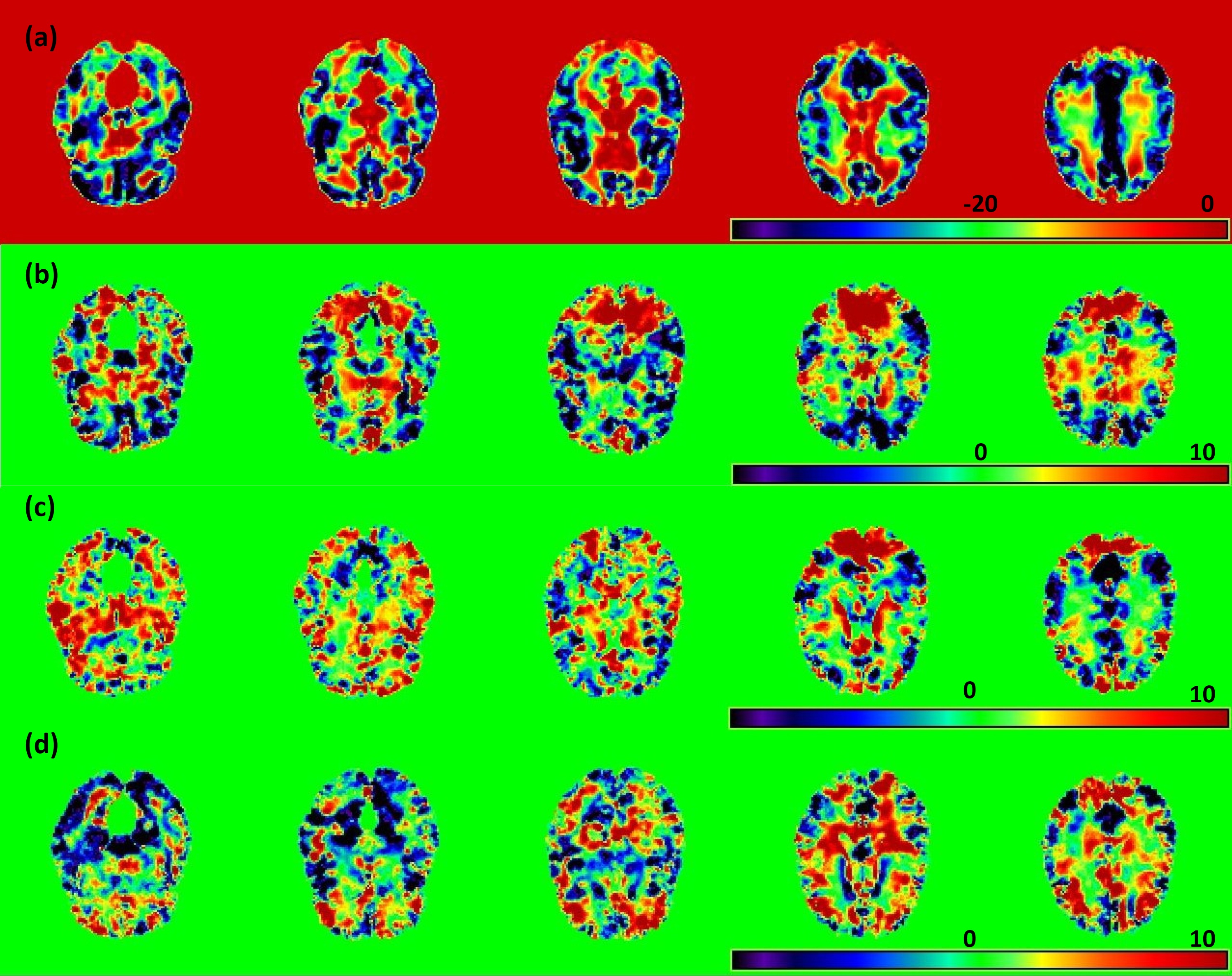}
\end{center}
\vspace*{0pt}
\caption{Results from the ADHD 200 data:  five selected slices of  the four  estimated eigenfunctions  corresponding to the first four largest eigenvalues of  $\hat\Sigma_{\eta}(\cdot, \cdot)$: (a) $\hat\psi_1({\bf d})$; (b) $\hat\psi_2({\bf d})$; (c) $\hat\psi_3({\bf d})$; and ({d}) $\hat\psi_4({\bf d})$.}
\label{fig9}
\end{figure}

We fitted model ~\eqref{SVCMeq1} to the RAVEN  images calculated from the NYU data set. Specifically, we set 
$\ve\beta({\bf d}_0)=(\beta_1({\bf d}_0), \ldots, \beta_8({\bf d}_0))^T$ and  
${\bf x}_i=(1, \mbox{G}_i, \mbox{A}_i, \mbox{D}_i, \mbox{WBV}_i, \mbox{A}_i\times\mbox{D}_i, \mbox{G}_i\times\mbox{D}_i,\mbox{A}_i\times\mbox{G}_i)^T, $
 where $\mbox{G}_i$, $\mbox{A}_i$, $\mbox{D}_i$, and $\mbox{WBV}_i$, respectively, represent gender, age, diagnosis (1 for NC and  0 for ADHD), and
 whole brain volume.
 We applied the three-stage estimation procedure described in Section 2.2.
 In MASS,  we  set $h_s=1.1^s$  for  $s=1, \ldots, 10$.
We are interested in assessing  the age and diagnosis interaction and the gender and diagnosis interaction. Specifically, we tested
  $H_{0}({\bf d}_0):~\beta_6({\bf d}_0)=0$ against $H_1({\bf d}_0): \beta_6({\bf d}_0)\not =0$  for the age$\times$diagnosis interaction   across all voxels.  
  Moreover, we also tested $H_0({\bf d}_0):~\beta_7({\bf d}_0)=0$ against $H_1({\bf d}_0): \beta_7({\bf d}_0)\not=0$
  for the gender$\times$diagnosis interaction, but we  present the associated results in the supplementary document.  
Furthermore,  as shown in the supplementary document,   the largest estimated eigenvalue is much larger than all other estimated  eigenvalues, which
  decrease very slowly to zero, and explains 22$\%$ of  variation in data after accounting for ${\bf x}_i$.
  Inspecting Figure ~\ref{fig9} reveals that  the   estimated eigenfunction corresponding to the largest estimated eigenvalue   captures
  the dominant morphometric variation.

\begin{figure}[ht!]
\vspace*{0pt}
\begin{center}
\includegraphics[width=.95\textwidth,height=.56\textheight]{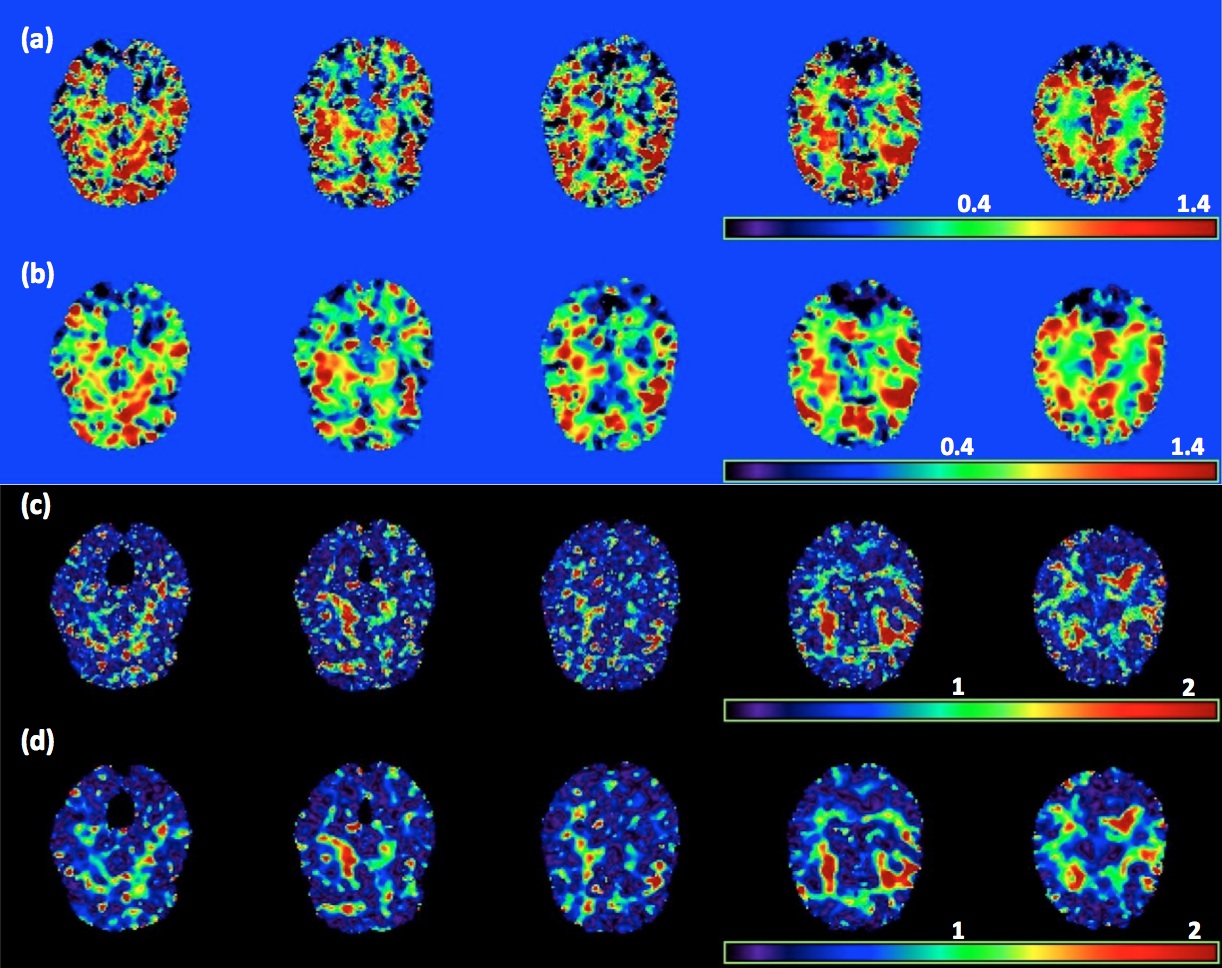}
\end{center}
\vspace*{0pt}
\caption{Results from the ADHD 200 data: five selected slices of  (a) $\hat\beta_6({\bf d}_0)$, (b) $\hat\beta_6({\bf d}_0; h_{10})$, the  $-\log_{10}(p)$ images for testing  $H_0: \beta_6({\bf d}_0)=0$ (c)   at scale $h_0$ and (d) at scale $h_{10}$, where $\beta_6({\bf d}_0)$ is the regression coefficient associated with the age$\times$diagnostic interaction.}
\label{fig7}
\end{figure}

As   $s$ increases from 0 to 10,  MASS shows an advantage
   in smoothing effective signals  within  relatively homogeneous ROIs, while preserving the edges of these ROIs  (Fig. \ref{fig7}  (a)-(d)). 
  Inspecting  Figure ~\ref{fig7} (c) and (d) reveals that  it is much easier to   identify significant ROIs in the $-\log_{10}(p)$ images at scale $h_{10}$, which   are  much smoother    than those at scale $h_0$.  To formally detect significant ROIs, we used
  a cluster-form of threshold of $5\%$  with a minimum voxel clustering value of 50 voxels. 
     We were able to  detect  26    significant clusters 
     across the brain. Then, we overlapped these clusters with the 96 predefined ROIs in the Jacob template and were able to detect several predefined ROIs for each cluster.  As shown in the supplementary document, 
    we were able to detect  several major ROIs, such as the frontal    lobes and the right parietal lobe. 
     The anatomical disturbance in the frontal lobes and the right parietal lobe has been consistently revealed in the literature and  may 
     produce difficulties with inhibiting prepotent responses and decreased brain activity during inhibitory tasks in children with ADHD
      \citep{Bush2011}.
These ROIs comprise the main components of the cingulo-frontal-parietal   cognitive-attention network. These areas, along with striatum, premotor areas, thalamus and cerebellum have been identified as nodes within parallel networks of attention and cognition       \citep{Bush2011}.

\begin{figure}[ht!]
\vspace*{0pt}
\begin{center}
\includegraphics[width=.95\textwidth,height=.68\textheight]{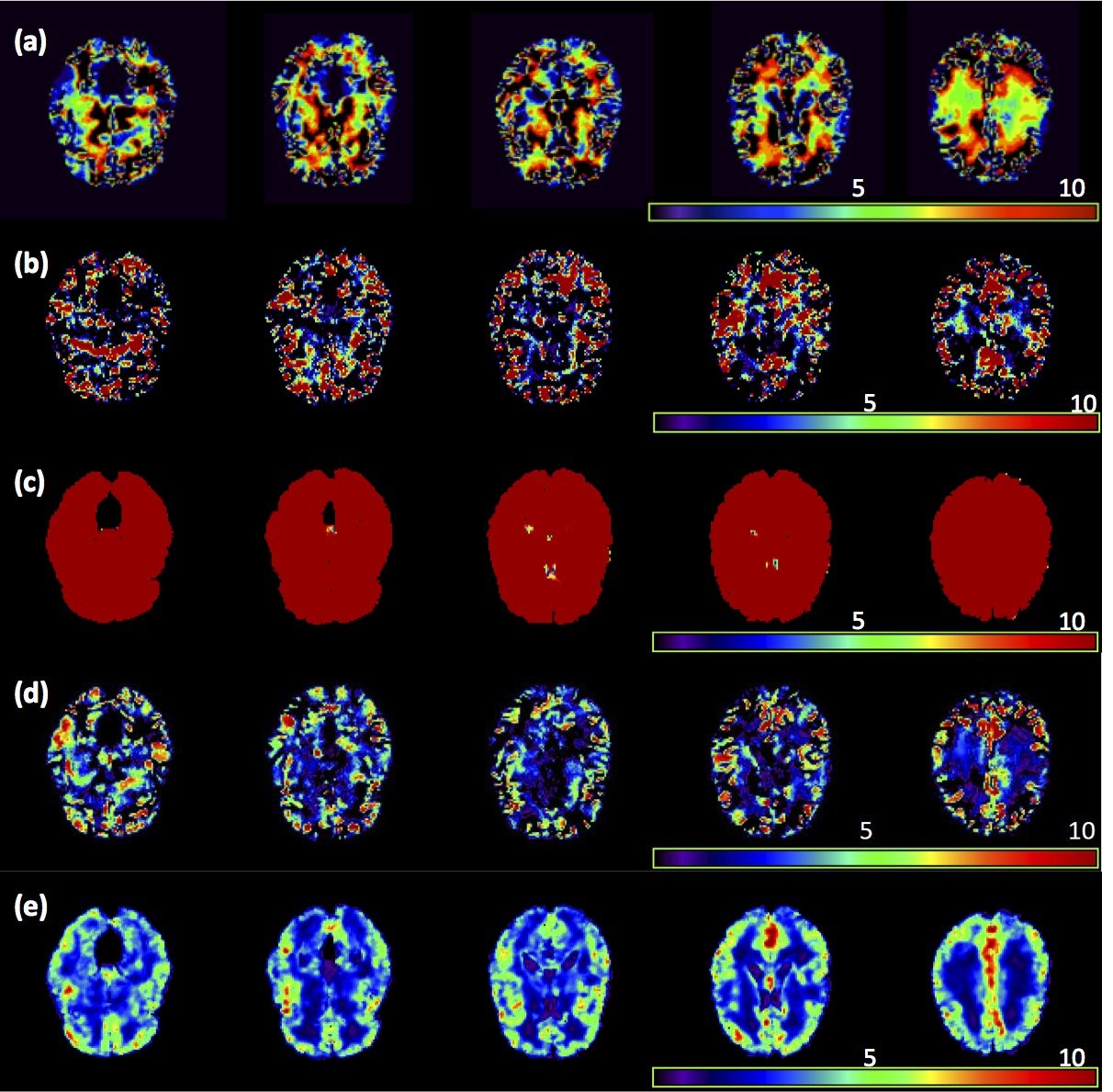}
\end{center}
\vspace*{0pt}
\caption{Results from the ADHD 200 data: The raw RAVENS image for a selected subject with ADHD (a), mean ((b) GLM and (d) SVCM) and standard error ((c) GLM and ({  e}) SVCM) of the errors to predict the RAVENS image in (a), where GLM denotes general linear model.}
\label{fig12}
\end{figure}

To evaluate the prediction accuracy of SVCM, we randomly selected one subject with ADHD from the NYU data set and predicted his/her RAVENS image  by using both model ~\eqref{SVCMeq1} and a standard linear model with normal noise. In both models, we used the same set of covariates, but different  covariance structures. Specifically, in the standard linear model, an independent correlation structure was used and the least squares estimates of $\ve\beta({\bf d}_0)$ were calculated. For SVCM, the functional principal component analysis model was used and $\hat{\ve\beta}({\bf d}_0; h_{10})$ were calculated.  After fitting both models to all subjects except the selected one, we used the fitted models to predict the RAVEN image of the selected subject and  then calculated the prediction error based on the difference between the true and predicted RAVEN images. We  repeated the prediction procedure 50 times and calculated the mean and standard deviation images of these prediction error images (Figure ~\ref{fig12}).  Inspecting Figure \ref{fig12} reveals the advantage and accuracy  of model ~\eqref{SVCMeq1} over the standard linear model   for the ADHD data.

%\medskip

%\end{document}
%%%%%%%%%%%%%%%%%%%%%%%%%%%%%%%%%%%%%%%%%%%%%%%%%%%%%%%%%%%%%
      %>>>> equivalent to \section*{ACKNOWLEDGMENTS}
%This work was supported in part by NSF grants SES-06-43663 and
%BCS-08-26844  and NIH grants   UL1-RR025747-01 and
% R21AG033387  to Dr.
%Zhu,     NIH grants GM 70335  and CA 74015 to Dr. Ibrahim,  NIH
%grant R01NS055754 to Dr. Lin, and NIH grant 1R01EB006733 to Dr.
% Shen.

%%%%%%%%%%%%%%%%%%%%%%%%%%%%%%%%%%%%%%%%%%%%%%%%%%%%%%%%%%%%%
%%%%% References %%%%%

\section{Discussion}

This article studies the idea of using SVCM  for
   the spatial and adaptive analysis of neuroimaging data with jump discontinuities, while explicitly modeling spatial dependence in neuroimaging data.
  We have developed a three-stage estimation procedure to
  carry out statistical inference under SVCM.
    MASS integrates
three methods including propagation-separation,  functional principal component analysis, and  jumping surface model   for neuroimaging data from multiple subjects.
We have developed  a fast and accurate estimation method for independently updating each of effect images, while consistently estimating their standard deviation  images.
Moreover,
we have derived the asymptotic properties of the estimated eigenvalues and eigenfunctions and
 the parameter estimates.

Many issues still merit further research. The basic setup of SVCM can be extended
 to more complex data structures (e.g.,
longitudinal, twin and family) and other parametric and
semiparametric models. For instance,   we  may
  develop  a    spatial varying coefficient mixed effects model for longitudinal neuroimaging
data.
It is also feasible to include
nonparametric components in SVCM.  More research is needed for   weakening  regularity assumptions and for developing   adaptive-neighborhood methods to
determine multiscale neighborhoods that adapt  to the pattern of
imaging data at each voxel. It is also interesting to examine the efficiency of our adaptive estimators obtained from MASS for different kernel functions and coefficient functions.  
An  important issue is that   SVCM and other voxel-wise methods  do not account for the errors caused by registration method.
We may need   to explicitly model the measurement errors caused by  the registration method, and integrate them with  smoothing method  and SVCM
  into a unified framework.

\section{Technical Conditions}

\subsection{Assumptions}

Throughout the paper, the following assumptions are needed to
facilitate the technical details, although they may not be the
weakest conditions.
We do not distinguish the
differentiation and continuation at the boundary points from those
in the interior of ${\mathcal D}$.

 %  Let $N(\mu, \Sigma)$ be a normal random vector with mean $\mu$ and covariance
% $\Sigma$.

% We   define  the fourth moments of $\eta_{i,j} ({\bf d})$ to be
%$\gamma_{jj'll'}(s_1, t_1, s_2, t_2)=E[\eta_{i,j} (s_1)\eta_{i,
%j'}(t_1)\eta_{i, l}(s_2)\eta_{i, l'}(t_2)]$ for any $j, j', l$, and
%$l'$.

%Finally, for $i=1, \ldots, n_1$ and $i'_1+1, \ldots, n$, we define
%\begin{eqnarray*}
%&&\gamma_{M,v}(s, t)=\mbox{Cov}\{[\sum_{j=4}^5{\bf 1}_3 \eta_{ik} (s )+\tilde{\bf x}_i
%v_{i3}({\bf d})], [\sum_{j=4}^5{\bf 1}_3 \eta_{ik} (t)+\tilde{\bf x}_i
%v_{i3}(t)]  \},  \\
%&&
%\gamma_{M,\epsilon}(s, t)=\mbox{Cov}\{[\sum_{j=4}^5{\bf 1}_3 \epsilon_{ik}(s )+\tilde{\bf x}_i
%\epsilon_{i3}({\bf d})], [\sum_{j=4}^5{\bf 1}_3 \epsilon_{ik}(t)+\tilde{\bf x}_i
%\epsilon_{i3}(t)]  \},   \\
%&&\gamma_{D,v}(s, t)=\mbox{Cov}\{[\sum_{j=4}^5{\bf 1}_3 v_{i'k}(s )+\tilde{\bf x}_{i'}
%v_{i'3}({\bf d})], [\sum_{j=4}^5{\bf 1}_3 v_{i'k}(t)+\tilde{\bf x}_{i'}
%v_{i'3}(t)]  \},  \\
%&&
%\gamma_{D,\epsilon}(s, t)=\mbox{Cov}\{[\sum_{j=4}^5{\bf 1}_3 \epsilon_{i'k}(s )+\tilde{\bf x}_{i'}
%\epsilon_{i'3}({\bf d})], [\sum_{j=4}^5{\bf 1}_3 \epsilon_{i'k}(t)+\tilde{\bf x}_{i'}
%\epsilon_{i'3}(t)]  \}.
%\end{eqnarray*}

\begin{description}

\item{\sl Assumption C1}. The number of parameters $p$ is finite. Both $N_D$ and $n$ increase to infinity such that
 $\lim_{n\rightarrow \infty} C_n/n=\lim_{n\rightarrow\infty}
C_n^{-1}\log (N_D)=\lim_{n\rightarrow\infty}
C_n^{-1}=0$.

\item{\sl Assumption C2}.    ${ \epsilon}_{i}({\bf d})$   are  identical and  independent  copies of    $\mbox{SP}(0, \Sigma_\epsilon)$  and
 ${\epsilon}_{i}({\bf d})$ and ${\epsilon}_{i}({\bf d}')$
 are independent for ${\bf d}\not = {\bf d}' \in {\mathcal D}$. Moreover, $\epsilon_i({\bf d})$ are, uniformly in $d$, sub-Gaussian such that
 $K_\epsilon^2[ E\exp(|{\epsilon}_i({\bf d})|^2/K_\epsilon)-1]\leq C_\epsilon$
 for all ${\bf d}\in {\mathcal D}$ and  some positive constants $K_\epsilon$ and $C_\epsilon$.

 \item{\sl Assumption C3}.   The covariate vectors ${\bf x}_i$s are independently and identically
distributed with $E{\bf x}_i=\mu_x$ and $||{\bf x}_{i}||_\infty<\infty$. Moreover, $E({\bf x}_{i}^{\otimes 2}) =\Omega_X$ is invertible.
 The ${\bf x}_i$,  ${ \epsilon}_i({\bf d})$, and ${ \eta}_i({\bf d})$ are mutually independent of each other.

\item{\sl Assumption C4}.
  Each component of    $\{ \eta({\bf d}): {\bf d}\in \mathcal D\}$, $\{ \eta({\bf d})\eta({\bf d}')^T: ({\bf d}, {\bf d}')\in {\mathcal D}^2\}$
and $\{ {\bf x}\eta^T({\bf d}): {\bf d}\in \mathcal D\}$ are Donsker classes.
     Moreover, $\min_{{\bf d}\in {\mathcal D}}\Sigma_\eta({\bf d}, {\bf d})>0$ and  $E[\sup_{{\bf d}\in \mathcal D}||{\eta}({\bf d})||_2^{2r_1}]<\infty$   for  some $r_1\in (2, \infty)$, where $||\cdot||_2$ is the Euclidean norm.   All components of $\Sigma_\eta({\bf d}, {\bf d}')$ have  continuous second-order partial derivatives with respect to  $({\bf d}, {\bf d}')\in {\mathcal D}^2$.

\item{\sl Assumption C5}.   The grid points ${\mathcal D}_0=\{{\bf d}_m, m=1, \ldots, N_D\}$ are independently and identically distributed with density function $\pi({\bf d})$, which has the bounded support $\mathcal D$. Moreover, $\pi({\bf d})>0$ for all ${\bf d}\in \mathcal D$  and $\pi({\bf d})$ has continuous second-order  derivative.

% \item{\sl Assumption C5b}.   The grid points ${\mathcal D}_0=\{{\bf d}_m, m=1, \ldots, N_D\}$ are independently and identically distributed with density function $\pi({\bf d})$, which has the bounded support $\mathcal D$. Moreover, $\pi({\bf d})>0$ for all ${\bf d}\in \mathcal D$  and $\pi({\bf d})$ has continuous second-order  derivative.

%\item{\sl Assumption C3}. The grid points ${\mathcal S}=\{s_m, m=1, \ldots, M\}$ are independently and identically distributed with density function $\pi({\bf d})$, which has the bounded support $[0, L_0]$.
% For some   constants $\pi_L$ and $\pi_U\in (0, \infty)$ and any $s\in [0, L_0]$, $\pi_L\leq \pi({\bf d}) \leq \pi_U$  and $\pi({\bf d})$ has continuous second-order  derivative.

\item{\sl Assumption C6}.  The kernel functions $K_{loc}(t)$ and $K_{st}(t)$ are  Lipschitz continuous and  symmetric density functions, while $K_{loc}(t)$ has  a compact
support $[-1, 1]$.
Moreover, they   are
continuously decreasing functions of $t\geq 0$ such that
$K_{st}(0)=K_{loc}(0)>0$ and  $\lim_{t\rightarrow \infty}
K_{st}(t)=0$.
%, and
%$\lim_{u\rightarrow\infty} u^{1/2}K_{st}(u)=0$.
% Moreover,
%$0<\inf_{h>0, s\in [0, L_0]}\lambda_{\min}(\Omega_1(h, s))\leq
%\sup_{h>0, s\in [0, L_0]}\lambda_{\max}(\Omega_1(h, s))<\infty$
%where $\lambda_{\min}(A)$ and $\lambda_{\max}(A)$, respectively,
%denote the smallest and largest eigenvalues of matrix $A$, and
%$\Omega_1(h, s)$ is defined as
%$$
%\Omega_1(h, s)=\int_0^{L_0}   \left(
%\begin{array}{cc}
%1 &  h^{-1}(u-s)\\
% h^{-1}(u-s)   &  h^{-2}(u-s)^2
%\end{array}
%\right)K_{h}(u-s)\pi(u)du.
%$$

\item{\sl Assumption C7}.    $h$ converges to zero such that
$$h\geq c(\log N_D/N_D)^{1-2/q_1} ~~\mbox{and}~~  h^{-12}(\log n/n)^{1-1/q_2}=o(1), $$
 where  $c>0$ is a fixed constant and $\min(q_1, q_2)>2$.

%\item{\sl Assumption C9}.
%$E[|\epsilon_{i,j}(s_m)|^{q_2}]<\infty$ for some $q_2\in (4,
%\infty)$ and all $j$;
% $\max_{j}h_{j}^{(2)}=o(1)$,  $Mh_{j}^{(2)}\rightarrow\infty$,
%    for $j=1, \ldots, J$.

%\noindent {\sl Assumption C7 (ii)}.       $Mh_{j}^{(1)3}$  for all $k$ are bounded.

\item{\sl Assumption C8}. There is a positive integer $E<\infty$ such that $\lambda_{1}>\ldots>\lambda_{E}\geq 0$.

\item{\sl Assumption C9}.
 For each  $j$,  the three assumptions of the jumping surface model  hold,
 each $D_{j, l}^o$ is  path-connected, and  $\beta_{j*}({\bf d})$ is a Lipschitz function of ${\bf d}$ with a common Lipschitz constant $K_j>0$ in each ${\mathcal D}_{j, l}^o$ such that
 $|\beta_{j*}({\bf d})-\beta_{j*}({\bf d}')|\leq K_j ||{\bf d}-{\bf d}'||_2$ for any ${\bf d}, {\bf d}'\in  {\mathcal D}_{j, l}^o$.  Moreover,    $\sup_{{\bf d}\in {\mathcal D}}|\beta_{j*}({\bf d})|<\infty$, and
   $\max(K_j, L_j)<\infty$.

\item{\sl Assumption C10}.   For piecewise constant $\beta_{j*}({\bf d})$,      $o({\ve u}^{(j)}(h_s))=\sqrt{\log(1+N_D)/n}$ and
$N_Dh_s^3K_{st}(C_n^{-1}n{\ve u}^{(j)}(h_s)^2/(3S_y))=o(\sqrt{\log(1+N_D)/n})$ holds uniformly  for    $h_0=0< \cdots<h_S$,
where $S_y=\max_{{\bf d}_0\in \mathcal D_0}\Sigma_y({\bf d}_0, {\bf d}_0)$ and ${\ve u}^{(j)}(h_s)$ is the smallest  absolute value  of   all possible  jumps at scale $h_s$ and given by  
$$
{\ve u}^{(j)}(h_s)= \min\{ |\beta_{j*}({\bf d}_0)-\beta_{j*}({\bf d}_0')|:  ({\bf d}_0, {\bf d}_0')\in \mathcal D_0^2,  \beta_{j*}({\bf d}_0)\not=\beta_{j*}({\bf d}_0'), 
 {\bf d}_0'\in B({\bf d}_0, h_s)\}. $$

\item{\sl Assumption C11}.   For piecewise continuous $\beta_{j*}({\bf d})$,  $\cup_{{\bf d}\in \mathcal D_0} [P_j({\bf d}_0, h_S)^c\cap I_j({\bf d}_0, \delta_L, \delta_U)]$  is an empty set and
$h_0=0<h_1<\cdots<h_S$ is  a sequence of bandwidths such that $\delta_L=O(\sqrt{\log(1+N_D)/n})=o(1)$, $\delta_U=\sqrt{C_n/n}M_n=o(1)$, in which $\lim_{n\rightarrow\infty}M_n=\infty$, 
$h_S=O(\sqrt{\log(1+N_D)/n})$ and $N_Dh_S^3K_{st}(M_n^{2}/(3S_y))=o(\sqrt{\log(1+N_D)/n})$.

% \noindent {\sl Assumption C7}.  For each $k$, let $\psi(u)=c_1+c_2u$ be a linear function of $u$, where $c_1$ and $c_2$ are two constants,
% we  define the function class
%\begin{equation}
%{\mathcal F}_{j}=\{ ({\bf x}, \eta(\cdot))\rightarrow {\bf x}(Mh)^{-1}\sum_{m=1}^{M}
% K(\frac{s_m-s}{h})\psi(\frac{s_m-s}{h}) \eta(s_m): s\in [a, b], h\in R\}
 %\end{equation}
 %with the envelop function
  %$F_{j}({\bf x}, \eta(\cdot))$ given by
 %\begin{equation}
 %F_{j}({\bf x}, \eta(\cdot))=C||{\bf x}||
 % \sup_{s\in [a, b], (Mh)^{1/5}\in [c_1, c_2]} \{(Mh)^{-1}\sum_{m=1}^{M} \eta_j(s_m)^2 {\bf 1}(|s_m-s|\leq h)\}^{1/2}.
%\end{equation}
%Let $L_2(Q)$ be $L_2(Q)$-metric,  $||F_{j}||_{Q, 2}=\sqrt{\int F_{j}^2 dQ}$,  and
 %$N(\epsilon ||F_{j}||_{Q, 2}, {\mathcal F}_{j}, L_2(Q))$  to be the minimal number of balls $\{f: \int [f-f']^2dQ<\epsilon^2\}$ needed to cover
 %${\mathcal F}_{j}$.
%For each $k$, the function class ${\mathcal F}_{j}$
%  has bounded uniform entropy integral with  the envelope $F_{j}({\bf x}, \eta(\cdot))$, that is
%$
 %\lim\sup_{n\rightarrow\infty}\sup_Q \int_0^\delta\sqrt{\log N(\epsilon ||F_{j}||_{Q, 2}, {\mathcal F}_{M}, L_2(Q))}d\epsilon<\infty,
%$

\end{description}

% Assumption (C1) requests that
%$\log (N({\mathcal D}))<<C_n<<n$. In
%neuroimaging data, although $N({\mathcal D})$ is much larger than
%the sample size $n$, Assumption (C1) claims that we just need a
%relatively large sample size compared to $\log(N({\mathcal D}))$.
%For instance, in most neuroimaging data, $N({\mathcal D})\approx
%100^3$ and  $\log (10^3)=14$. Therefore, relative to $\log(N_D)$, a
%  sample size such as $n=100$ and a value of $C_n$, such as $C_n=40$,  may be reasonable  in making statistical inferences for    SVCM.
{\sc Remark 5}. 
Assumption (C2) is needed to invoke Hoeffding inequality  \citep{Buhlmann2011,  VaartWellner1996}  in order to establish the uniform bound for $\hat\beta({\bf d}_0; h_s)$.
  In practice, since most neuroimaging data are often bounded, the sub-Gaussian assumption is reasonable.
  The  bound   assumption on $||{\bf x}||_\infty$
in Assumption (C3) is not essential and can be removed if we put
a restriction on the tail of the distribution ${\bf x}$.
Moreover, with some additional efforts, all results are valid even for  the case with fixed design predictors.
  Assumption (C4)   avoids   smoothness conditions on the sample path  $ \eta({\bf d})$, which are commonly assumed
in the literature
\citep{MR2278365}.  The assumption on
the moment of $\sup_{{\bf d}\in {\mathcal D}}||\eta({\bf d})||^{2r_2}_2$   is similar to the conditions used in \citep{LiHsing2010}.
 Assumption   (C5) on the stochastic grid points   is not essential and   can be modified to accommodate  the case for fixed grid points with some additional complexities.

{\sc Remark 6}.
   %  Assumptions (C6)   on $K_{st}(\cdot)$ and $K_{loc}(\cdot)$ are very standard and
The bounded support restriction on $K_{loc}(\cdot)$ in  Assumption  (C6)
can be weaken to
a restriction on the tails of $K_{loc}(\cdot)$.
% Assumption  (C7)  on
%bandwidths is similar to the conditions used in \citep{LiHsing2010,
%EinmahlMason2000}.
%Assumption (C8) is only needed to investigate the asymptotic properties of  the first $E$ eigenvalues and eigenfunctions.
Assumption (C9) requires   smoothness and shape conditions on the image of $\beta_{j*}({\bf d})$ for each $j$.
For piecewise constant $\beta_{j*}({\bf d})$,  assumption (C10) requires    conditions on the amount of changes at   jumping points relative to $n$, $N_D$, and $h_S$.
If $K_{st}(t)$ has a compact support, then $K_{st}({\ve u}^{(j)2}/C)=0$ for relatively large $ {\ve u}^{(j)2}$. In this case, $h_S$ can be very large.
However, for piecewise   continuous $\beta_{j*}({\bf d})$,  assumption (C11) requires the convergence rate of   $h_S$ and
    the amount of changes at   jumping points.

\bibliographystyle{asa}

\bibliography{HongtuNeuroimaging}

\newpage

%%
%%\begin{figure}[ht!]
%%\vspace*{0pt}
%%\begin{center}
%%\includegraphics[width=.95\textwidth,height=.56\textheight]{adhdPvalue67.JPG}
%%\end{center}
%%\caption{Results from the ADHD 200 data:   selected slices of  the  $-\log_{10}(p)$ images for testing  $H_0: \beta_6({\bf d})=0$  at scale $h_0$ in panel (a) and  at scale $h_{10}$ in panel (b), and those for testing $H_0: \beta_7({\bf d})=0$  at scale $h_0$ in panel (c) and  at scale $h_{10}$) in panel ({\bf d}).}
%%\label{fig10}
%%\end{figure}

\newpage

\begin{table}
\begin{center}
\caption{ Simulation results: Average Bias ($\times 10^{-2}$), RMS, SD, and RE  of ${\beta}_2({\bf d}_0)$ parameters in the five ROIs  at 3 different
scales ($h_0, h_5, h_{10}$), $N(0, 1)$ and $\chi(3)^2-3$ distributed noisy data,  and 2 different sample sizes
($n=60,  80$).    BIAS denotes the bias of the mean of   estimates;   RMS denotes the
root-mean-square error; SD denotes the mean of the standard
deviation estimates; RE denotes the ratio of RMS over SD.   For each case, 200 simulated
data sets were used.
 }

{\small
\begin{tabular}{|cc|cccccc|cccccc|}
\hline\hline
&&\multicolumn{6}{c}{$\chi^2(3)-3$}& \multicolumn{6}{|c|}{$N(0, 1)$}\\
&&   \multicolumn{3}{c}{$n=60$} & \multicolumn{3}{c|}{$n=80$}& \multicolumn{3}{c}{$n=60$}& \multicolumn{3}{c|}{$n=80$}\\ \hline
\multicolumn{2}{|c|}{$\beta_2({\bf d}_0)$} 		&$h_0$		&$h_5$		&$h_{10}$		 &$h_0$		 &$h_5$		&$h_{10}$		&$h_0$		&$h_5$		&$h_{10}$		 &$h_0$		 &$h_5$		 &$h_{10}$	\\ \hline
0.0	&	BIAS	&	-0.03	&	0.36	&	0.61	&	0.00	&	0.34	 &	0.56	&	 -0.01	&	0.17	&	0.22	&	0.01	&	0.16	&	 0.20	\\	
	&	RMS	&	0.18	&	0.13	&	0.13	&	0.15	&	0.10	&	 0.10	&	 0.14	&	0.07	&	0.07	&	0.12	&	0.06	&	0.06	 \\	
	&	SD	&	0.18	&	0.13	&	0.12	&	0.15	&	0.11	&	 0.11	&	 0.14	&	0.07	&	0.07	&	0.12	&	0.06	&	0.06	 \\	
	&	RE	&	1.03	&	1.00	&	1.04	&	1.00	&	0.94	&	 0.98	&	 0.99	&	0.94	&	1.03	&	1.00	&	0.95	&	1.04	 \\	\hline
0.2	&	BIAS	&	0.72	&	0.37	&	0.38	&	0.15	&	-0.35	 &	-0.39	&	 -0.04	&	-0.55	&	-0.66	&	0.10	&	-0.48	&	 -0.61	\\	
	&	RMS	&	0.19	&	0.14	&	0.13	&	0.16	&	0.11	&	 0.11	&	 0.14	&	0.07	&	0.07	&	0.12	&	0.06	&	0.06	 \\	
	&	SD	&	0.18	&	0.14	&	0.13	&	0.16	&	0.12	&	 0.11	&	 0.14	&	0.08	&	0.07	&	0.12	&	0.07	&	0.06	 \\	
	&	RE	&	1.02	&	0.99	&	1.03	&	1.00	&	0.96	&	 0.99	&	 0.99	&	0.96	&	1.04	&	1.00	&	0.97	&	1.06	 \\	\hline
0.4	&	BIAS	&	-0.40	&	-0.55	&	-0.68	&	-0.10	&	-0.15	 &	-0.24	&	 0.04	&	0.12	&	0.13	&	-0.10	&	0.05	&	 0.08	\\	
	&	RMS	&	0.19	&	0.14	&	0.14	&	0.16	&	0.12	&	 0.12	&	 0.14	&	0.07	&	0.07	&	0.12	&	0.07	&	0.07	 \\	
	&	SD	&	0.18	&	0.14	&	0.13	&	0.16	&	0.12	&	 0.12	&	 0.14	&	0.08	&	0.07	&	0.12	&	0.07	&	0.06	 \\	
	&	RE	&	1.02	&	1.00	&	1.03	&	1.00	&	0.96	&	 1.00	&	 0.99	&	0.96	&	1.04	&	1.00	&	0.97	&	1.06	 \\	\hline
0.6	&	BIAS	&	0.42	&	-1.14	&	-1.93	&	0.05	&	-1.20	 &	-1.89	&	 0.03	&	-0.55	&	-0.69	&	-0.01	&	-0.43	&	 -0.54	\\	
	&	RMS	&	0.18	&	0.13	&	0.13	&	0.15	&	0.11	&	 0.11	&	 0.14	&	0.07	&	0.07	&	0.12	&	0.06	&	0.06	 \\	
	&	SD	&	0.18	&	0.13	&	0.13	&	0.15	&	0.11	&	 0.11	&	 0.14	&	0.08	&	0.07	&	0.12	&	0.07	&	0.06	 \\	
	&	RE	&	1.02	&	1.00	&	1.04	&	1.00	&	0.95	&	 0.99	&	 0.99	&	0.97	&	1.05	&	1.00	&	0.97	&	1.05	 \\	\hline
0.8	&	BIAS	&	-1.04	&	-2.95	&	-4.09	&	-0.13	&	-1.71	 &	-2.70	&	 -0.11	&	-0.82	&	-1.03	&	-0.03	&	-0.59	&	 -0.77	\\	
	&	RMS	&	0.19	&	0.15	&	0.15	&	0.16	&	0.12	&	 0.12	&	 0.14	&	0.08	&	0.07	&	0.12	&	0.07	&	0.07	 \\	
	&	SD	&	0.19	&	0.15	&	0.14	&	0.16	&	0.13	&	 0.12	&	 0.14	&	0.08	&	0.07	&	0.12	&	0.07	&	0.06	 \\	
	&	RE	&	1.02	&	1.00	&	1.03	&	1.00	&	0.96	&	 0.99	&	 0.99	&	0.94	&	1.01	&	1.00	&	0.95	&	1.02	 \\	
\hline
\hline
\end{tabular}
}
\end{center}
\end{table}

\begin{table}
\begin{center}
\caption{ Simulation Study for $W_\beta({\bf d}_0; h)$:  estimates (ES) and standard errors (SE) of
rejection rates for pixels inside the five ROIs   were reported at 2 different
scales ($h_0,   h_{10}$), $N(0, 1)$ and $\chi^2(3)-3$ distributed data,  and 2 different sample sizes
($n=60,  80$) at $\alpha=5\%$.  For each case, 200 simulated
data sets were used.
 }

{\small
 \begin{tabular}{|cc|cccc|cccc|}\hline\hline
   & &  \multicolumn{4}{c|}{$\chi^2(3)-3$}& \multicolumn{4}{c|}{$N(0, 1)$}\\
  & &  \multicolumn{2}{c}{$n=60$} & \multicolumn{2}{c|}{$n=80$}& \multicolumn{2}{c}{$n=60$}& \multicolumn{2}{c|}{$n=80$}\\ \hline
$\beta_{2}({\bf d}_0)$	&	s	&	ES	&	SE	&	ES	&	SE	&	ES	&	SE	&	 ES	&	SE	\\	 \hline
0.0	&	$h_0$	&	0.056	&	0.016	&	0.049	&	0.015	&	0.048	 &	0.015	&	 0.050	&	0.016	\\	
	&	$h_{10}$	&	0.055	&	0.016	&	0.042	&	0.015	&	 0.036	&	0.016	 &	0.040	&	0.019	\\	\hline
0.2	&	$h_0$	&	0.210	&	0.043	&	0.245	&	0.039	&	0.282	 &	0.033	&	 0.370	&	0.035	\\	
	&	$h_{10}$	&	0.358	&	0.126	&	0.413	&	0.139	&	 0.777	&	0.107	 &	0.870	&	0.081	\\	\hline
0.4	&	$h_0$	&	0.556	&	0.072	&	0.692	&	0.054	&	0.794	 &	0.030	&	 0.895	&	0.024	\\	
	&	$h_{10}$	&	0.792	&	0.129	&	0.894	&	0.078	&	 0.994	&	0.006	 &	0.998	&	0.003	\\	\hline
0.6	&	$h_0$	&	0.907	&	0.040	&	0.966	&	0.022	&	0.988	 &	0.008	&	 0.998	&	0.003	\\	
	&	$h_{10}$	&	0.986	&	0.023	&	0.997	&	0.009	&	 1.000	&	0.001	 &	1.000	&	0.000	\\	\hline
0.8	&	$h_0$	&	0.978	&	0.016	&	0.997	&	0.004	&	1.000	 &	0.001	&	 1.000	&	0.000	\\	
	&	$h_{10}$	&	0.997	&	0.006	&	1.000	&	0.001	&	 1.000	&	0.000	 &	1.000	&	0.000	\\	\hline
\hline
\end{tabular}
}
\end{center}
\end{table}

\end{document}

\newcommand{\ve}[1]{{\mbox{\boldmath ${#1}$}}} 
\renewcommand{\baselinestretch}{1.2}
%\lhead[\fancyplain{} \leftmark]{}
%\chead[]{}
%\rhead[]{\fancyplain{}\rightmark}
%\cfoot{}
%\headrulewidth=0pt
\markright{
%\hbox{\footnotesize\rm Statistica Sinica
%{\footnotesize\bf ??}(200?), 000-000}\hfill
}

 \makeatletter 
\renewcommand{\thefigure}{S\@arabic\c@figure}
\renewcommand{\thetable}{S\@arabic\c@table}
\@addtoreset{table}{1}
\makeatother

\begin{titlepage}
\title{
{\Large  \bf Supplementary Document of  ``Spatially Varying Coefficient Model   for Neuroimaging Data with Jump Discontinuities"}
 \author{
 Hongtu Zhu$^\star$,
 Department  of Biostatistics   \\
 and Biomedical Research Imaging Center \\
 University of North Carolina at Chapel Hill\\
 Chapel Hill, NC 27599, USA\\
 Jianqing Fan$^\dag$,
 \\Department of Oper Res and Fin. Eng \\
 Princeton University, 
 Princeton, NJ 08540\\
 and
 Linglong Kong$^\ddag$
 \\
 Department of Mathematical and Statistical Sciences \\ 
  University of Alberta, 
  Edmonton, AB Canada T6G 2G1 
 } 
}
\date{}
\maketitle 
\thispagestyle{empty}
 \end{titlepage}

\section*{Theoretical Results with Detailed Discussions}

To investigate the asymptotic properties of $\hat{\beta}_j({\bf d}_0;  h_s)$, we need to  characterize points close to  and far from the boundary set $\partial {\mathcal D}^{(j)}$.
For a given bandwidth $h_s$, we first  define $h_s$-boundary sets:
\begin{equation}
\partial {\mathcal
D}^{(j)}(h_s)=\{ {\bf d}\in {\mathcal D}:  B({\bf d};  h_s)\cap   \partial {\mathcal
D}^{(j)}\not =\emptyset\}~~\mbox{and}~~
\partial {\mathcal
D}^{(j)}_0(h_s)=\partial {\mathcal
D}^{(j)}(h_s)\cap {\mathcal  D}_0.  \label{ThemEq1}
\end{equation}
Thus,  $\partial {\mathcal
D}^{(j)}(h_s)$ can be regarded as a band with radius $h_s$ covering the boundary set $\partial {\mathcal
D}^{(j)}$, while $\partial {\mathcal
D}^{(j)}_0(h_s)$ contains all grid points within such band.
  It is easy to show that for a sequence of bandwidths $h_0=0<h_1<\cdots<h_S$, we have
\begin{equation}
\partial {\mathcal
D}^{(j)}(h_0)=\partial {\mathcal
D}^{(j)}\subset    \cdots \subset \partial {\mathcal
D}^{(j)}(h_S) ~~~\mbox{and}~~~
\partial {\mathcal
D}^{(j)}_0(h_0)\subset    \cdots \subset \partial {\mathcal
D}^{(j)}_0(h_S).  \label{ThemEq2}
 \end{equation}
 Therefore, for a fixed bandwidth $h_s$, any point ${\bf d}\in\mathcal D$ belongs to either $ {\mathcal D}\setminus \partial {\mathcal
D}^{(j)}(h_s)$ or $\partial {\mathcal
D}^{(j)}(h_s)$. For each ${\bf d}_0\in {\mathcal D}\setminus \partial {\mathcal
D}^{(j)}(h_s)$,  there exists one and only one ${\mathcal D}_{j, l}$ such that
\begin{equation}
B({\bf d}_0;  h_0)\subset \cdots\subset B({\bf d}_0;  h_s)\subset {\mathcal D}_{j, l}^o.
\label{ThemEq3}
\end{equation}

 For any  ${\bf d}_0\in\partial {\mathcal
D}^{(j)}(h_s)$, it follows from the local patch assumption   that  $B({\bf d}_0, h_s)=P_j({\bf d}_0, h_s)\cup P_j({\bf d}_0, h_s)^c$ and  $P_j({\bf d}_0, h_s)^c$ contains all possible jump  points.
The performance of MASS strongly depends on  $K_{st}(D_{{\beta}_j}({\bf d}_0, {\bf d}_0';  h_{s-1})/C_{n})$ and the degree of jumps as  $\beta_{j*}({\bf d}_0')$ varies in $P_j({\bf d}_0, h_s)^c$ relative to $\beta_{j*}({\bf d}_0)$. To have a better understanding of MASS,  we examine the behavior of $K_{st}(D_{{\beta}_j}({\bf d}_0, {\bf d}_0';  h_{s-1})/C_{n})$
as $s=1$.  Let $\hat\Delta_j({\bf d}_0)=\hat\beta_j({\bf d}_0)-\beta_{j*}({\bf d}_0)$ and $\Delta_{j*}({\bf d}_0, {\bf d}_0')=\beta_{j*}({\bf d}_0)-\beta_{j*}({\bf d}_0')$. It follows from Theorem 1 that
$D_{\beta_j}({\bf d}_0, {\bf d}_0';  h_{0})/C_n$ can be written as
 \begin{eqnarray}
D_{\beta_j}({\bf d}_0, {\bf d}_0';  h_{0})/C_n&=&C_n^{-1}n\{\hat\Delta_j({\bf d}_0)-\hat\Delta_j({\bf d}_0')+\beta_{j*}({\bf d}_0)-\beta_{j*}({\bf d}_0')\}^2/\Sigma_n(\sqrt{n}\hat{\beta}_j({\bf d}_0)) \nonumber  \\
&=&O_p(\{\sqrt{\log(1+N_D)/C_n}+ \Delta_{j*}({\bf d}_0, {\bf d}_0')\sqrt{n/C_n}\}^2).   \label{ThemEq5}
\end{eqnarray}
That is,  $D_{\beta_j}({\bf d}_0, {\bf d}_0';  h_{0})/C_n$ is determined by the  size  of $\hat\Delta_j({\bf d}_0)-\hat\Delta_j({\bf d}_0')$ relative to $\Delta_{j*}({\bf d}_0, {\bf d}_0')$.
If  $\log(1+N_D)=o(C_n)$, $C_n=o(n)$, and $\lim_{u\rightarrow\infty} K_{st}(u)=0$,  then $K_{st}(D_{\beta_j}({\bf d}_0, {\bf d}_0';  h_{0})/C_n)$ converges to $0$, when $\Delta_{j*}({\bf d}_0, {\bf d}_0')\sqrt{n/C_n}$ diverges. Therefore,  if the jump $\Delta_{j*}({\bf d}_0, {\bf d}_0')$ is an order larger than
$\sqrt{C_n/n}$,     the voxel  ${\bf d}_0'\in P_j({\bf d}_0, h_s)^c$ has a small impact on $\hat\beta_j({\bf d}_0;  h_1)$.
%In contrast, if $\Delta_{j*}({\bf d}_0, {\bf d}_0')\sqrt{n/C_n}$ is   small, then $D_{\beta_j}({\bf d}_0, {\bf d}_0';  h_{0})/C_n$ and $K_{st}(D_{\beta_j}({\bf d}_0, {\bf d}_0';  h_{0})/C_n)$ converge to $0$ and
%$K_{st}(0)>0$, respectively.
  We will show below that the above discussions are also valid even for $s>1$.

Due to the discontinuity of $\beta_{j*}({\bf d}_0)$ in   $\partial {\mathcal
D}^{(j)}(h_s)$, we need   a better refinement (or decomposition) of  $  {\mathcal
D}$ according to the value of  $\beta_{j*}({\bf d}_0)$. Specifically, for each ${\bf d}\in {\mathcal D}$ and $\delta_2> \delta_1\geq 0$,
 we define a $(\delta_1, \delta_2)$-neighborhood set of $\beta_{j*}({\bf d})$ as follows:
 \begin{equation}
 I_j({\bf d}, \delta_1, \delta_2)=\{{\bf d}': {\bf d}'\in {\mathcal D}, \delta_1\leq |\beta_{j*}({\bf d})-\beta_{j*}({\bf d}')|< \delta_2\}.  \label{ThemEq4}
  \end{equation}
% The set $I_j({\bf d}_0, \delta_1, \delta_2)$ includes all points ${\bf d}_0'\in {\mathcal D}$, whose values $\beta_j({\bf d}_0')$ are in a  neighborhood of $\beta_{j*}({\bf d}_0)$, even though   some points ${\bf d}_0'\in {\mathcal D}$ may be far from $d$.
If $\delta_1=0$ and $\delta_2=\infty$, then $ I_j({\bf d}_0, \delta_1, \delta_2)={\mathcal D}$.
For ${\bf d}_0\in  {\mathcal D}_{j, l}^o$, since $\beta_{j*}({\bf d}_0)$ is  a smooth  function in $ {\mathcal D}_{j, l}^o$,
 there always exists a sufficiently small bandwidth $h>0$ such that $B({\bf d}_0, h)\subset I_j({\bf d}_0, 0,  \delta)$ for a given  $\delta>0$.   Particularly, if  $\beta_{j*}({\bf d}_0)$ is constant in
 ${\mathcal D}_{j, l}^o$, then ${\mathcal D}_{j, l}^o\subset I_j({\bf d}_0, 0,  \delta)$ for any $\delta>0$.

To further delineate the structure of $P_j({\bf d}_0, h_s)^c$, we introduce a lower threshold and an upper threshold, which are denoted by $\delta_{L}$ and $\delta_U$, respectively.
For  any $0\leq \delta_L<\delta_U$, $P_j({\bf d}_0, h_s)^c$  is a union of  three sets including
$P_j({\bf d}_0, h_s)^c\cap  I_j({\bf d}_0, 0,  \delta_L)$,    $P_j({\bf d}_0, h_s)^c\cap I_j({\bf d}_0, \delta_L, \delta_U)$, and    $P_j({\bf d}_0, h_s)^c\cap I_j({\bf d}_0, \delta_U,  \infty).
$  We consider $\delta_L=O(\sqrt{\log(1+N_D)/n})=o(1)$ and $\delta_U=\sqrt{C_n/n}M_n=o(1)$, in which $\lim_{n\rightarrow\infty}M_n=\infty$.
For any ${\bf d}_0'\in  I_j({\bf d}_0, 0,  \delta_L)$ and ${\bf d}_0''\in I_j({\bf d}_0, \delta_U,  \infty)$, it follows from (\ref{ThemEq5}) that
\begin{eqnarray*}
&&K_{st}(D_{\beta_j}({\bf d}_0, {\bf d}_0';  h_{s})/C_n)=K_{st}(O_p(\log(1+N_D)/C_n))\approx K_{st}(0)>0,  \\
&&K_{st}(D_{\beta_j}({\bf d}_0, {\bf d}_0''; h_{s})/C_n)=K_{st}(O_p(M_n^2))\approx K_{st}(\infty)=0.
\end{eqnarray*}
For any ${\bf d}_0'''\in I_j({\bf d}_0, \delta_L, \delta_U)$, we have $K_{st}(D_{\beta_j}({\bf d}_0, {\bf d}_0'''; h_{s})/C_n)\in [0, K_{st}(0)]$.
Generally,  MASS    discards almost all information contained in voxels in
$P_j({\bf d}_0, h_s)^c\cap I_j({\bf d}_0, \delta_U,  \infty)$, whereas it  incorporates almost all information contained in
voxels in $P_j({\bf d}_0, h_s)^c\cap  I_j({\bf d}_0, 0,  \delta_L)$ and partial information contained in voxels in $P_j({\bf d}_0, h_s)^c\cap I_j({\bf d}_0, \delta_L, \delta_U)$.
For  voxels in $P_j({\bf d}_0, h_s)^c\cap  I_j({\bf d}_0, 0,  \delta_U)$, MASS has difficulty in preventing  biases
in estimating  $\beta_{j*}({\bf d}_0)$. In practice, $\delta_U$ can be regarded as the {\it sensitivity} (or {\it capability}) of MASS to respond to jumps in $\partial {\mathcal D}^{(j)}$.

 We first investigate the asymptotic behavior of $\hat\beta_j({\bf d}_0;  h_s)$ when $\beta_{j*}({\bf d}_0)$ is piecewise constant.
%That is, $\beta_{j*}({\bf d}_0)$ is constant in
% ${\mathcal D}_{j, l}^o$   and for any ${\bf d}_0'\in \partial {\mathcal D}^{(j)}$,  there exists a $d\in \cup_{l=1}^{L_j} {\mathcal D}_{j, l}^o$ such that
% $\beta_{j*}({\bf d}_0)=\beta_{j*}({\bf d}_0')$.
Let $\Delta_{j*}({\bf d}_0, {\bf d}_0')=\beta_{j*}({\bf d}_0)-\beta_{j*}({\bf d}_0')$ and $  \tilde \beta_{j*}({\bf d}_0;  h_s)= \sum_{{\bf d}_m\in B({\bf d}_0, h_s)}\tilde \omega_j({\bf d}_0, {\bf d}_m; h_s) \beta_{j*}({\bf d}_m)$ be the pseudo-true value of $\beta_j({\bf d}_0)$ at scale $h_s$ in voxel ${\bf d}_0$.
For all ${\bf d}_0\in {\mathcal D}\setminus \partial {\mathcal
D}^{(j)}(h_S)$,  we have $  \tilde \beta_{j*}({\bf d}_0;  h_s)=\cdots=\tilde\beta_{j*}({\bf d}_0;  h_0)=\beta_{j*}({\bf d}_0)$ due to (\ref{ThemEq3}). In contrast, for ${\bf d}_0\in \partial {\mathcal
D}^{(j)}(h_S)$, let ${\ve u}^{(j)}(h_s)= \min_{({\bf d}_0, {\bf d}_0'): \Delta_{j*}({\bf d}_0, {\bf d}_0')\not=0, {\bf d}_0'\in B({\bf d}_0, h_s)}|\Delta_{j*}({\bf d}_0, {\bf d}_0')| $ be the smallest  absolute value  of   all possible  jumps at scale $h_s$.
In this case,   $  \tilde \beta_{j*}({\bf d}_0;  h_s)$ may vary from $h_0$ to $h_S$. However,  we can show below that $\tilde\Delta_{j*}({\bf d}_0;  h_s)=  \tilde \beta_{j*}({\bf d}_0;  h_s)- \beta_{j*}({\bf d}_0)=o_p(\sqrt{\log(1+N_D)/n})$ under some mild conditions on $h_s$  and ${\ve u}^{(j)}(h_s)$, which will be detailed below. A remarkable property of MASS is that  $h_S$ is not required to converge to zero when ${\ve u}^{(j)}(h_s)$ is relatively large and $K_{st}(t)$ satisfies certain tail property.

For  a fixed  $S>0$ and   piecewise constant $\beta_{j*}({\bf d}_0)$, we can   establish  several important theoretical results to
characterize the asymptotic behavior of $\hat{\ve\beta}({\bf d}_0;  h_s)$.
We need to introduce some additional notation as follows:
   \begin{eqnarray}
&&  \omega_{j}^{(0)}({\bf d}_0, {\bf d}_0';  h_s)= K_{loc}(||{\bf d}_0-{\bf d}_0'||_2/h_{s})K_{st}(0) {\bf 1}(\Delta_{j*}({\bf d}_0, {\bf d}_0')=0),  \label{Th3Eq1}\\
&& \omega_j^{(1)}({\bf d}_0, {\bf d}_0';  h_s)=K_{loc}(||{\bf d}_0-{\bf d}_0'||_2/h_s)K_{st}(0){\bf 1}({\bf d}_0'\in P_j({\bf d}_0, h_s)\cup I_j({\bf d}_0, 0,  \delta_L)),  \nonumber \\
&& \tilde\omega_{j}^{(k)}({\bf d}_0, {\bf d}_0';  h_s)= {\omega_{j}^{(k)}({\bf d}_0, {\bf d}_0';  h_s)}/{\sum_{{\bf d}_m\in B({\bf d}_0, h_s)\cap \mathcal D_0} \omega_{j}^{(k)}({\bf d}_0, {\bf d}_m; h_s)},
 \nonumber  \\
 && \hat\Sigma^{(k)}(\sqrt{n}\hat\beta_{j}({\bf d}_0;  h_s))= {\bf e}_{j, p}^T\Omega_{X, n}^{-1}{\bf e}_{j, p}\sum_{{\bf d}_m, {\bf d}_m'\in B({\bf d}_0, h_s)\cap \mathcal D_0} \tilde\omega_j^{(k)}({\bf d}_0, {\bf d}_m; h_s) \tilde\omega_j^{(k)}({\bf d}_0, {\bf d}_m'; h_s)\hat\Sigma_y({\bf d}_m, {\bf d}_m'), \nonumber  \\
  && \Sigma_j^{(k)}({\bf d}_0;  h_s)= {\bf e}_{j, p}^T\Omega_{X}^{-1}{\bf e}_{j, p}\sum_{{\bf d}_m, {\bf d}_m'\in B({\bf d}_0, h_s)\cap\mathcal D_0} \omega_{j}^{(k)}({\bf d}_0, {\bf d}_m; h_s) \omega_{j}^{(k)}({\bf d}_0, {\bf d}_m'; h_s)\Sigma_y({\bf d}_m, {\bf d}_m'). \nonumber
  \end{eqnarray}

\noindent   {\bf Theorem 3.}      {\it Under assumptions (C1)-(C10) in Section 6 for piecewise constant $\{\beta_{j*}({\bf d}): d\in {\mathcal D}\}$,
 we have    the following results for all $0\leq s\leq S$:

(i) $\sup_{d\in {\mathcal D}_0}|\tilde\Delta_{j*}({\bf d}_0;  h_s)|=o_p(\sqrt{\log(1+N_D)/n})$, where $\tilde\Delta_{j*}({\bf d}_0;  h_s)=  \tilde \beta_{j*}({\bf d}_0;  h_s)- \beta_{j*}({\bf d}_0)$;

(ii) $\hat{\beta}_j({\bf d}_0;  h_s)-\beta_{j*}({\bf d}_0)= \sum_{{\bf d}_m\in B({\bf d}_0, h_s)\cap \mathcal D_0}   \tilde \omega_{j}^{(0)}({\bf d}_0, {\bf d}_m; h_s) \hat\Delta_j({\bf d}_m) [1+o_p(1)]$;

 (iii)
$\sup_{{\bf d}_0\in {\mathcal D}_0}|\hat\Sigma(\sqrt{n}\tilde\beta_{j*}({\bf d}_0;  h_s))-\Sigma_{j}^{(0)}({\bf d}_0;  h_s)|=o_p(1); $

(iv) $\sqrt{n}\{\hat{\beta}_j({\bf d}_0;  h_s)-\beta_{j*}({\bf d}_0)\}$ converges in distribution to a normal distribution with mean zero and variance $\Sigma_{j}^{(0)}({\bf d}_0;  h_s)$ as $n\rightarrow\infty$.

 }

%    Theorem 3 shows that MASS has remarkable features for a piecewise constant   function $\beta_{j*}({\bf d}_0)$.
% Theorem 3 (i) quantifies the maximum absolute difference (or bias) between the true value
%${\beta}_{j*}({\bf d}_0, h_s)$ and the pseudo true value  $\tilde \beta_{j*}({\bf d}_0;  h_s)$ across all $d\in {\mathcal D}_0$ for any $s$.  Since $\tilde\Delta_{j*}({\bf d}_0;  h_s)=0$ for
%$d\in  {\mathcal D}\setminus \partial {\mathcal
%D}^{(j)}(h_s)$,  this result delineates the potential biase    for voxels  $d$ in $ \partial {\mathcal
%D}^{(j)}(h_s)$.
%Theorem 3 (ii) indicates the asymptotic equivalence between $\hat{\beta}_j({\bf d}_0;  h_s)-\beta_{j*}({\bf d}_0)$ and $\sum_{d_m\in B({\bf d}_0, h_s)}   \tilde \omega_{j}^{(0)}(d, d_m; h_s) \hat\Delta_j(d_m)$.
%Theorem 3 (iii) ensures that
%$\hat\Sigma(\sqrt{n}\tilde\beta_{j*}({\bf d}_0;  h_s))$ is a uniform consistent estimator of $\Sigma_{j}({\bf d}_0;  h_s)$  across $d\in {\mathcal D}_0$.
%Theorem 3 (iv) ensures that $\sqrt{n}[\hat{\beta}_j({\bf d}_0;  h_s)-\beta_{j*}({\bf d}_0)]$ is   asymptotically normally
%distributed.
%It also indicates that
% the Wald test statistic $W_\mu({\bf d}_0;  h)$ is asymptotically $\chi^2(r)$
%distributed under the null hypothesis $R_1\ve\beta({\bf d}_0)={\bf b}_0$.

  We now consider  a much complex  scenario  when $\beta_{j*}({\bf d}_0)$ is piecewise smooth.
In this case,   $  \tilde \beta_{j*}({\bf d}_0;  h_s)$ may vary from $h_0$ to $h_S$ for all voxels ${\bf d}_0\in   {\mathcal
D}$ regardless whether ${\bf d}_0$ belongs to  $\partial {\mathcal D}^{(j)}(h_s)$ or not.
In this case, if  $\beta_{j*}({\bf d}_0)$ is Lipschitz  continuous for  each piece,
 it will be shown below that the bias of $\tilde \beta_{j*}({\bf d}_0;  h_s)$ is always at the order of $h_s$ for ${\bf d}_0\in {\mathcal D}\setminus\partial {\mathcal D}^{(j)}(h_s)$.
 However, for ${\bf d}_0\in \partial {\mathcal D}^{(j)}(h_s)$, only when $P_j({\bf d}_0, h_s)^c\cap I_j({\bf d}_0, \delta_L, \delta_U)$ is an empty set,
  we   can control the bias of $\tilde \beta_{j*}({\bf d}_0;  h_s)$ to be  at the order of $h_s$.
 Therefore, to control the bias of $\tilde \beta_{j*}({\bf d}_0;  h_s)$ across all voxels,  $h_s$ must  converge to zero. Moreover, as shown below,
 we can only establish the asymptotic normality of $\hat{\beta}_j({\bf d}_0;  h_s)$ relative to $\tilde\beta_{j*}({\bf d}_0;  h_s)$, not $\beta_{j*}({\bf d}_0)$.
These results  differ  significantly from those for piecewise smooth
 $\beta_{j*}({\bf d}_0)$.
 Generally, for  a fixed  $S>0$, we can   establish important theoretical results to
characterize the asymptotic behavior of $\hat{\ve\beta}({\bf d}_0;  h_s)$
  as follows.

\noindent   {\bf Theorem 4.}      {\it
  Suppose assumptions (C1)-(C9) and (C11) in Section 6 hold for piecewise continuous $\{\beta_{j*}({\bf d}): d\in {\mathcal D}\}$. For all $0\leq s\leq S$,  we have    the following results:

(i) $\sup_{{\bf d}_0\in {\mathcal D}_0}|\tilde\Delta_{j*}({\bf d}_0;  h_s)|=O_p(h_s)$;

(ii) $\hat{\beta}_j({\bf d}_0;  h_s)-\tilde\beta_{j*}({\bf d}_0;  h_s)= \sum_{{\bf d}_m\in B({\bf d}_0, h_s) \cap \mathcal D_0}   \tilde \omega_{j}^{(1)}({\bf d}_0, {\bf d}_m; h_s) \hat\Delta_j({\bf d}_m) [1+o_p(1)]$;

 (iii)
$\sup_{{\bf d}_0\in {\mathcal D}_0}|\hat\Sigma(\sqrt{n}\tilde\beta_{j*}({\bf d}_0;  h_s))-\Sigma_{j}^{(1)}({\bf d}_0;  h_s)|=o_p(1). $

(iv) $\sqrt{n}\{\hat{\beta}_j({\bf d}_0;  h_s)-\tilde\beta_{j*}({\bf d}_0;  h_s)\}$ converges in distribution to a normal distribution with mean zero and variance $\Sigma_{j}^{(1)}({\bf d}_0;  h_s)$ as $n\rightarrow\infty$.

 }

    Theorem 4 characterizes several key features of  MASS   for a piecewise continuous function $\beta_{j*}({\bf d}_0)$.
 Theorem 4 (i) quantifies the  bias of the pseudo true value  $\tilde \beta_{j*}({\bf d}_0;  h_s)$ relative to  the true value
${\beta}_{j*}({\bf d}_0)$ across all ${\bf d}_0\in {\mathcal D}_0$ for a fixed $s$.  Even for  voxels inside  the smooth areas of $\beta_{j*}({\bf d}_0)$, the bias
 $O_p(h_s)$ is still much higher than the standard bias at the rate of $h_s^2$ due to the presence of
$K_{st}(D_{{\beta}_j}({\bf d}_0, {\bf d}_0';  h_{s-1})/C_{n})$.  If we set $K_{st}(u)={\bf 1}(u\in [0, 1])$ and $\beta_{j*}({\bf d}_0)$ is twice differentiable,  then
 the  bias of    $\tilde \beta_{j*}({\bf d}_0;  h_s)$ relative to ${\beta}_{j*}({\bf d}_0)$  may be reduced to be close to $O_p(h_s^2)$.
  Theorem 4 (ii) establishes the asymptotic equivalence between $\hat{\beta}_j({\bf d}_0;  h_s)-\tilde \beta_{j*}({\bf d}_0;  h_s)$ and $\sum_{{\bf d}_m\in B({\bf d}_0, h_s)\cap \mathcal D_0}   \tilde \omega_{j}^{(1)}({\bf d}, {\bf d}_m; h_s) \hat\Delta_j({\bf d}_m)$.
Theorem 4 (iii) ensures that
$\hat\Sigma(\sqrt{n}\tilde\beta_{j*}({\bf d}_0;  h_s))$ is a uniform consistent estimator of $\Sigma_{j}^{(1)}({\bf d}_0;  h_s)$  across ${\bf d}_0\in {\mathcal D}_0$.
Theorem 4 (iv) ensures that $\sqrt{n}\{\hat{\beta}_j({\bf d}_0;  h_s)-\tilde\beta_{j*}({\bf d}_0;  h_s)\}$ is   asymptotically normally
distributed.
%It also indicates that
% the Wald test statistic $W_\mu({\bf d}_0;  h)$ is asymptotically $\chi^2(r)$
%distributed under the null hypothesis $R_1\ve\beta({\bf d}_0)={\bf b}_0$.

 Finally, we    delineate   the asymptotic variance of $\hat{\beta}_j({\bf d}_0;  h_s)$. For   simplicity, we  focus on  ${\bf d}_0\in {\mathcal D}\setminus\partial {\mathcal D}^{(j)}(h_s)$ and do not distinguish the piecewise constant case and
 the piecewise continuous one. Let $\tilde K_{loc}(||{\bf d}_0-{\bf d}_m||_2/h)=K_{loc}(||{\bf d}_0-{\bf d}_m||_2/h)/[\sum_{{\bf d}_m' \in B({\bf d}_0;  h)\cap \mathcal D_0} K_{loc}(||{\bf d}_0-{\bf d}_m'||_2/h)]$.  It follows from (\ref{Th3Eq1})  that for $k=0$ and $1$,   $\Sigma_{j}^{(k)}({\bf d}_0;  h_s)/{\bf e}_{j, p}^T\Omega_{X}^{-1}{\bf e}_{j, p}$ equals the sum of terms (T1) and  (T2), which are, respectively,  given by
 \begin{eqnarray}
   (\mbox{T1})&=&{ \sum_{{\bf d}_m, {\bf d}_m'\in B({\bf d}_0;  h_s)\cap \mathcal D_0}\tilde K_{loc}(||{\bf d}_0-{\bf d}_m||^2/h_s)\tilde K_{loc}(||{\bf d}_0-{\bf d}_m'||^2/h_s) \Sigma_\eta({\bf d}_m, {\bf d}_m')}  \nonumber  \\
   &=& \sum_{l=1}^\infty \lambda_l [\sum_{{\bf d}_m\in B({\bf d}_0;  h_s)\cap \mathcal D_0}  \tilde K_{loc}(||{\bf d}_0-{\bf d}_m||^2/h_s)\psi_l({\bf d}_m)]^2,  \\
  (\mbox{T2}) &=&{ \sum_{{\bf d}_m\in B({\bf d}_0;  h_s)\cap \mathcal D_0}\tilde K_{loc}(||{\bf d}_0-{\bf d}_m||^2/h_s)^2 \Sigma_\epsilon({\bf d}_m, {\bf d}_m)}. \nonumber
 \end{eqnarray}
 If $h_s\rightarrow 0$ and $N_Dh_s^{3/2}\rightarrow\infty$,  it can be shown that (T1) and (T2), respectively, converge to $\Sigma_\eta({\bf d}_0, {\bf d}_0)$ and $0$. Thus,
both  $\Sigma_{j}^{(0)}({\bf d}_0;  h_s)$ and $\Sigma_{j}^{(1)}({\bf d}_0;  h_s)$ converge  to  ${\bf e}_{j, p}^T\Omega_{X}^{-1}{\bf e}_{j, p}\Sigma_\eta({\bf d}_0, {\bf d}_0)$, which is smaller than
the asymptotic variance of the raw estimate $\hat{\beta}_j({\bf d}_0)$.
In general, for relatively small $h_s$,  MASS leads to smaller standard deviations for  estimating $\beta_j({\bf d}_0)$.

\section*{Proofs}

  {\it Proof of Theorem 1.} The proof of Theorem 1 (i) can be easily proved by using the standard
  asymptotic arguments \citep{VaartWellner1996}, so we omit repeating them here.
To prove Theorem 1(ii), we will show
 \begin{equation} \label{Th1Step1}
\sup_{{\bf d}_0\in {\mathcal D}_0} ||\hat{\ve\beta}({\bf d}_0)- {\ve\beta}_*({\bf d}_0)||_2=O_p(n^{-1/2}\sqrt{\log(1+N_D)}).
\end{equation}
It is easy to show that 
 \begin{eqnarray}
 \hat{\ve\beta}({\bf d}_0)&=&(\sum_{i=1}^n {\bf
x}_i^{\otimes 2})^{-1}\sum_{i=1}^n {\bf
x}_iy_i({\bf d}_0)={\ve\beta}_*({\bf d}_0)+A_{n, \eta}({\bf d}_0)+A_{n, \epsilon}({\bf d}_0) \nonumber \\
&=& {\ve\beta}_*({\bf d}_0)+(\sum_{i=1}^n {\bf x}_i^{\otimes
2})^{-1}\sum_{i=1}^n {\bf x}_i{\eta}_i({\bf d}_0)+(\sum_{i=1}^n {\bf x}_i^{\otimes
2})^{-1}\sum_{i=1}^n {\bf x}_i{\epsilon}_i({\bf d}_0).  \label{Th1Eq0}
\end{eqnarray}
It follows from the law of the large number and assumption (C3) that $n^{-1}\sum_{i=1}^n {\bf x}_i^{\otimes 2}$ converges to $\Omega_X$ almost surely.
It follows from assumption (C4) that $\{ {\bf x} \eta({\bf d}): {\bf d}\in {\mathcal D}\}$ is a Donsker class and
$\sum_{i=1}^n {\bf x}_i{\eta}_i({\bf d})/\sqrt{n}$ converges to a Gaussian process with zero mean and covariance function
$\Omega_X \Sigma_\eta({\bf d}, {\bf d}')$ as $n\rightarrow \infty$ \citep{VaartWellner1996}.  Thus,  we have
\begin{equation} \label{Th1Eq1}
\sup_{{\bf d}\in {\mathcal D}} |\sum_{i=1}^n {\bf x}_i{\eta}_i({\bf d})|=O_p(\sqrt{n}).
\end{equation}
It follows from assumptions (C2) and (C3) that
$$
P(|\sum_{i=1}^n {x}_{ij}{\epsilon}_i({\bf d}_0)|>t)\leq C  \exp(- \frac{C  t^2}{n   ||{\bf x}||_\infty^2\max_{{\bf d}_0\in {\mathcal D}_0} ||\epsilon_i({\bf d}_0)||_{\psi_2}^2 }),
$$
where $C$ is a generic constant and $||\cdot||_{\psi_l}$ denotes the Orlicz norm for $\psi_l(x)=\exp(x^l)-1$.
Then, we can apply Lemma 2.2.10 in  \cite{VaartWellner1996}  to get
$$
||\max_{{\bf d}_0\in {\mathcal D}_0} |\sum_{i=1}^n {x}_{ij}{\epsilon}_i({\bf d}_0)|||_{\psi_1}\leq C\{ \sqrt{n}  (||{\bf x}||_\infty C)\sqrt{\log (1+N_D)}\}.
$$
 Finally, we get
\begin{equation}\label{Th1Eq2}
\max_{{\bf d}_0\in {\mathcal D}_0} |\sum_{i=1}^n {x}_{ij}{\epsilon}_i({\bf d}_0)|=O_p(\sqrt{n\log(1+N_D)}).
\end{equation}
By combining (\ref{Th1Eq0})-(\ref{Th1Eq2}), we can finish the proof of (\ref{Th1Step1}).

   We define some notation as follows:
 \begin{eqnarray*}  \label{Th2Eq1}
&& \Delta_{j*}({\bf d}, {\bf d}')= \beta_{j*}({\bf d})-\beta_{j*}({\bf d}'),  \\ 
 && 
 I_j({\bf d}, \delta_1, \delta_2)=\{{\bf d}':  {\bf d}'\in {\mathcal D}, \delta_1\leq |\Delta_{j*}({\bf d}, {\bf d}')|< \delta_2\}~~~\mbox{for}~~~
 j=1, \ldots, p, \\ 
&&\tilde K_h^0({\bf d}_m,  {\bf d}) =(1, 0, 0, 0)\{\sum_{{\bf d}_m }K_{h}({\bf d}_m-{\bf d}){\bf
z}_{h}({\bf d}_m-{\bf d})^{\otimes 2}\}^{-1}  K_{h}({\bf d}_m-{\bf d}){\bf z}_{h}({\bf d}_m-{\bf d}), \\
&&\hat \eta_i({\bf d})=(1, 0, 0, 0)\hat C_i({\bf d})=\sum_{{\bf d}_m}\tilde K^0_h ({\bf d}_m;  {\bf d})\{{
y}_{i}({\bf d}_m)-{\bf x}_i^T\hat {\ve\beta}({\bf d}_m)\},   \\ 
  &&\overline \epsilon_{i}({\bf d})= \sum_{{\bf d}_m} \tilde K_{h}^0({\bf d}_{m}, {\bf d})\epsilon_{i}({\bf d}_m),~~~~~~~
 \Delta { \eta}_{i}({\bf d})= \sum_{{\bf d}_m} \tilde K_{h}^0({\bf d}_{m}, {\bf d})[{\eta}_{i}({\bf d}_m)-{\eta}_{i}({\bf d})], \\
 &&  \Delta_{i}({\bf d})=\overline \epsilon_{i}({\bf d})+\Delta \eta_{i}({\bf d})+{\bf x}_i^T \Delta {\ve\beta}({\bf d}),~~
 \Delta {\ve\beta}({\bf d})=\sum_{{\bf d}_m} \tilde  K_{h}^0({\bf d}_{m},  {\bf d})
 [{\ve\beta}_*({\bf d}_m)-\hat {\ve\beta}({\bf d}_m)], 
 \nonumber
 \end{eqnarray*}
 where $\delta_2>0$ and $\delta_1\geq 0$ are non-negative scalars.  
 Moreover,
  $\Pi_{N_D}(\cdot)$ is   the sampling distribution function
based on ${\mathcal D}_0$,
 and $\Pi(\cdot)$ is the distribution function of ${\bf d}$.
 We need the following lemmas to prove Theorem 2.

 \noindent {\it Lemma 1}. {\it
   Under Assumptions (C1)-(C7),   we have the following results:  }
\begin{eqnarray} % \label{Lemma5Eq1}
&& \sup_{{\bf d}\in \mathcal D}\left|\int
K_{h}({\bf u}-{\bf d})\frac{ \prod_{k=1}^3(u_k- d_k)^r}{h^{3r}}d[\Pi_{N_D}({\bf u})-\Pi({\bf u})]\right| \nonumber \\
&&
~~~~~~~~=O_p((N_Dh^3)^{-1/2}\max(3|\log h|, \log\log N_D)^{1/2}), \label{Lemma1Eq0}  \\
 && \sup_{{\bf d}\in \mathcal D}n^{-1}
 |\sum_{i=1}^n\overline \epsilon_{i}({\bf d}){\bf x}_{i}|=o_p(n^{-1/2}), \label{Lemma1Eq1} \\
 && \sup_{({\bf d}, {\bf d}')\in \mathcal D^2}n^{-1}
 |\sum_{i=1}^n\overline \epsilon_{i}({\bf d})\Delta\eta_{i}({\bf d}')|=O_p(n^{-1/2}(\log n)^{1/2}), \label{Lemma1Eq2}
 \\
&&\sup_{({\bf d}, {\bf d}')\in \mathcal D^2}n^{-1}
 |\sum_{i=1}^n\overline\epsilon_{i}({\bf d}) \overline\epsilon_{i}({\bf d}')|=O_p((N_Dh^{3})^{-1}  + (\log n/n)^{1/2}). \label{Lemma1Eq3}
 \end{eqnarray}

\noindent  {\it Proof of Lemma 1.}  Equation (\ref{Lemma1Eq0}) follows directly from Theorem 1 of \cite{EinmahlMason2000}.
 It follows from (\ref{Lemma1Eq0}) that   for large enough $N_D$, there exists a constant $C_1>1$ such that
\[
 \sup_{{\bf d}\in \mathcal D}n^{-1}|\sum_{i=1}^n\overline \epsilon_{i}({\bf d}){\bf x}_{i}| \leq  n^{-1/2}C_1  \sup_{{\bf d}\in \mathcal D} |N_D^{-1}\pi({\bf d})^{-1}\sum_{m=1}^{N_D}
K_{h}({\bf d}_m-{\bf d})  F_n({\bf d}_m)|,
\]
where
  $F_n({\bf d}_m)=
 n^{-1/2}\sum_{i=1}^n {\bf x}_i\epsilon_{i}({\bf d}_m)$. By following the arguments in \cite{EinmahlMason2000}, we can show that
 $$
 \sup_{{\bf d}\in \mathcal D} \pi({\bf d})^{-1} |N_D^{-1}\sum_{m=1}^{N_D}
K_{h}({\bf d}_m-{\bf d})  F_n({\bf d}_m)|
 =O_p((N_Dh^3)^{-1/2}|\log h|^{1/2})=o_p(1),
$$
which yields (\ref{Lemma1Eq1}).

By following Lemmas 1-4 of \citet{LiHsing2010}, we can   prove (\ref{Lemma1Eq2}) and (\ref{Lemma1Eq3}).   Let's consider (\ref{Lemma1Eq3}) as an illustration.
  We define  $\Delta_{n, \epsilon\epsilon}({\bf d}, {\bf d}')
 = \sum_{i=1}^n\overline \epsilon_{i}({\bf d})\overline\epsilon_{i}({\bf d}')$ and
 \begin{eqnarray*}
\Delta_{n, \epsilon\epsilon}^{(1)}({\bf d}, {\bf d}')&=&\Delta_{n, \epsilon\epsilon}^{(1,1)}({\bf d}, {\bf d}')+\Delta_{n, \epsilon\epsilon}^{(1,2)}({\bf d}, {\bf d}')\\
 &=&n^{-1} \sum_{i=1}^n \frac 1{N_D^2 \pi({\bf d})\pi({\bf d}')}\sum_{m=1}^{N_D}
  K_{h}({\bf d}_m-{\bf d})K_{h}({\bf d}_{m}-{\bf d}')  [\epsilon_{i}({\bf d}_m)^2-\Sigma_\epsilon({\bf d}_m, {\bf d}_m)]\\
  &+&n^{-1} \sum_{i=1}^n \frac 1{N_D^2 \pi({\bf d})\pi({\bf d}')}\sum_{m=1}^{N_D}
  K_{h}({\bf d}_m-{\bf d})K_{h}({\bf d}_{m}-{\bf d}')  \Sigma_\epsilon({\bf d}_m, {\bf d}_m),
  \end{eqnarray*}
 \begin{eqnarray*}
  \Delta_{n, \epsilon\epsilon}^{(2)}({\bf d}, {\bf d}')
 &=&n^{-1} \sum_{i=1}^n \frac 1{N_D^2 \pi({\bf d})\pi({\bf d}')}\sum_{m\not= m'}
  K_{h}({\bf d}_m-{\bf d})K_{h}({\bf d}_{m'}-d') \epsilon_{i}({\bf d}_m)\epsilon_{i}({\bf d}_{m'}).
%%  &&V_{n, \epsilon\epsilon}(s,  v_1,  t, v_2)\\
%%& =&|\frac 1n \sum_{i=1}^n \frac 1{M^2 \pi(s)\pi(t)}\sum_{m,
%%m'=1}^{M} \epsilon_{ij}(s_m)\epsilon_{ij}(s_{m'})
%%{\bf 1}(s\leq s_m\leq s+v_1){\bf 1}(t\leq s_{m'}\leq t+v_2)|.
  \end{eqnarray*}
For  large enough $N_D$, there exists a constant $C_1>1$ such that
  \begin{eqnarray*}
&& \sup_{({\bf d}, {\bf d}')\in \mathcal D^2} n^{-1} |\Delta_{n,  \epsilon\epsilon}({\bf d}, {\bf d}')| \\
&& \leq C_1\sup_{({\bf d}, {\bf d}')\in  \mathcal D^2}\left |n^{-1} \sum_{i=1}^n \frac
1{N_D^2 \pi({\bf d})\pi({\bf d}')}\sum_{m, m'=1}^{N_D}
  K_{h}({\bf d}_m-{\bf d})K_{h}({\bf d}_{m'}-d) \epsilon_{i}({\bf d}_m)\epsilon_{i}({\bf d}_{m'})
\right|  \nonumber
\\
&&\leq C_1\{\sup_{({\bf d}, {\bf d}')\in  \mathcal D^2}\left|\Delta_{n,
\epsilon\epsilon}^{(1)}({\bf d}, {\bf d}')  \right|+\sup_{({\bf d}, {\bf d}')\in  \mathcal D^2}\left|\Delta_{n, \epsilon\epsilon}^{(2)}({\bf d}, {\bf d}')  \right|\}.
 \end{eqnarray*}
Similar to the arguments in  Lemmas 3 and 4 of \citet{LiHsing2010},
we have
$$
\sup_{({\bf d}, {\bf d}')\in  \mathcal D^2}\left|\Delta_{n,
\epsilon\epsilon}^{(2)}({\bf d}, {\bf d}')  \right|=O(\sqrt{\log
n/n})~~\mbox{a.s}.
$$
Thus,
we only  need to consider $\sup_{({\bf d}, {\bf d}')\in  \mathcal D^2}\left|\Delta_{n,
\epsilon\epsilon}^{(1)}({\bf d}, {\bf d}')  \right|$. Similar to the arguments in  Lemmas 1 and 2 of \citet{LiHsing2010}, we can obtain
\begin{eqnarray*}
 \sup_{({\bf d}, {\bf d}')\in  \mathcal D^2}\left |\Delta_{n, \epsilon\epsilon}^{(1,1)}({\bf d}, {\bf d}')  \right|
%\leq C_2 (Mh_{2j})^{-1}
% \sup_{s\in [0, 1]} \frac 1n \sum_{i=1}^n \frac 1{M  \pi(s)}\sum_{m=1}^{M}
%  K_{h_{2j}}(s_m-s)  \epsilon_{ij}(s_m)^2 \\
%  && \leq C_3 (Mh_{2j})^{-1}  + C_4\sup_{s\in [0, 1]} \left| \frac 1n \sum_{i=1}^n \frac 1{M  \pi(s)}\sum_{m=1}^{M}
%  K_{h_{2j}}(s_m-s)  [\epsilon_{ij}(s_m)^2-\Sigma_\epsilon(s_m, s_m)]\right| \\
=O_p( (N_Dh^3)^{-1}(\log n/n)^{1/2}),~~
 \sup_{({\bf d}, {\bf d}')\in  \mathcal D^2}\left |\Delta_{n, \epsilon\epsilon}^{(1,2)}({\bf d}, {\bf d}')  \right|=O_p((N_Dh^3)^{-1}),
\end{eqnarray*}
which yield (\ref{Lemma1Eq3}).
This completes the proof of Lemma 1.

 \noindent {\it Lemma 2}. {\it
    Under Assumptions (C1)-(C7),  we have the following results:  }
\begin{eqnarray} % \label{Lemma5Eq1}
 && \sup_{{\bf d}_0\in \mathcal D_0} n^{-1}|\sum_{i=1}^n  {\bf x}_i^T \Delta {\ve\beta}({\bf d}_0)  \epsilon_i({\bf d}_0)|=O_p(n^{-1}\log(1+N_D)),
  \label{Lemma2Eq1} \\
 && \sup_{{\bf d}_0\in \mathcal D_0} n^{-1}|\sum_{i=1}^n  \Delta \eta_{i}({\bf d}_0)\epsilon_i({\bf d}_0)|=O_p(\sqrt{h^2+n^{-1/2}}),   \label{Lemma2Eq2} \\
 && \sup_{{\bf d}_0\in \mathcal D_0} n^{-1}|\sum_{i=1}^n  \overline \epsilon_{i}({\bf d}_0)\epsilon_i({\bf d}_0)|=O_p(\sqrt{(N_Dh^{3})^{-1}  + (\log n/n)^{1/2}}).    \label{Lemma2Eq3}
\end{eqnarray}

 \noindent  {\it Proof of Lemma 2.}  It follows from Theorem 1 and the Cauchy-Schwarz inequality that
 $$
 \sup_{{\bf d}_0\in \mathcal D_0} n^{-1}|\sum_{i=1}^n  {\bf x}_i^T \Delta {\ve\beta}({\bf d}_0)  \epsilon_i({\bf d}_0)| \leq  \sup_{{\bf d}_0\in \mathcal D_0} n^{-1}||\sum_{i=1}^n  {\bf x}_i   \epsilon_i({\bf d}_0)||_2
 ||\Delta {\ve\beta}({\bf d}_0)||_2=O_p(n^{-1}\log(1+N_D)).
 $$
 It follows from   the Cauchy-Schwarz inequality  that
\begin{eqnarray*}
 \sup_{{\bf d}_0\in \mathcal D_0}\{ n^{-1}|\sum_{i=1}^n  \Delta \eta_{i}({\bf d}_0)\epsilon_i({\bf d}_0)|\}^2 &\leq&   \sup_{{\bf d}_0\in \mathcal D_0} \{n^{-1}\sum_{i=1}^n  \Delta \eta_{i}({\bf d}_0)^2\}  \sup_{{\bf d}_0\in \mathcal D_0} \{n^{-1}\sum_{i=1}^n  \epsilon_{i}({\bf d}_0)^2\}, \\
 \sup_{{\bf d}_0\in \mathcal D_0} \{n^{-1}|\sum_{i=1}^n  \overline \epsilon_{i}({\bf d}_0)\epsilon_i({\bf d}_0)|\}^2 &\leq&
 \sup_{{\bf d}_0\in \mathcal D_0} \{n^{-1}\sum_{i=1}^n  \overline \epsilon_{i}({\bf d}_0)^2\}  \sup_{{\bf d}_0\in \mathcal D_0} \{n^{-1}\sum_{i=1}^n  \epsilon_{i}({\bf d}_0)^2\}.
 \end{eqnarray*}
Let $\Delta^{(2)}_{\Sigma, \eta}({\bf d}_0, {\bf d}_m, {\bf d}_m')=\Sigma_\eta({\bf d}_0, {\bf d}_m')+\Sigma_\eta({\bf d}_m, {\bf d}_0)-\Sigma_\eta({\bf d}_m, {\bf d}_m')-\Sigma_\eta({\bf d}_0, {\bf d}_0)$. Based on assumption (C4), we have
\begin{eqnarray}
&& \sup_{{\bf d}_0\in \mathcal D_0} |n^{-1} \sum_{i=1}^n\Delta \eta_{i}({\bf d}_0)^2|\leq    O_p(h^2+n^{-1/2})=  \label{Lemma2PF1}  \\
 &&
 \sup_{{\bf d}_0, {\bf d}_0'}\{ |\sum_{{\bf d}_m, {\bf d}_m'}\tilde K_h^0({\bf d}_m, {\bf d}_0)\tilde K_h^0({\bf d}_m', {\bf d}_0)\Delta^{(2)}_{\Sigma, \eta}({\bf d}_0, {\bf d}_m, {\bf d}_m')|
 +
 \sum_{{\bf d}_m, {\bf d}_m'}|\tilde K_h^0({\bf d}_m, {\bf d}_0)\tilde K_h^0({\bf d}_m', {\bf d}_0')|\times  \nonumber \\
 && \sup_{ {\bf d}_m, {\bf d}_m'} ||n^{-1}\sum_{i=1}^n [\eta_i({\bf d}_m)-\eta_i({\bf d}_0)]
[\eta_i({\bf d}_m')-\eta_i({\bf d}_0)]-
 \Delta^{(2)}_{\Sigma, \eta}({\bf d}_0, {\bf d}_m, {\bf d}_m')
||\}.
\nonumber
\end{eqnarray}
Let $\hat\Sigma_\epsilon({\bf d}_0)=n^{-1}\sum_{i=1}^n  \epsilon_{i}({\bf d}_0)^2$ and $\lambda(K_\epsilon, n, N_D)=K_\epsilon\log (2N_D)/n+\sqrt{2\log(2N_D)/n}$.
It follows assumption (C2) and Lemma  14.13 of \cite{Buhlmann2011} that
\begin{equation}
 P(\sup_{{\bf d}_0\in\mathcal D_0}|\hat\Sigma_\epsilon({\bf d}_0)-\Sigma_\epsilon({\bf d}_0, {\bf d}_0)|\geq 2K_\epsilon^2t+2K_\epsilon C_\epsilon \sqrt{2t} +2K_\epsilon C_\epsilon \lambda(K_\epsilon, n, N_D))\leq \exp(-nt),  \label{Lemma2PF2}
\end{equation}
which yields that $\sup_{{\bf d}\in\mathcal D_0}|\hat\Sigma_\epsilon({\bf d})-\Sigma_\epsilon({\bf d}_0, {\bf d}_0)|=o_p(1)$.
Combining (\ref{Lemma2PF1}) and (\ref{Lemma2PF2}) yields (\ref{Lemma2Eq2}).
Similarly, we can prove (\ref{Lemma2Eq3}).

\noindent  {\it Proof of Theorem 2.}
We have 
 \begin{equation}\label{Th2Eq2}
 \hat \eta_{i}({\bf d})-\eta_{i}({\bf d})=\Delta_{i}({\bf d})=\overline \epsilon_{i}({\bf d})+\Delta \eta_{i}({\bf d})+{\bf x}_i^T \Delta \ve\beta({\bf d}).
 \end{equation}
  Therefore, we have
 \begin{eqnarray} \label{Th2Eq3}
 n^{-1}\sum_{i=1}^n  \hat \eta_{i}({\bf d}) \hat \eta_{i}({\bf d}')&=&n^{-1}
 \sum_{i=1}^n\Delta_{i}({\bf d}) \Delta_{i}({\bf d}')+ n^{-1}
 \sum_{i=1}^n\eta_{i}({\bf d}) \Delta_{i}({\bf d}') \\
 &&+n^{-1}\sum_{i=1}^n\Delta_{i}({\bf d}) \eta_{i}({\bf d}')+n^{-1}\sum_{i=1}^n\eta_{i}({\bf d}) \eta_{i}({\bf d}').  \nonumber
 \end{eqnarray}
This proof  of Theorem 2 (i) consists of three steps as follows.
\begin{itemize}
\item{}
  Show the uniform convergence of $n^{-1}\sum_{i=1}^n\eta_{i}({\bf d}) \eta_{i}({\bf d}')$ to $\Sigma_\eta({\bf d}, {\bf d}')$ over
  $({\bf d}, {\bf d}')\in {\mathcal D}^2$ in probability.
 \item{}
 Show that   $n^{-1}
 \sum_{i=1}^n\eta_{i}({\bf d}) \Delta_{i}({\bf d}')+n^{-1}\sum_{i=1}^n\Delta_{i}({\bf d}) \eta_{i}({\bf d}')$  converges  to zero  uniformly for all $({\bf d}, {\bf d}')\in {\mathcal D}^2$ in probability.
  \item{}
 Show that   $n^{-1}
 \sum_{i=1}^n\Delta_{i}({\bf d}) \Delta_{i}({\bf d}')$  converges  to zero  uniformly for all $({\bf d}, {\bf d}')\in {\mathcal D}^2$ in probability.
\end{itemize}

In the first step,  it follows from assumption (C4) that
 \begin{equation} \label{Th2Eq4}
\sup_{({\bf d}, {\bf d}')\in \mathcal D^2} |n^{-1}\sum_{i=1}^n[ \eta_{i}({\bf d})
\eta_{i}({\bf d}')-\Sigma_{\eta}({\bf d}, {\bf d}')]|=O_p(n^{-1/2}).
 \end{equation}

In the second step, we can   show  that
\begin{equation} \label{Th2Eq5}
 \sup_{({\bf d}, {\bf d}')\in \mathcal D^2}n^{-1}
 |\sum_{i=1}^n\Delta_{i}({\bf d}) \eta_{i}({\bf d}')|=o_p(h^2+(\log
n/n)^{1/2}+n^{-1}\sqrt{\log(1+N_{\mathcal D})}).
 \end{equation}
With some simple calculations, we have
 \begin{equation} \label{Th2Eq6}
\sum_{i=1}^n\Delta_{i}({\bf d}) \eta_{i}({\bf d}') \leq   \{
|\sum_{i=1}^n\overline \epsilon_{i}({\bf d})\eta_{i}({\bf d}')|+ |\sum_{i=1}^n{\bf
x}_i^T \Delta \ve\beta({\bf d}) \eta_{i}({\bf d}')|+
|\sum_{i=1}^n\Delta \eta_{i}({\bf d})\eta_{i}({\bf d}')|\}.
\end{equation}
Thus,
it is sufficient to focus on the three terms on the right-hand side
of (\ref{Th2Eq6}).
First,  it follows from Lemma 1 that  $\sup_{({\bf d}, {\bf d}')}n^{-1}
\{ |\sum_{i=1}^n\overline \epsilon_{i}({\bf d})\eta_{i}({\bf d}')|=O((\log
n/n)^{1/2})$.
Secondly,
since $\{{\bf x} \eta({\bf d}): {\bf d}\in {\mathcal D}\}$ is a Donsker class and $\sup_{d} || \Delta \ve\beta({\bf d})||_2=O_p(n^{-1/2}\sqrt{\log(1+N_{\mathcal D})})$, we have
 $$
 n^{-1} |\sum_{i=1}^n{\bf x}_i^T \Delta \ve\beta({\bf d}) \eta_{i}({\bf d}')|\leq  \sup_{d} || \Delta \ve\beta({\bf d})||_2  ||n^{-1} \sum_{i=1}^n {\bf x}_i^T \eta_{i}({\bf d}')||_2=O_p(n^{-1}\sqrt{\log(1+N_{\mathcal D})}).
 $$
Thirdly, based on the definition of $\Delta\eta_i({\bf d})$, we have
\begin{eqnarray}
 &&n^{-1} \sum_{i=1}^n\Delta \eta_{i}({\bf d})\eta_{i}({\bf d}')
 =\{\sum_{{\bf d}_m}\tilde K_h^0({\bf d}_m, {\bf d})\Sigma_\eta({\bf d}_m, {\bf d}')-\Sigma_\eta({\bf d}, {\bf d}')\}  \label{Th2Eq7}
 \\
 &&
 + \sum_{{\bf d}_m}\tilde K_h^0({\bf d}_m, d)n^{-1}\{\sum_{i=1}^n[\eta_i({\bf d}_m)\eta_i({\bf d}')-\Sigma_\eta({\bf d}_m, {\bf d}')-\eta_i({\bf d})\eta_i({\bf d}')+\Sigma_\eta({\bf d}, {\bf d}')]\}.   \nonumber
 \end{eqnarray}
It follows from assumption (C4) that   the first term on the right hand side of (\ref{Th2Eq7}) is $O_p(h^2)$ and
the second one is
$O_p(n^{-1/2}).$

The third step is to show that
  \begin{equation} \label{Th2Eq8}
 \sup_{({\bf d}, {\bf d}')}n^{-1}
 |\sum_{i=1}^n\Delta_{i}({\bf d}) \Delta_{i}({\bf d}')|=O_p((N_Dh^3)^{-1}  + (\log n/n)^{1/2}+  h^2+n^{-1}\log(1+N_{\mathcal D})).
\end{equation}
With some calculations, we have
\begin{eqnarray}
|\sum_{i=1}^n\Delta_{i}({\bf d}) \Delta_{i}({\bf d}')| &\leq& C_1 \sup_{(d,d')} [  |\sum_{i=1}^n\overline \epsilon_{i}({\bf d}) \overline\epsilon_{i}({\bf d}')|+
|\sum_{i=1}^n\overline \epsilon_{i}({\bf d}) \Delta \eta_{i}({\bf d}')|  \nonumber
\\
&+&
|\sum_{i=1}^n\overline \epsilon_{i}({\bf d}){\bf x}_i^T\Delta \ve\beta({\bf d}')|+ |\sum_{i=1}^n\Delta \eta_{i}({\bf d}'){\bf x}_i^T\Delta \ve\beta({\bf d})| \label{Th2Eq8}\\
&+&   |\sum_{i=1}^n\Delta \eta_{i}({\bf d}) \Delta\eta_{i}({\bf d}')|+|\sum_{i=1}^n{\bf x}_i^T\Delta\ve\beta({\bf d})\Delta \ve\beta({\bf d}'){\bf x}_i|], \nonumber
\end{eqnarray}
for a positive constant $C_1$.
 It follows from Lemma 1 that
  the first three terms on the right hand side of (\ref{Th2Eq8}) uniformly converge to zero.
We only need to
consider the last three terms on the right hand side of (\ref{Th2Eq8}) as follows:
 \begin{eqnarray*}
n^{-1} |\sum_{i=1}^n\Delta \eta_{i}({\bf d}'){\bf x}_i^T\Delta \ve\beta({\bf d})|&\leq & ||\Delta\ve\beta({\bf d})||_2n^{-1} ||\sum_{i=1}^n\Delta \eta_{i}({\bf d}'){\bf x}_i^T||_2 \\
&\leq &  ||\Delta\ve\beta({\bf d})||_2
\sum_{{\bf d}_m}|\tilde K_h^0({\bf d}_m-{\bf d}')|||n^{-1}\sum_{i=1}^n {\bf x}_i\eta_i({\bf d}_m)-n^{-1}\sum_{i=1}^n {\bf x}_i\eta_i({\bf d})||_2\\
&=&O_p(n^{-1}\sqrt{\log(1+N_{\mathcal D})}), \\
n^{-1} \sum_{i=1}^n\Delta \eta_{i}({\bf d}) \Delta\eta_{i}({\bf d}')&=&\sum_{{\bf d}_m, {\bf d}_m'}\tilde K_h^0({\bf d}_m, {\bf d})\tilde K_h^0({\bf d}_m', {\bf d}')n^{-1}\sum_{i=1}^n [\eta_i({\bf d}_m)-\eta_i({\bf d})]
[\eta_i({\bf d}_m')-\eta_i({\bf d}')]\\
&=&O_p(h^2+n^{-1/2}),
 \\
 n^{-1}\sum_{i=1}^n{\bf x}_i^T\Delta {\ve\beta}({\bf d})\Delta {\ve\beta}({\bf d}')^T{\bf x}_i&=&\mbox{tr}\{
 \Delta {\ve\beta}({\bf d})\Delta {\ve\beta}({\bf d}')^T n^{-1}\sum_{i=1}^n{\bf x}_i^{\otimes 2}
 \}=
 O_p(n^{-1}\log(1+N_{\mathcal D})).
 \end{eqnarray*}

To prove Theorem 2 (ii), we note  that
$
\hat\epsilon_i({\bf d}_m)=\epsilon_i({\bf d}_m)-{\bf x}_i^T[\hat{\ve\beta}({\bf d}_m)-{\ve\beta}({\bf d}_m)]-\Delta_i({\bf d}_m)
$
holds for all ${\bf d}_m\in \mathcal D_0$. It
  yields that
\begin{eqnarray}
&&\sup_{{\bf d}_m\in \mathcal D_0}|n^{-1}\sum_{i=1}^n \hat\epsilon_i({\bf d}_m)^2- n^{-1}\sum_{i=1}^n \epsilon_i({\bf d}_m)^2|\leq   \nonumber\\
&&2\sup_{{\bf d}_m\in \mathcal D_0} n^{-1}|\sum_{i=1}^n \Delta_i({\bf d}_m)\epsilon_i({\bf d}_m)|+
  2\sup_{{\bf d}_m\in \mathcal D_0}|n^{-1}\sum_{i=1}^n \epsilon_i({\bf d}_m){\bf x}_i^T[\hat{\ve\beta}({\bf d}_m)-{\ve\beta}({\bf d}_m)]| \label{Th2Eq9}\\
&&+n^{-1}\sup_{{\bf d}_m\in \mathcal D_0}\sum_{i=1}^n\Delta_i({\bf d}_m)^2 + \sup_{{\bf d}_m\in \mathcal D_0} [\hat{\ve\beta}({\bf d}_m)-{\ve\beta}({\bf d}_m)]^Tn^{-1}\sum_{i=1}^n{\bf x}_i^{\otimes 2}[\hat{\ve\beta}({\bf d}_m)-{\ve\beta}({\bf d}_m)]. \nonumber
\end{eqnarray}
It follows from Lemma 2 that the first two terms on the right hand side of (\ref{Th2Eq9}) are at the order of
$O_p(n^{-1}\log(1+N_D)+\sqrt{h^2+n^{-1/2}}+\sqrt{(N_Dh^3)^{-1}+(\log n/n)^{1/2}})$, while
it follows from (\ref{Th2Eq8}) and Theorem 1 that the last two terms on the right hand side of above inequality
 converge to zero uniformly for all ${\bf d}_m\in {\mathcal D}_0$ in probability.  This completes the proof of Theorem 2 (ii).

  Theorem 2 (iii)  directly follows from   the same arguments in Lemma 6 of \citet{LiHsing2010}. So, we omit the details.

 \noindent   {\it Proof of Theorem 3.}
 For $s\geq 1$, we   define
 \begin{eqnarray}
&& F_1({\bf d}_0, h_s)=\frac{\sum_{{\bf d}_m\in B({\bf d}_0, h_s)\cap \mathcal D_0} [\omega_j^{(0)}({\bf d}_0, {\bf d}_m; h_s)-\omega_{j}({\bf d}_0, {\bf d}_m; h_s)]}{ \sum_{{\bf d}_m\in B({\bf d}_0, h_s)\cap \mathcal D_0} \omega_{j}({\bf d}_0, {\bf d}_m; h_s)}, \label{Th3Eq1}   \\
&&  F_2({\bf d}_0, h_s)=\frac{\sum_{{\bf d}_m\in B({\bf d}_0, h_s)\cap \mathcal D_0} [\omega_j({\bf d}_0, {\bf d}_m; h_1)-\omega_{j}^{(0)}({\bf d}_0, {\bf d}_m; h_s)]\hat\Delta_j({\bf d}_m)}{ \sum_{{\bf d}_m\in B({\bf d}_0, h_s)\cap \mathcal D_0} \omega_{j}({\bf d}_0, {\bf d}_m; h_s)}.  \nonumber   \end{eqnarray}
  For $0\leq s\leq S$, we have the following results:
\begin{eqnarray*}
&\mbox{(R.1)} &  F_1({\bf d}_0, h_s)=o_p(1),~~   F_2({\bf d}_0, h_s)=o_p(\sqrt{\log(1+N_D)/n}), ~~  \nonumber \\
&\mbox{(R.2)} & \tilde\Delta_{j*}({\bf d}_0; h_s)=O_p(1)   N_Dh_s^3 K_{st}( 0.5C_n^{-1}n{\ve u}^{(j)}(h_s)^2/\hat\Sigma_n(\sqrt{n}\hat{\beta}_j({\bf d}_0;  h_{s-1})))=o_p(\sqrt{\log(1+N_D)/n}), \\
&\mbox{(R.3)} & \hat\beta_j({\bf d}_0; h_s)-   \beta_{j*}({\bf d}_0)= F_0({\bf d}_0, h_s)[1+o_p(1)]=O_p(\sqrt{\log(1+N_D)/n}), \\
&\mbox{(R.4)} &  \sup_{{\bf d}_0\in {\mathcal D}_0}|\hat\Sigma_n(\sqrt{n} \hat\beta_j({\bf d}_0; h_s))- \Sigma_{j}^{(0)}({\bf d}_0; h_s)|=o_p(1).   \nonumber
\end{eqnarray*}

It follows from Theorem 1 that (R.1)-(R.4) hold for $s=0$.
For $s=1$,
we  consider two different cases including (i) $ {\bf d}_0\in {\mathcal D}\setminus \partial {\mathcal
D}^{(j)}(h_S)$ and (ii) ${\bf d}_0\in \partial {\mathcal
D}^{(j)}(h_S)$.
Since $\Delta_{j*}({\bf d}_0, {\bf d}_0')=0$ for ${\bf d}_0\in {\mathcal D}_{j, l}^o$ and ${\bf d}_0'\in B({\bf d}_0; h_1)$,
$D_{\beta_j}({\bf d}_0, {\bf d}_0'; h_{0})/C_n$ can be written as
 \begin{eqnarray} \label{Th3Eq3}
D_{\beta_j}({\bf d}_0, {\bf d}_0'; h_{0})/C_n&=& C_n^{-1}n\{\hat{\beta}_j({\bf d}_0)  -
\hat{\beta}_j({\bf d}_0')\}^2/\hat\Sigma_n(\sqrt{n}\hat{\beta}_j({\bf d}_0))\\
&=&C_n^{-1}n\{\hat\Delta_j({\bf d}_0)-\hat\Delta_j({\bf d}'_0)\}^2/\hat\Sigma_n(\sqrt{n}\hat{\beta}_j({\bf d}_0))=O_p(\log(1+N_D)/C_n). \nonumber
\end{eqnarray}
Note that $O_p(1)$ in above inequality is independent of ${\bf d}_0$ and ${\bf d}_0'$.
Therefore, we have
\begin{equation}
{|K_{st}(D_{{\beta}_j}({\bf d}_0, {\bf d}_0'; h_{0})/C_{n})- K_{st}(0)|}
\leq  O_p(1)\log(1+N_D)C_n^{-1},      \label{Th3Eq4}
\end{equation}
 which yields
 \begin{equation} \label{Th3Eq5}
 F_1({\bf d}_0, h_s)=O_p(1)\log(1+N_D)C_n^{-1}=o_p(1).
\end{equation}
It follows from (\ref{Th3Eq4}) and (\ref{Th3Eq5}) that
   \begin{eqnarray}
  F_2({\bf d}_0, h_1)&= &
  \frac{\sum_{{\bf d}_m\in B({\bf d}_0, h_s)\cap \mathcal D_0} [\omega_j({\bf d}_0, {\bf d}_m; h_1)-\omega_{j}^{(0)}({\bf d}_0, {\bf d}_m; h_s)]\hat\Delta_j({\bf d}_m)}{ \sum_{{\bf d}_m\in B({\bf d}_0, h_s)\cap \mathcal D_0} \omega_{j}^{(0)}({\bf d}_0, {\bf d}_m; h_s)}
   [1+F_1({\bf d}_0, h_1)]^{-1}    \nonumber  \\
   &\leq & \{\log(1+N_D)\}^{3/2}/(C_n\sqrt{n})O_p(1).   \label{Th3Eq6}
  \end{eqnarray}
Since $\Delta_{j*}({\bf d}_0, {\bf d}_0')=0$ for ${\bf d}_0\in {\mathcal D}_{j, l}^o\cap \mathcal D_0$ and ${\bf d}_0'\in B({\bf d}_0; h_1)\cap \mathcal D_0$,  we have $\tilde \beta_{j*}({\bf d}_0; h_s)=\beta_{j*}({\bf d}_0)$ for all $s=1, \ldots, S$, which yields 
(R.2).
 It follows from (\ref{Th3Eq4})-(\ref{Th3Eq6}) that
\begin{equation}
\hat\beta_j({\bf d}_0; h_1)-   \beta_{j*}({\bf d}_0) =F_2({\bf d}_0, h_1)+F_0({\bf d}_0, h_1)[1+F_1({\bf d}_0, h_1)]^{-1}= F_0({\bf d}_0, h_1) [1+o_p(1)],   \label{Th3Eq7}
\end{equation}
which yields (R.3).

To prove (R.4), we only need some notation as follows:
\begin{eqnarray}
&&T_1(h_s)=\sup_{{\bf d}_0\in {\mathcal D}_0}\left|
 \sum_{{\bf d}_m, {\bf d}_m'\in B({\bf d}_0, h_s)\cap \mathcal D_0}  \tilde\omega_{j}({\bf d}_0, {\bf d}_m; h_s)\tilde \omega_{j}({\bf d}_0, {\bf d}_m'; h_s)
\{\hat \Sigma_{y}({\bf d}_m, {\bf d}_m')-\Sigma_{y}({\bf d}_m, {\bf d}_m') \}
 \right|,\nonumber \\
 &&T_2(h_s)=\sup_{{\bf d}_0\in {\mathcal D}_0}\left| {\sum_{{\bf d}_m, {\bf d}_m'\in B({\bf d}_0, h_s)\cap \mathcal D_0}  \{\tilde\omega_{j}({\bf d}_0, {\bf d}_m; h_s)-\tilde \omega_{j}^{(0)}({\bf d}_0, {\bf d}_m; h_s)\}\tilde\omega_{j}({\bf d}_0, {\bf d}_m'; h_s)
 \Sigma_{y}({\bf d}_m, {\bf d}_m')
 } \right|, \nonumber  \\
  &&T_3(h_s)=\sup_{{\bf d}_0\in {\mathcal D}_0}\left| {\sum_{{\bf d}_m, {\bf d}_m'\in B({\bf d}_0, h_s)\cap \mathcal D_0}  \{\tilde\omega_{j}({\bf d}_0, {\bf d}_m; h_s)-\tilde \omega_{j}^{(0)}({\bf d}_0, {\bf d}_m; h_s)\}\tilde\omega_{j}^{(0)}({\bf d}_0, {\bf d}_m'; h_s)
 \Sigma_{y}({\bf d}_m, {\bf d}_m')
 } \right|.  \nonumber
\end{eqnarray}
A sufficient condition of (R.4) is $|T_1(h_1)|+ |T_2(h_1)|+|T_3(h_1)|=o_p(1)$.
It follows from Theorem 1 that
$T_1(h_1)\leq  \sup_{{\bf d}_m, {\bf d}_m'\in {\mathcal D}_0} |\hat \Sigma_{y}({\bf d}_m, {\bf d}_m')-\Sigma_{y}({\bf d}_m, {\bf d}_m') |  =o_p(1). $
Moreover, $
\tilde \omega_{j}({\bf d}_0, {\bf d}_m; h_s)-\tilde\omega_{j}^{(0)}({\bf d}_0, {\bf d}_m; h_s) $ equals
\begin{equation}
\frac{ \omega_{j}({\bf d}_0, {\bf d}_m; h_s)- \omega_{j}^{(0)}({\bf d}_0, {\bf d}_m; h_s)}{\sum_{{\bf d}_m\in B({\bf d}_0, h_s)\cap \mathcal D_0} \omega_{j}({\bf d}_0, {\bf d}_m; h_s)}+F_1({\bf d}_0, h_1)\tilde \omega_{j}^{(0)}({\bf d}_0, {\bf d}_m; h_s).
 \label{Th3Eq8}
\end{equation}
Substituting (\ref{Th3Eq8}) into $T_2(h_1)$, we have
\begin{eqnarray*}
T_2(h_1) &\leq& C_1 \{|F_1({\bf d}_0, h_1)|+ \frac{\sum_{{\bf d}_m\in B({\bf d}_0, h_s)\cap \mathcal D_0} |\omega_j^{(0)}({\bf d}_0, {\bf d}_m; h_1)-\omega_{j}({\bf d}_0, {\bf d}_m; h_1)|}{ \sum_{{\bf d}_m\in B({\bf d}_0, h_1)\cap \mathcal D_0} \omega_{j}({\bf d}_0, {\bf d}_m; h_1)}   \} =o_p(1).
\end{eqnarray*}
Similar to the derivation of  $T_2(h_1)$, we can prove $T_3(h_1)=o_p(1)$.

For ${\bf d}_0\in  \partial {\mathcal
D}^{(j)}(h_S)$, we assume ${\bf d}_0\in \partial {\mathcal
D}^{(j)}(h_1)$ without loss of generality.  It follows from assumption (C10) that $B({\bf d}_0, h_1)$ is the union of $B({\bf d}_0, h_1)\cap \{{\bf d}_0':  {\bf d}_0'\in \mathcal D_0, \Delta_{j*}({\bf d}_0, {\bf d}_0')=0\}$ and
$B({\bf d}_0, h_1)\cap \{{\bf d}_0':  {\bf d}_0'\in \mathcal D_0, |\Delta_{j*}({\bf d}_0, {\bf d}_0')|\geq  {\ve u}^{(j)}(h_1)\}$.
For ${\bf d}_0'\in B({\bf d}_0, h_1)\cap  \{{\bf d}_0':  {\bf d}_0'\in \mathcal D_0, \Delta_{j*}({\bf d}_0, {\bf d}_0')=0\}$, it is easy to see that (\ref{Th3Eq4}) is true.
 For $ {\bf d}_0'\in B({\bf d}_0, h_1)\cap I_j({\bf d}_0,  {\ve u}^{(j)}(h_1), \infty)$, it follows from the inequality $2(a-b)^2+2b^2\geq a^2$ for any $a, b$  that
  \begin{eqnarray}
 D_{\beta_j}({\bf d}_0, {\bf d}_0'; h_{0})/C_n  &\geq&  C_n^{-1}n[0.5\Delta_{j*}({\bf d}_0, {\bf d}_0')^2 - \{\hat\Delta_j({\bf d}_0)-\hat\Delta_j({\bf d}_0')\}^2]/\hat\Sigma_n(\sqrt{n}\hat{\beta}_j({\bf d}_0))\nonumber \\
 &\geq&  [ 0.5C_n^{-1}n{\ve u}^{(j)}(h_1)^2-4C_n^{-1} n\sup_{{\bf d}_0}  \hat\Delta_j({\bf d}_0)^2 ]/\hat\Sigma_n(\sqrt{n}\hat{\beta}_j({\bf d}_0)).  \label{Th3Eq9}
  \end{eqnarray}
 Thus, we have
  \begin{eqnarray}
 K_{st}(D_{\beta_j}({\bf d}_0, {\bf d}_0'; h_{0})/C_n)  &\leq&  O_p(1)   K_{st}( 0.5 C_n^{-1}n{\ve u}^{(j)2}/\hat\Sigma_n(\sqrt{n}\hat{\beta}_j({\bf d}_0))),  \label{Th3Eq10}
  \end{eqnarray}
which yields that
  \begin{eqnarray}
  && \frac{ \sum_{{\bf d}_m  \in B({\bf d}_0, h_1)\cap \mathcal D_0\cap I_j({\bf d}_0, {\ve u}^{(j)}(h_1), \infty) } K_{loc}(||{\bf d}_0-{\bf d}_m||_2/h_{1})K_{st}(D_{\beta_j}({\bf d}_0, {\bf d}_m; h_{0})/C_n)  }{\sum_{{\bf d}_m\in   B(d, h_1)\cap \mathcal D_0} K_{loc}(||{\bf d}_0-{\bf d}_m||_2/h_{1})K_{st}(D_{\beta_j}({\bf d}_0, {\bf d}_m; h_{0})/C_n) }  \nonumber \\
  &\leq & O_p(1) K_{st}( 0.5 C_n^{-1}n{\ve u}^{(j)}(h_1)^2/\hat\Sigma_n(\sqrt{n}\hat{\beta}_j({\bf d}_0))) \times \nonumber \\
  &&
  \frac{ \sum_{{\bf d}_m  \in B({\bf d}_0, h_1)\cap I_j({\bf d}_0, {\ve u}^{(j)}(h_1)^2, \infty) \cap \mathcal D_0} K_{loc}(||{\bf d}_0-{\bf d}_m||_2/h_{1})   }{ K_{loc}(0)K_{st}(0) }  \nonumber  \\
  &\leq &  O_p(1)   N_Dh_1^3 K_{st}( 0.5 C_n^{-1}n{\ve u}^{(j)}(h_1)^2/\hat\Sigma_n(\sqrt{n}\hat{\beta}_j({\bf d}_0))).
   \label{Th3Eq11}
  \end{eqnarray}
Therefore, it follows from (\ref{Th3Eq10}) and (\ref{Th3Eq11}) that
 \begin{eqnarray}
 F_1({\bf d}_0, h_s)&=&O_p(1)\{\log(1+N_D)C_n^{-1}+  N_Dh_1^3 K_{st}(0.5C_n^{-1}n{\ve u}^{(j)}(h_1)^2/\hat\Sigma_n(\sqrt{n}\hat{\beta}_j({\bf d}_0)))\}=o_p(1),  \nonumber  \\
   F_2({\bf d}_0, h_1)
   &\leq & \{\log(1+N_D)\}^{3/2}/(C_n\sqrt{n})O_p(1)+\nonumber \\
   &&O_p(N_Dh_1^3 K_{st}( 0.5C_n^{-1}n{\ve u}^{(j)}(h_1)^2/\hat\Sigma_n(\sqrt{n}\hat{\beta}_j({\bf d}_0))))\sqrt{\log(1+N_D)/n} \nonumber\\
   &=&o_p( \sqrt{\log(1+N_D)/n}),   \label{Th3Eq11a}
  \end{eqnarray}
  which yield   (R.1).
  Furthermore, it follows from (\ref{Th3Eq11}) that
  \begin{eqnarray}
  |\tilde\Delta_{j*}({\bf d}_0; h_s)|&\leq &\sum_{{\bf d}_m\in B({\bf d}_0, h_1)\cap \mathcal D_0\cap  I_j({\bf d}_0, \sqrt{C_n/n}{\ve u}^{(j)}, \infty)} \tilde\omega_j({\bf d}_0, {\bf d}_m; h_s)|\beta_{j*}({\bf d}_m)-\beta_{j*}({\bf d}_0)| \nonumber \\
&\leq& O_p(1)   N_Dh_1^3 K_{st}( 0.5C_n^{-1}n{\ve u}^{(j)}(h_1)^2/\hat\Sigma_n(\sqrt{n}\hat{\beta}_j({\bf d}_0))).  \label{Th3Eq11b}
  \end{eqnarray}
 Furthermore, it follows from (\ref{Th3Eq4})-(\ref{Th3Eq6}) that
\begin{eqnarray}
\hat\beta_j({\bf d}_0; h_1)&=&\tilde\beta_{j*}({\bf d}_0; h_1)+F_2({\bf d}_0, h_1)+F_0({\bf d}_0, h_1)[1+F_1({\bf d}_0, h_1)]^{-1}    \label{Th3Eq12} \\
&=& \beta_{j*}({\bf d}_0)+\tilde\Delta_{j*}({\bf d}_0; h_1)+F_2({\bf d}_0, h_1)+F_0({\bf d}_0, h_1)[1+F_1({\bf d}_0, h_1)]^{-1}\nonumber\\
&=&\beta_{j*}({\bf d}_0)+ F_0({\bf d}_0, h_1)[1+o_p(1)]+o_p(\sqrt{\log(1+N_D)/n}),   \nonumber
\end{eqnarray}
which yields (R.3).
Similar to the arguments near (\ref{Th3Eq8}), we can easily prove (R.4) for ${\bf d}_0\in  \partial {\mathcal
D}^{(j)}(h_S)\cap \mathcal D_0$.

  Note that (R.2) and (R.3) are the key results used in deriving  (\ref{Th3Eq3})-(\ref{Th3Eq12}).
   Based on (R.1)-(R.4) for $s=1$, we can  use the same arguments from (\ref{Th3Eq3}) to (\ref{Th3Eq12}) to prove (R.1)-(R.4) for $s=2$.
   Generally, if (R.1)-(R.4) are true for any $s$, we can use the same arguments in (\ref{Th3Eq3})-(\ref{Th3Eq12}) to prove (R.1)-(R.4) for $s+1$.
 This finishes the proof of Theorem 3.

%With some algebraic calculations, we have
%\begin{eqnarray}
%&& |\beta_{j}({\bf d}_0; h_s)-\beta_{j}({\bf d})|\leq K_j h_s,
%\\
%&&  |\Delta_{j}({\bf d}_0, {\bf d}_0'; h_s)| \leq |\beta_{j}({\bf d}_0; h_s)-\beta_{j}({\bf d})|+|\beta_{j}({\bf d}_0'; h_s)-\beta_{j}({\bf d}')|+|\beta_{j}({\bf d})-\beta_{j}({\bf d}')|,
%\nonumber
%\\
%&&\Delta\tilde \beta_{j}({\bf d}_0; h_1) =-\beta_{j}({\bf d}_0; h_1)F_1(d, h_1)
%+  \frac{\sum_{{\bf d}_m\in B(d, h_1)}   [\omega_{j}(d, {\bf d}_m; h_1)-\omega_{j}(d, {\bf d}_m; h_1)] \beta_j({\bf d}_m)}{\sum_{{\bf d}_m\in B(d, h_1)}   %\omega_{j}(d, {\bf d}_m; h_1)  }.   \nonumber
%\end{eqnarray}

\noindent   {\it Proof of Theorem 4.}
For $s\geq 1$, we    define
 \begin{eqnarray}
 \hat\Delta_{j}({\bf d}_0; h_s)=\hat\beta_{j}({\bf d}_0; h_s)-\tilde \beta_{j*}({\bf d}_0; h_s), ~ \tilde \Delta_{j*}({\bf d}_0, {\bf d}_0'; h_s)=\tilde \beta_{j*}({\bf d}_0; h_s)-\tilde \beta_{j*}({\bf d}_0'; h_s). 
 \label{Th4Eq1}
 \end{eqnarray}
For $0\leq s\leq S$, we want to prove the following results  by introduction:
\begin{eqnarray*}
&\mbox{(R.1)} &  F_1({\bf d}_0, h_s)=o_p(1),~~   F_2({\bf d}_0, h_s)=o_p(\sqrt{\log(1+N_D)/n}), ~~  \nonumber \\
&\mbox{(R.2)} & \tilde\Delta_{j*}({\bf d}_0; h_s)=L_jh_s+\delta_L+O_p(1)   N_Dh_s^3 K_{st}( 0.5M_n^{2}/\hat\Sigma_n(\sqrt{n}\hat{\beta}_j({\bf d}_0))), \\
&\mbox{(R.3)} & \hat\beta_j({\bf d}_0; h_s)-   \tilde \beta_{j*}({\bf d}_0; h_s)= F_0({\bf d}_0, h_s)[1+o_p(1)]=O_p(\sqrt{\log(1+N_D)/n}), \\
&\mbox{(R.4)} &  \sup_{{\bf d}_0\in {\mathcal D}_0}|\hat\Sigma_n(\sqrt{n} \hat\beta_j({\bf d}_0; h_s))- \Sigma_{j}^{(1)}({\bf d}_0; h_s)|=o_p(1).   \nonumber
\end{eqnarray*}

It follows from Theorem 1 that (R.1)-(R.4) hold for $s=0$.
For $s=1$,
we  consider two different cases including (i) $ {\bf d}_0\in {\mathcal D}\setminus \partial {\mathcal
D}^{(j)}(h_S)$ and (ii) ${\bf d}_0\in \partial {\mathcal
D}^{(j)}(h_S)$.
For $ {\bf d}_0\in {\mathcal D}\setminus \partial {\mathcal
D}^{(j)}(h_S)$ and ${\bf d}_0'\in B({\bf d}_0, h_1)$,  it follows from assumption (C9) that
$D_{\beta_j}({\bf d}_0, {\bf d}_0'; h_{0})/C_n$ can be written as
 \begin{eqnarray} \label{Th4Eq2}
D_{\beta_j}({\bf d}_0, {\bf d}_0'; h_{0})/C_n
&=&C_n^{-1}n\{\hat\Delta_j({\bf d}_0)-\hat\Delta_j({\bf d}_0')+\Delta_{j*}({\bf d}_0, {\bf d}'_0)\}^2/\hat\Sigma_n(\sqrt{n}\hat{\beta}_j({\bf d}_0)) \nonumber \\
 &\leq&  2\hat\Sigma_n(\sqrt{n}\hat{\beta}_j({\bf d}_0))^{-1} \{\log(1+N_D)C_n^{-1}+K_j^2nC_n^{-1}||{\bf d}_0-{\bf d}_0'||^2_2\}O_p(1). \nonumber
 \end{eqnarray}
Therefore,  we have
\begin{eqnarray}
{|K_{st}(D_{{\beta}_j}({\bf d}_0, {\bf d}_0'; h_{0})/C_{n})- K_{st}(0)|}
&\leq &
  O_p(1)\hat\Sigma_n(\sqrt{n}\hat{\beta}_j({\bf d}_0))^{-1}   |\log(1+N_D)C_n^{-1}+K_j^2nC_n^{-1}h_1^2| \nonumber \\
  &=& O_p(1) \log(1+N_D)C_n^{-1}.     \label{Th4Eq3}
\end{eqnarray}
%which yields
%\begin{equation} \label{Th3Eq5}
%F_1(d, h_s)\leq O(1)\Sigma_n(\sqrt{n}\hat{\beta}_j({\bf d}))^{-1}   |O_p(1)\log(1+N_D)C_n^{-1}+K_j^2nC_n^{-1}||{\bf d}_0-{\bf d}_0'||^2_2|.
%\end{equation}
 For ${\bf d}_0\in  \partial {\mathcal
D}^{(j)}(h_S)$, we assume ${\bf d}_0\in \partial {\mathcal
D}^{(j)}(h_1)$ without loss of generality.  It follows from assumption (C10b) that $B({\bf d}_0, h_1)$ is the union of $P_j({\bf d}_0, h_1)$, $P_j({\bf d}_0, h_1)^c   \cap I_j({\bf d}_0, 0, \delta_L)$ and
$P_j({\bf d}_0, h_1)^c   \cap I_j({\bf d}_0,  \delta_U, \infty)$, and $P_j({\bf d}_0, h_1)^c   \cap I_j({\bf d}_0,  \delta_L, \delta_U)=\emptyset$.
For ${\bf d}_0'\in P_j({\bf d}_0, h_1)\cup[P_j({\bf d}_0, h_1)^c   \cap I_j({\bf d}_0, 0, \delta_L)]$, it is easy to see that (\ref{Th4Eq3}) is true.
 For $ {\bf d}_0'\in B({\bf d}_0, h_1)\cap I_j({\bf d}_0,  \delta_U, \infty)$, by using the same arguments in  (\ref{Th3Eq6})-(\ref{Th3Eq9}), we have
  \begin{eqnarray}
  && \frac{ \sum_{{\bf d}_m  \in B({\bf d}_0, h_1)\cap I_j({\bf d}_0, \delta_U, \infty) } K_{loc}(||{\bf d}_0-{\bf d}_m||_2/h_{1})K_{st}(D_{\beta_j}({\bf d}_0, {\bf d}_m; h_{0})/C_n)  }{\sum_{{\bf d}_m\in   B({\bf d}_0, h_1)} K_{loc}(||{\bf d}_0-{\bf d}_m||_2/h_{1})K_{st}(D_{\beta_j}({\bf d}_0, {\bf d}_m; h_{0})/C_n) }  \nonumber \\
  &\leq &  O_p(1)   N_Dh_1^3 K_{st}( 0.5M_n^{2}/\hat\Sigma_n(\sqrt{n}\hat{\beta}_j({\bf d}))).
   \label{Th4Eq4}
  \end{eqnarray}
Therefore, similar to (\ref{Th3Eq6}) and (\ref{Th3Eq11a}),  we have
 \begin{eqnarray}
 F_1({\bf d}_0, h_1)&=&O_p(1)\{\log(1+N_D)C_n^{-1}+  N_Dh_1^3 K_{st}(0.5M_n^{2}/\hat\Sigma_n(\sqrt{n}\hat{\beta}_j({\bf d}_0)))\}=o_p(1),
\nonumber \\
   F_2({\bf d}_0, h_1)
   &= & o_p(\sqrt{\log(1+N_D)/n}),     \label{Th4Eq5}
  \end{eqnarray}
  which yield   (R.1).

  We prove (R.2) as follows.
For  ${\bf d}_0\in \partial {\mathcal
D}^{(j)}(h_S)\cap \mathcal D_0$,
  it follows from (\ref{Th4Eq3}) and (\ref{Th4Eq4})  that
  \begin{eqnarray}
  |\tilde\Delta_{j*}({\bf d}_0; h_1)|&\leq &|\sum_{{\bf d}_m\in P_j({\bf d}_0, h_1)}   \tilde\omega_j({\bf d}_0, {\bf d}_m; h_s)[\beta_{j*}({\bf d}_m)-\beta_{j*}({\bf d}_0)]| \nonumber \\
  &+&\sum_{{\bf d}_m\in P_j({\bf d}_0, h_1)^c\cap I_j({\bf d}_0, 0, \delta_L)}  \tilde\omega_j({\bf d}_0, {\bf d}_m; h_s)|\beta_{j*}({\bf d}_m)-\beta_{j*}({\bf d}_0)|   \nonumber\\
  &+&
  \sum_{{\bf d}_m\in P_j({\bf d}_0, h_1)^c\cap I_j({\bf d}_0,  \delta_U, \infty)}   \tilde\omega_j({\bf d}_0, {\bf d}_m; h_s)|\beta_{j*}({\bf d}_m)-\beta_{j*}({\bf d}_0)|  \nonumber \\
&\leq& L_j h_1+\delta_L+  N_Dh_1^3 K_{st}( 0.5M_n^{2}/\hat\Sigma_n(\sqrt{n}\hat{\beta}_j({\bf d}_0))) O_p(1).  \label{Th4Eq6}
  \end{eqnarray}
However,
 for $ {\bf d}_0\in {\mathcal D}\setminus \partial {\mathcal
D}^{(j)}(h_S)$,    by using Taylor series expansion, we have
     \begin{eqnarray}
  |\tilde\Delta_{j*}({\bf d}_0; h_1)|&=& |\sum_{{\bf d}_m\in B(d, h_1)}   \tilde\omega_j({\bf d}_0, {\bf d}_m; h_s) [\beta_{j*}({\bf d}_m)-\beta_{j*}({\bf d}_0)] |\leq L_jh_1.   \label{Th4Eq7}
  \end{eqnarray}
This yields (R.2).
  Similar to the arguments in Theorem 3, we can easily prove (R.3) and (R.4) for $s=1$. So we omit the details.

%With some algebraic calculations, we have
%\begin{eqnarray}
%&& |\beta_{j}({\bf d}_0; h_s)-\beta_{j}({\bf d})|\leq K_j h_s,
%\\
%&&  |\Delta_{j}({\bf d}_0, {\bf d}_0'; h_s)| \leq |\beta_{j}({\bf d}_0; h_s)-\beta_{j}({\bf d})|+|\beta_{j}({\bf d}_0'; h_s)-\beta_{j}({\bf d}')|+|\beta_{j}({\bf d})-\beta_{j}({\bf d}')|,
%\nonumber
%\\
%&&\Delta\tilde \beta_{j}({\bf d}_0; h_1) =-\beta_{j}({\bf d}_0; h_1)F_1(d, h_1)
%+  \frac{\sum_{{\bf d}_m\in B(d, h_1)}   [\omega_{j}(d, {\bf d}_m; h_1)-\omega_{j}(d, {\bf d}_m; h_1)] \beta_j({\bf d}_m)}{\sum_{{\bf d}_m\in B(d, h_1)}   %\omega_{j}(d, {\bf d}_m; h_1)  }.   \nonumber
%\end{eqnarray}
For $s=2$, we only prove the result  (R.1).
$D_{\beta_j}({\bf d}_0, {\bf d}_0'; h_{1})$ can be written as
 \begin{eqnarray} \label{Th4Eq11}
D_{\beta_j}({\bf d}_0, {\bf d}_0'; h_{1})&=& n\{\hat{\beta}_j({\bf d}_0; h_1)  -
\hat{\beta}_j({\bf d}_0'; h_1)\}^2/\hat\Sigma_n(\sqrt{n}\hat{\beta}_j({\bf d}_0; h_1))\\
&=&n\{\hat\Delta_{j}({\bf d}_0; h_1)-\hat\Delta_j({\bf d}_0'; h_1)+\tilde \Delta_{j*}({\bf d}_0, {\bf d}_0'; h_1)\}^2/\hat\Sigma_n(\sqrt{n}\hat{\beta}_j({\bf d}_0; h_1)). \nonumber
\end{eqnarray}
We first consider the cases with  ${\bf d}_0'\in P_j(d, h_2)$ for  $ {\bf d}_0\in {\mathcal D}\setminus \partial {\mathcal
D}^{(j)}(h_S)$ and ${\bf d}_0'\in B({\bf d}_0, h_2)$ for  ${\bf d}_0\in \partial {\mathcal
D}^{(j)}(h_S)$. It follows from (R.2) and (R.3) that
\begin{eqnarray*}
&& nC_n^{-1}\{\hat\Delta_{j}({\bf d}_0; h_1)-\hat\Delta_j({\bf d}_0'; h_1)+\tilde \Delta_{j}({\bf d}_0, {\bf d}_0'; h_1)\}^2   \nonumber \\
&&\leq   2nC_n^{-1}\{\hat\Delta_j({\bf d}_0; h_1)-\hat\Delta_j({\bf d}_0'; h_1)\}^2+2nC_n^{-1}\{ \tilde \Delta_{j*}({\bf d}_0; h_1)- \tilde \Delta_{j*}({\bf d}_0'; h_1)+\Delta_{j*}({\bf d}_0, {\bf d}_0')\}^2\\
&& \leq O_p(1)\{\log(1+N_D)C_n^{-1}+nC_n^{-1}(h_1^2+h_2^2)\},  \nonumber
\end{eqnarray*}
which yields
$
F_1({\bf d}_0, h_2)\leq O_p(1) | \log(1+N_D)C_n^{-1}+nC_n^{-1}(h_1^2+h_2^2)|.
$

For $ {\bf d}_0'\in P_j({\bf d}_0, h_1)^c\cap I_j({\bf d}_0,  \delta_U, \infty)$, by using the same arguments in  (\ref{Th3Eq6})-(\ref{Th3Eq9}),  we  have
   \begin{eqnarray*}
&&\{\hat{\beta}_j({\bf d}_0; h_1)  -
\hat{\beta}_j({\bf d}_0'; h_1)\}^2\\
&\geq &0.5\Delta_{j*}({\bf d}_0, {\bf d}_0')^2- \{\tilde\Delta_{j*}({\bf d}_0; h_1)-\tilde\Delta_{j*}({\bf d}_0'; h_1)+ \hat\Delta_j({\bf d}_0; h_1)-\hat\Delta_j({\bf d}_0'; h_1)\}^2\nonumber \\
 &\geq&  0.5 n^{-1}C_nM_n^2-2 \{\tilde\Delta_{j*}({\bf d}_0; h_1)-\tilde\Delta_{j*}({\bf d}_0'; h_1)\}^2-2\{ \hat\Delta_j({\bf d}_0; h_1)-\hat\Delta_j({\bf d}_0'; h_1)\}^2  \\
 &\geq& 0.5 n^{-1}C_nM_n^2-O_p(h_1^2+n^{-1}\log(1+N_D)).
  \end{eqnarray*}
 Thus, we have
  \begin{eqnarray}
  && \frac{ \sum_{{\bf d}_m\in  P_j({\bf d}_0, h_2)^c\cap I_j({\bf d}_0, \delta_U, \infty)} K_{loc}(||{\bf d}_0-{\bf d}_m||_2/h_{2})K_{st}(D_{\beta_j}({\bf d}_0, {\bf d}_m; h_{1})/C_n)  }
  {\sum_{{\bf d}_m\in  B({\bf d}_0, h_2)} K_{loc}(||{\bf d}_0-{\bf d}_m||_2/h_{2})K_{st}(0) }  \nonumber \\
  &\leq &   O_p(1) N_Dh_2^3 K_{st}(0.5M_n^{2}/\hat\Sigma_n(\sqrt{n}\hat{\beta}_j({\bf d}_0; h_1)))\label{Th4Eq8}
  \end{eqnarray}
Therefore,  by using the similar arguments in (\ref{Th3Eq6}) and (\ref{Th3Eq11a}), we can get
   \begin{eqnarray}
  F_1({\bf d}_0, h_2)&\leq & O_p(1) | \log(1+N_D)C_n^{-1}+ nC_n^{-1}(h_1^2+h_2^2)|  \label{Th4Eq9} \\
  &+&
 O_p(1)N_Dh_2^3 K_{st}(0.5M_n^{2}/\hat\Sigma_n(\sqrt{n}\hat{\beta}_j({\bf d}_0; h_1))), \nonumber  \\
  F_2({\bf d}_0, h_2)&=&o_p(\sqrt{\log(1+N_D)/n}).  \nonumber
   \end{eqnarray}
    Generally, if (R.1)-(R.4) are true for any $s$, we can use the same arguments in (\ref{Th4Eq1})-(\ref{Th4Eq9}) to prove (R.1)-(R.4) for $s+1$.
 This finishes the proof of Theorem 4.

\section*{Simulation Studies}

\subsection*{Additional Simulation Results}

\begin{figure}[ht!]
\vspace*{0pt}
\begin{center}
\includegraphics[width=.95\textwidth,height=.34\textheight]{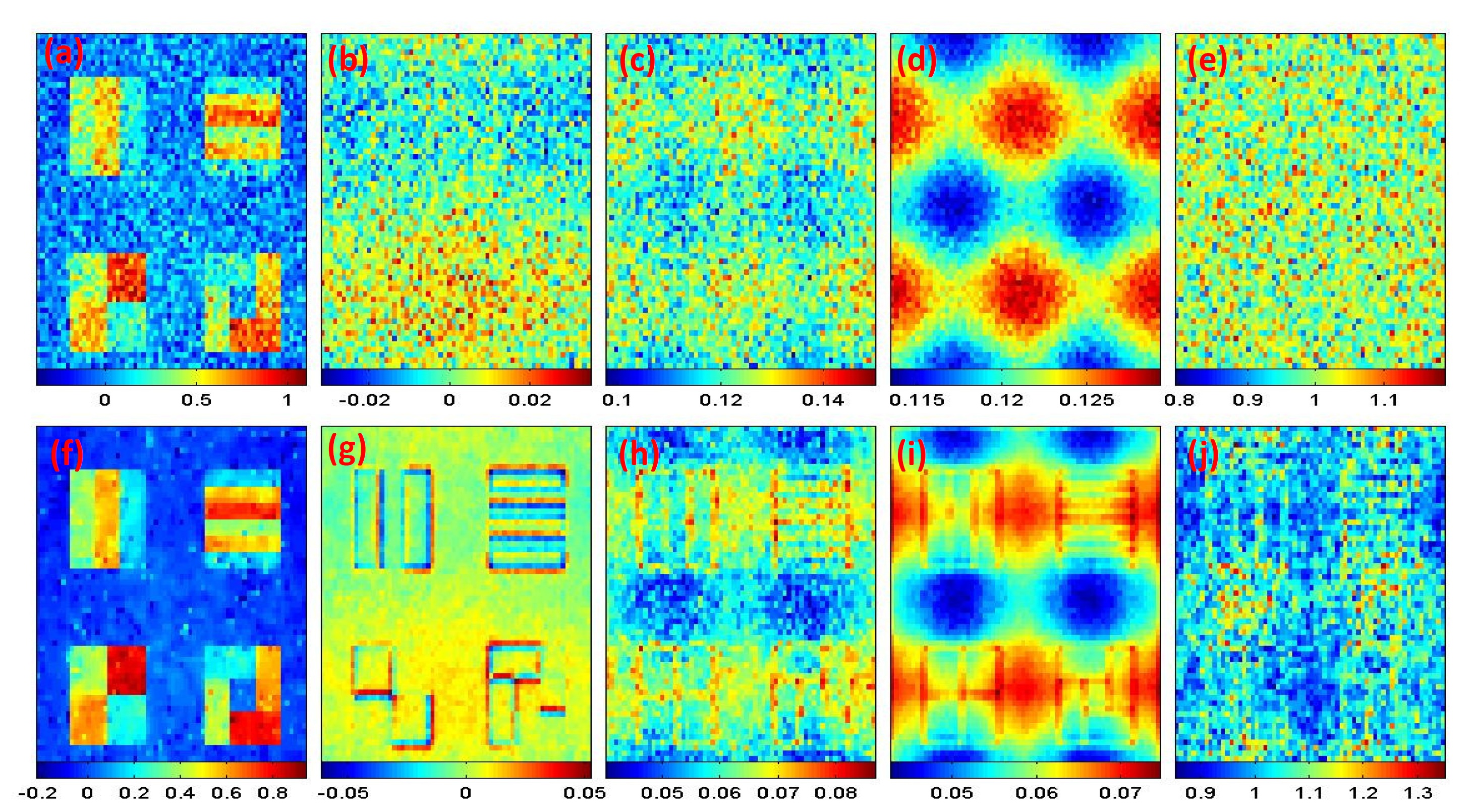}
\end{center}
\vspace*{0pt}
\caption{Simulation results:  a selected slice of (a) and (f)  $\hat\beta_3({\bf d}_0; h_s)$; (b) and (g) the biases of $\hat\beta_3({\bf d}_0; h_s)$; (c) and (h) the root-mean-square errors (RMSs) of $\hat\beta_3({\bf d}_0; h_s)$;  ({d}) and (i) the standard deviation estimates  (SDs) of $\hat\beta_3({\bf d}_0; h_s)$; and (e) and (f)  the ratios of RMS over SD.
 Upper panels and lower panels correspond to  $h_0$ and $h_{10}$, respectively.
  }
\label{svcm_s2}

\end{figure}

We present some additional results obtained from the simulation studies in the main paper. 
Figure \ref{svcm_s2} shows  some selected results based on $\hat\beta_3({\bf d}_0, h_0)$ and $\hat\beta_3({\bf d}_0; h_{10})$  with $N(0, 1)$ distributed  data and $n=60$ from the 200 simulated data sets. The biases    slightly increase   from $h_0$  to $h_{10}$ (Figure ~\ref{svcm_s2} (b) and (g)), whereas 
the
root-mean-square errors (RMSs) 
 and standard deviations (SDs) at $h_{10}$ are much smaller than those at $h_0$ (Figure ~\ref{svcm_s2} (c), (d),  (h), and (i)). In addition, the RMSs and their corresponding SDs are relatively close to each other
at all scales for both the normal  (Figure ~\ref{svcm_s2} (e) and (j)) and Chi-square distributed data (not shown here).
Moreover,  SDs in these voxels of  regions of interest (ROIs) with $\beta_3({\bf d}_0)>0$ are  larger than SDs in those voxels of
 ROI with $\beta_3({\bf d}_0)=0$ (the last column in the lower row of  Figure ~\ref{svcm_s2}), because the interior of ROI with $\beta_3({\bf d}_0)=0$ contains more pixels (Figure 3 (c)). Moreover, both the SDs at steps $h_0$ and $h_{10}$ show clear spatial patterns caused by spatial correlations (Figure ~\ref{svcm_s2} (d) and (i)). The RMSs also show some evidence of spatial patterns (Figure ~\ref{svcm_s2} (c) and (h)).  All these results confirm the conclusions that we make  based on  Table 1 in the main paper.

\begin{figure}[ht!]
\vspace*{0pt}
\begin{center}
\includegraphics[width=.95\textwidth,height=.52\textheight]{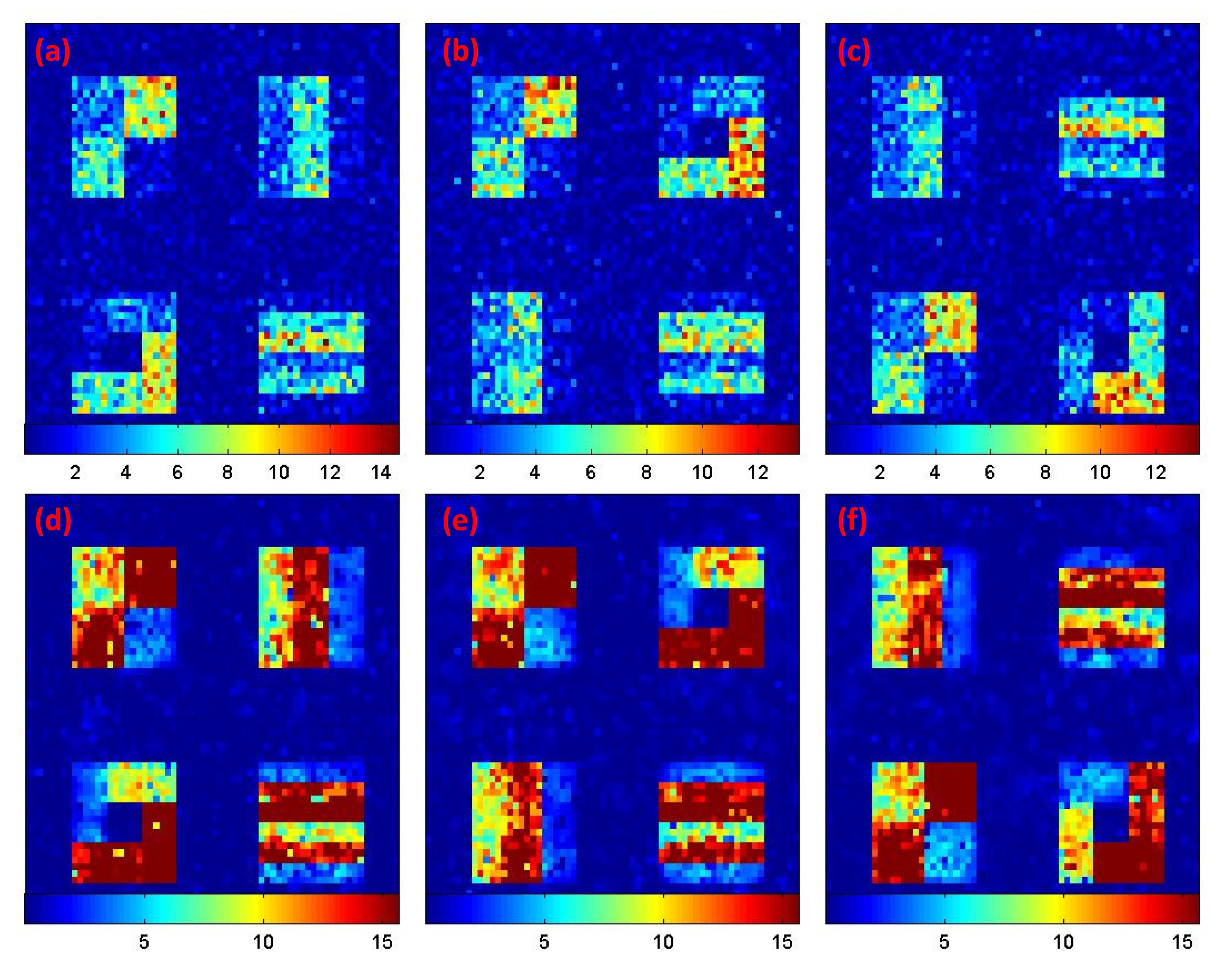}
\end{center}
\vspace*{0pt}
\caption{Simulation results:  a  selected slice of the $-\log_{10}(p)$ images for testing  (a) and ({\bf d}) $H_0({\bf d}_0): \beta_1({\bf d}_0)=0$; (b) and (e)  $H_0({\bf d}_0): \beta_2({\bf d}_0)=0$; and (c) and (f) $H_0({\bf d}_0): \beta_3({\bf d}_0)=0$.
 Upper panels and lower panels correspond to  $h_0$ and $h_{10}$, respectively.
}
\label{svcm_s3}
\end{figure}

We test  the hypotheses $ H_{0}({\bf d}_0):~\beta_j({\bf d}_0)=0$ versus $H_{1}({\bf d}_0): \beta_j({\bf d}_0)\neq 0$ for $j=1,2,3$ across all ${\bf d}_0\in \mathcal{D}_0$ using the MASS procedure at  scales $h_0$ and $h_{10}$.  The $-\log_{10}(p)$ values  on some selected slices  are shown in Figure ~\ref{svcm_s3}.  The values that are greater than $1.3$ indicate a significant effect at $5\%$ significance level and a highly significant effect at $1\%$ significance level if they are greater than $3$. The results are consistent with that from Table 2. In  the  lower panels of Figure ~\ref{svcm_s3} at scale $h_{10}$,  all the nonzero regions of $\beta_j({\bf d}_0)$ are detected as significant at $5\%$ significance level, while most of them are even identified as highly significant and the boundaries between different regions are fairly identifiable. In contrast,  in the upper  panels of Figure ~\ref{svcm_s3},  at scale $h_0$,  many voxels in  ROIs with $\beta_2({\bf d}_0)\not=0$ are  significant at  $\alpha =5\%$ significance level, while the boundaries of ROIs are blurred.

\subsection*{ Local Constant Estimation}

As suggested by one of the referees, we compare SVCM with another estimation method, called  local constant estimation (LCE). Specifically, we  calculate
 the least squares estimate  $\hat{\beta}_j({\bf d}_0)$ and then  use  local constant method based on  the Epanechnikov kernel function, $K(u) = 3/4(1-u^2)I(|u|\leq 1)$,  to directly  smooth  the initial estimate image, which leads to a new estimate, denoted as $\tilde{\beta}_j({\bf d}_0; h)$, at each voxel.  
Subsequently, we use  the method in Stage (III) of SVCM  to compute the standard errors of   $\tilde{\beta}_j({\bf d}_0; h)$ and construct  a  Wald type test. 
   We consider   small   ($h_s = 1.1$), moderate   ($h_m=2$), and  
 large bandwidths ($h_l=4$).

\begin{figure}[ht!]
\vspace*{0pt}
\begin{center}
\includegraphics[width=.95\textwidth]{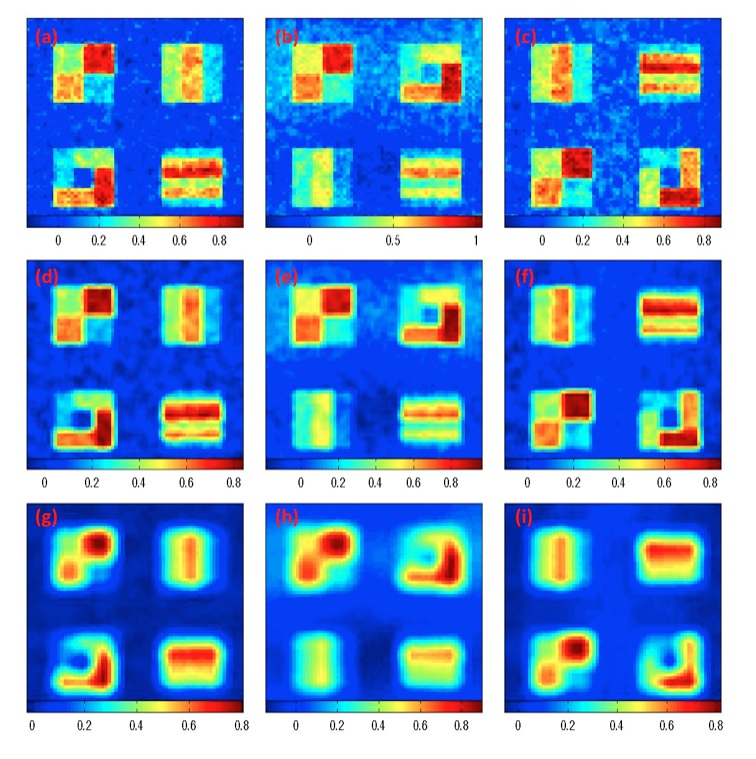}
\end{center}
\vspace*{0pt}
\caption{Simulation results from LCE:  a selected slide of  (a)  $\tilde{\beta}_1({\bf d}_0; h_s)$; (b) $\tilde{\beta}_2({\bf d}_0; h_s)$; and  (c)  $\tilde{\beta}_3({\bf d}_0; h_s)$ with small bandwidth $h_s$; 
 (d)  $\tilde{\beta}_1({\bf d}_0; h_m)$; (e) $\tilde{\beta}_2({\bf d}_0; h_m)$; and (f)  $\tilde{\beta}_3({\bf d}_0; h_m)$ with mediate bandwidth $h_m$; (g)  $\tilde{\beta}_1({\bf d}_0; h_l)$; (h) $\tilde{\beta}_2({\bf d}_0; h_l)$; and (i)  $\tilde{\beta}_3({\bf d}_0; h_l)$ with large bandwidth $h_l$; }
\label{lce_fig1}
\end{figure}

\begin{figure}[ht!]
\vspace*{0pt}
\begin{center}
\includegraphics[width=.95\textwidth,height=.51\textheight]{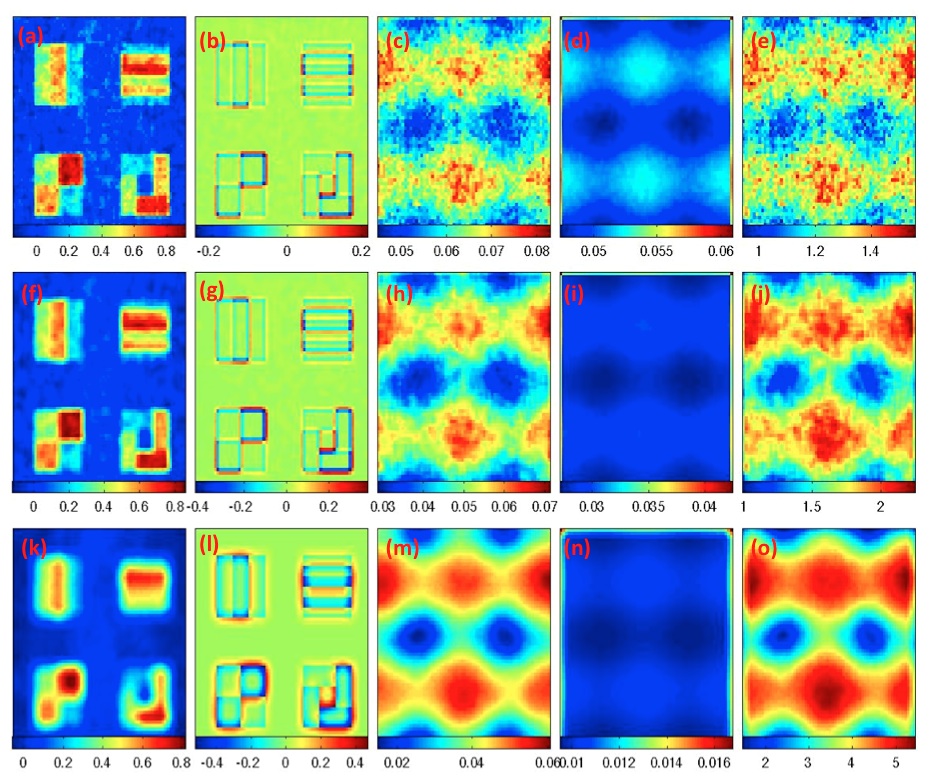}
\end{center}
\vspace*{0pt}
\caption{Simulation results from LCE: a selected slice of (a) , (f) and (k)  $\tilde\beta_3({\bf d}_0; h)$; (b), (g) and (l) the biases of $\tilde\beta_3({\bf d}_0; h)$; (c),  (h) and (m) the root-mean-square errors (RMSs) of $\tilde\beta_3({\bf d}_0; h)$;  (d), (i) and (n) the standard deviation estimates  (SDs) of $\tilde\beta_3({\bf d}_0;h)$; and (e), (j) and (o)  the ratios of RMS over SD.
 Upper, middle and lower panels correspond to  bandwidths $h_s$, $h_m$ and $h_l$, respectively. 
  }
\label{lce_fig2}
\end{figure}

\begin{figure}[ht!]
\vspace*{0pt}
\begin{center}
\includegraphics[width=.95\textwidth]{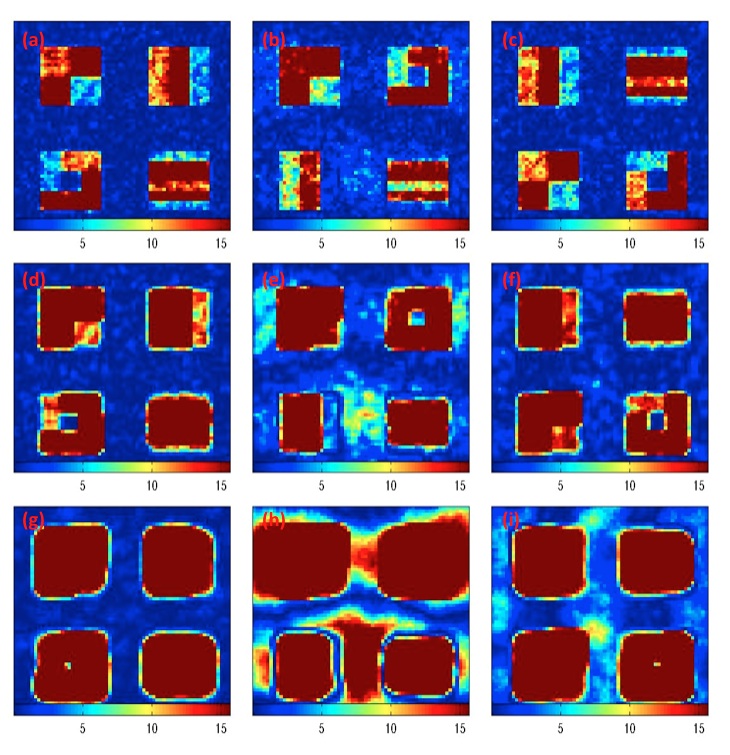}
\end{center}
\vspace*{0pt}
\caption{Simulation results from LCE:  a  selected slice of the $-\log_{10}(p)$ images for testing  (a), (d) and (g) $H_0({\bf d}_0): \beta_1({\bf d}_0)=0$; (b), (e) and (h) $H_0({\bf d}_0): \beta_2({\bf d}_0)=0$; and (c), (f) and (i) $H_0({\bf d}_0): \beta_3({\bf d}_0)=0$.
 Upper, middle and lower panels correspond to  bandwidths $h_s$, $h_m$ and $h_l$, respectively. 
}
\label{lce_fig3}
\end{figure}

Figure  \ref{lce_fig1} presents the LCE estimates obtained from the three different bandwidths based on one selected simulated data set.  For the small bandwidth, effect ROIs cannot be clearly detected.  
As bandwidth increases,   the $-\log_{10}(p)$ plots in Figure \ref{lce_fig3} reveal  that  the coefficients near the boundaries of all ROIs are easily   oversmoothed and  the edges of all ROIs are blurred  
  at moderate and large bandwidths. In addition, Figure \ref{lce_fig3} shows that   false positive rates are high for moderate and large bandwidths, as confirmed in Table \ref{lce_t2}. The failure of detecting edges by LCE is also observed from the bias plots in Figure \ref{lce_fig2} (panels (b), (g), and (l)). As shown in Figure \ref{lce_fig2} (panels (e), (j), and (o)), the ratios of RMS over SD are uniformly greater than 1, which indicates that the SDs are  underestimated. 

We repeated the simulation 200 times at the three different
bandwidths with $N(0, 1)$ and $\chi^2(3)-3$ distributed data for two different sample sizes
($n=60$ and  $ 80$) as we did in the main paper.  For the sake of space, we only report  the results for $\beta_2({\bf d}_0)$ in  Table \ref{lce_t1}. The bias in ROIs with no or weak signals ($\beta_2({\bf d}_0)=0 ~ \mbox{or}~  0.2$) is positive,  whereas the bias is   negative for ROIs with median or   strong signals ($\beta_2({\bf d}_0)=0.4,~ 0.6~ \mbox{or}~ 0.8$). It indicates that  weak signals are overestimated,   whereas mediate and strong signals are underestimated mainly due to the burring edges of LCE. Table \ref{lce_t1} reveals that 
  the SDs are  underestimated. We also calculated the rejection rates for  testing $H_0: \beta_2({\bf d}_0)=0$ in all voxels and include them in Table \ref{lce_t2}. The effect sizes 
(or false positive rates) are much larger  than the preselected significant level $\alpha = 5\%$,   and thus  the Wald test is invalid even though it is very powerful for  detecting  relatively weak signals. Such large false positive rates may be due to  positive bias in ROIs with none or weak signals 
and underestimated  SDs.  

%Similarly, the detecting power may come from over smoothing and underestimating SDs.

\subsection*{Gaussian Kernel Smoothing}

As suggested by one of the referees,  we  compare SVCM  with a standard voxel-wise method, called Gaussian Kernel Smoothing (GKS)  hereafter. The GKS  consists of two steps including a smoothing step to smooth the simulated raw imaging data and an inference step to    calculate the least squares estimate of $\ve\beta({\bf d}_0)$, denoted as $\tilde {\ve\beta}^o({\bf d}_0; h)$, and test hypothesis of interest at each voxel.  
In the smoothing step, we use the Gaussian kernel smoothing function   and consider three different bandwidths including 
a small bandwidth ($h_s = 1.1$), a moderate bandwidth ($h_m=2$), and  
a large bandwidth ($h_l=4$).  

\begin{figure}[ht!]
\vspace*{0pt}
\begin{center}
\includegraphics[width=.95\textwidth]{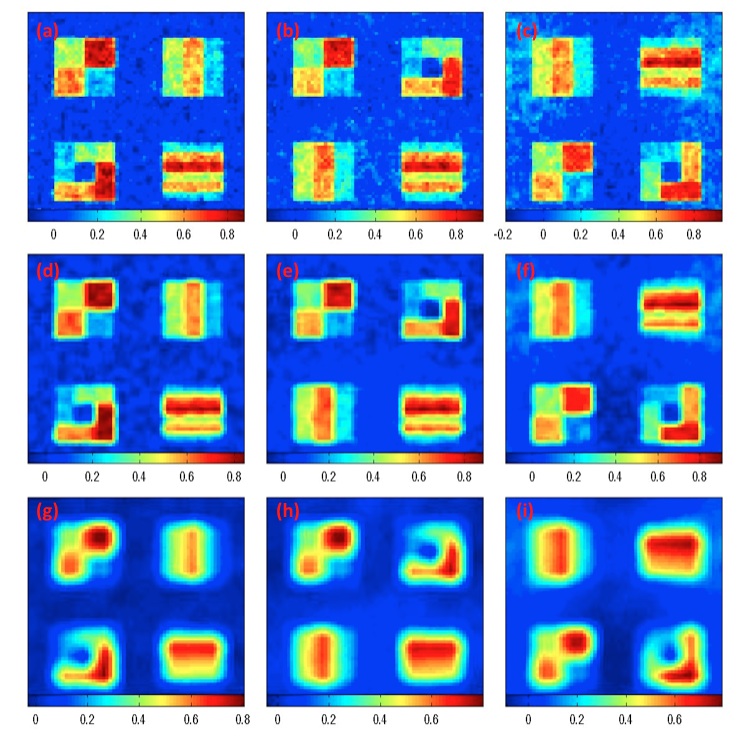}
\end{center}
\vspace*{0pt}
\caption{Simulation results from GKS:  a selected slide of  (a)  $\tilde{\beta}_1^o({\bf d}_0; h_s)$; (b) $\tilde{\beta}^o_2({\bf d}_0; h_s)$; and  (c)  $\tilde{\beta}^o_3({\bf d}_0; h_s)$ with small bandwidth $h_s$; 
 (d)  $\tilde{\beta}^o_1({\bf d}_0; h_m)$; (e) $\tilde{\beta}^o_2({\bf d}_0; h_m)$; and (f)  $\tilde{\beta}^o_3({\bf d}_0; h_m)$ with mediate bandwidth $h_m$; (g)  $\tilde{\beta}^o_1({\bf d}_0; h_l)$; (h) $\tilde{\beta}^o_2({\bf d}_0; h_l)$; and (i)  $\tilde{\beta}^o_3({\bf d}_0; h_l)$ with large bandwidth $h_l$; }
\label{gks_fig1}
\end{figure}

\begin{figure}[ht!]
\vspace*{0pt}
\begin{center}
\includegraphics[width=.95\textwidth,height=.51\textheight]{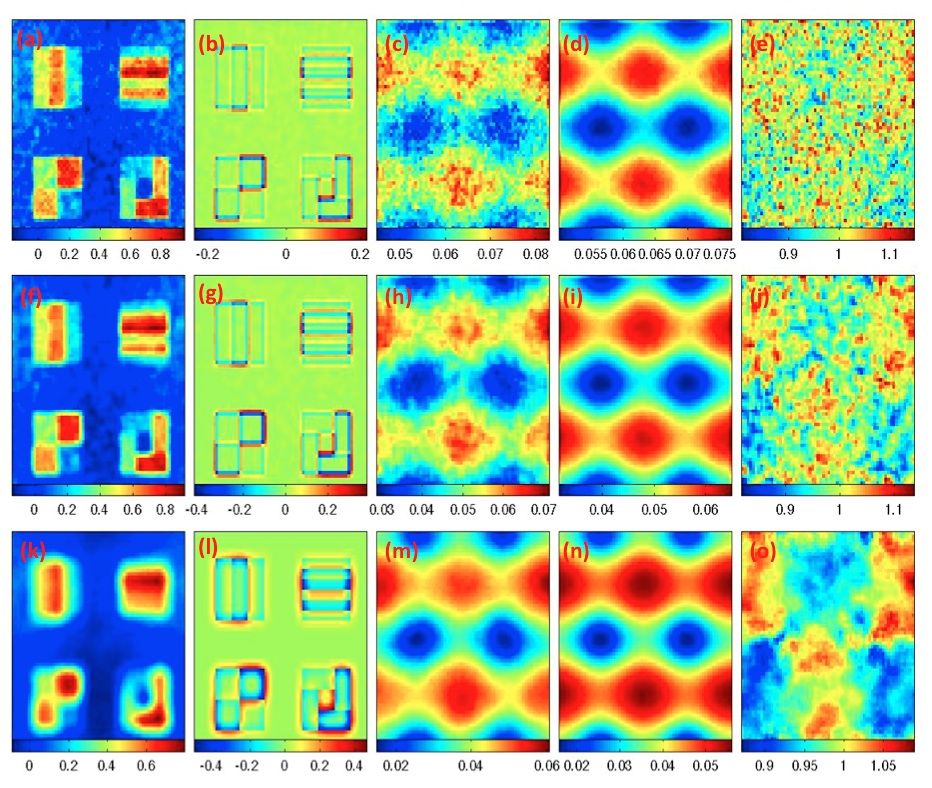}
\end{center}
\vspace*{0pt}
\caption{Simulation results from GKS: a selected slice of (a) , (f) and (k)  $\tilde\beta^o_3({\bf d}_0; h)$; (b), (g) and (l) the biases of $\tilde\beta^o_3({\bf d}_0; h)$; (c),  (h) and (m) the root-mean-square errors (RMSs) of $\tilde\beta^o_3({\bf d}_0; h)$;  (d), (i) and (n) the standard deviation estimates  (SDs) of $\tilde\beta^o_3({\bf d}_0; h)$; and (e), (j) and (o)  the ratios of RMS over SD.
 Upper, middle and lower panels correspond to  bandwidths $h_s$, $h_m$ and $h_l$, respectively. 
  }
\label{gks_fig2}
\end{figure}

\begin{figure}[ht!]
\vspace*{0pt}
\begin{center}
\includegraphics[width=.95\textwidth]{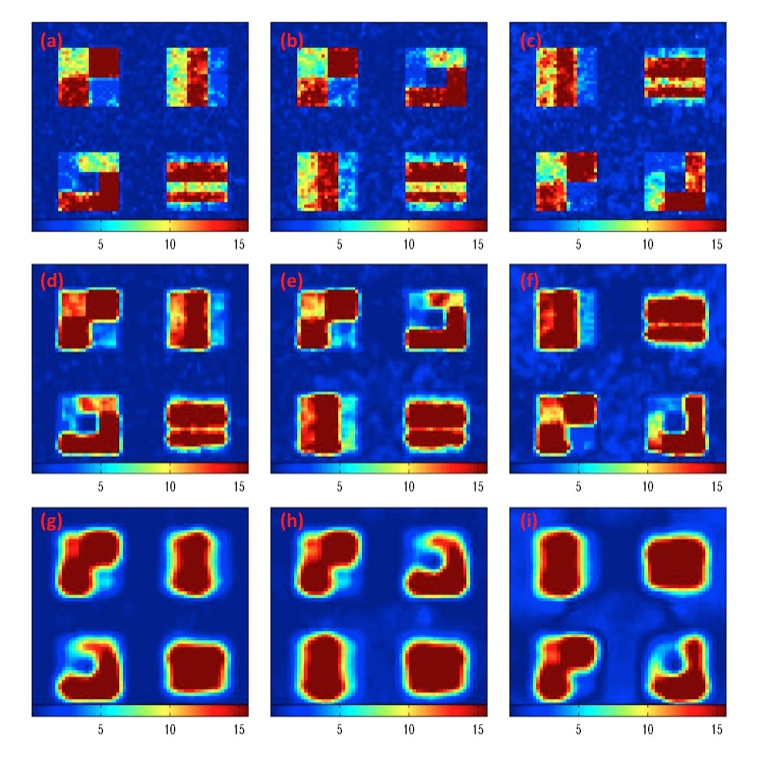}
\end{center}
\vspace*{0pt}
\caption{Simulation results from GKS:  a  selected slice of the $-\log_{10}(p)$ images for testing  (a), (d) and (g) $H_0({\bf d}_0): \beta_1({\bf d}_0)=0$; (b), (e) and (h) $H_0({\bf d}_0): \beta_2({\bf d}_0)=0$; and (c), (f) and (i) $H_0({\bf d}_0): \beta_3({\bf d}_0)=0$.
 Upper, middle and lower panels correspond to  bandwidths $h_s$, $h_m$ and $h_l$, respectively. 
}
\label{gks_fig3}
\end{figure}

Figure  \ref{gks_fig1} presents the GSK estimates obtained from the three different bandwidths based on one selected simulated data set.  
Similar to LCE, small bandwidth does not increase signal detection  especially near the boundaries of  ROIs (Figure \ref{gks_fig1} (a)-(c)),  while moderate 
and large bandwidths oversmooth the coefficient images and blur the boundaries of ROIs (Figure \ref{gks_fig1} (d)-(i)). It indicates that GKS is not capable of effectively estimating functions with potential jumps and edges.   Figure \ref{gks_fig3} shows that the false positive rates are high for moderate and large bandwidths. See also Table \ref{gks_t2}. The bias plots in Figure \ref{gks_fig2} (panels (b), (g), and (l)) show  strong blurred edges. It further confirms the limitation of GKS in preserving boundaries.

We repeated the simulation 200 times at the three different
bandwidths with $N(0, 1)$ and $\chi^2(3)-3$ distributed data for two different sample sizes
($n=60,  80$) as we did in the main paper.   For the sake of space, we only report  the results for $\beta_2({\bf d}_0)$ in  Table \ref{gks_t1}.  
 Inspecting   Table \ref{gks_t1} reveals that  bias in ROIs with weak signals ($\beta_2({\bf d}_0)=0 ~ \mbox{or}~  0.2$  is positive,   whereas   bias  in ROIs with mediate and strong signals ($\beta_2({\bf d}_0)=0.4,~ 0.6~ \mbox{or}~ 0.8$) are negative. The rejection rate results for testing $H_0({\bf d}_0): \beta_2({\bf d}_0)=0$ are shown in Table \ref{gks_t2}. The effect sizes 
(false positive rates) are   much larger than the preselected significant level $\alpha = 0.05$.  

\section*{ADHD 200} 

\subsection*{Image Processing}

The image processing is performed as follows. 
First, we do an AC-PC (anterior commissure - posterior commissure) correction on all images using MIPAV software \citep{MIPAV}, and then resampled the MRI images to  256$\times$256$\times$256. To correct the intensity inhomogeneity, we use N3 algorithm \citep{Sled1998}.  An accurate and robust skull stripping method \citep{Wang2011b} was performed, and the skull stripping results were further manually reviewed to ensure clean skull and dura removal. After the skull-stripping, we used N3 algorithm again to correct for  intensity inhomogeneity. Then the cerebellum is removed based on registration, in which we use a manually labeled cerebellum as a template. After intensity inhomogeneity correction, we use FAST in FSL \citep{Zhang2001TMI} to segment the human brain into three different tissues: grey matter (GM), white matter (WM),  and Cerebrospinal fluid (CSF).
We use HAMMER \citep{Shen2002} to do the registration. After registration, we get the subject-labeled image based on the Jacob template \citep{Kabani1998}, which is manually labeled into 93 ROIs. For each of the 93 ROIs in the labeled image of one subject, we compute the GM/WM/CSF tissue volumes in this ROI region combining the segmentation result of this subject.

  To quantify the local volumetric group differences, we generate the RAVENS  maps \citep{Goldszal1998,Davatzikos2001} for whole brain and for GM, WM and CSF, respectively, by using the deformation field that we get in registration. RAVENS methodology is based on a volume-preserving spatial transformation, which ensures that no volumetric information is lost during the process of spatial normalization, since this process changes an individual\rq{}s brain morphology to conform it to the morphology of a template. A physical analog is the squeezing of a rubber object, which changes the density of the rubber, to maintain the same total mass in the object. Regional volumetric measurements are then performed via the resulting tissue density maps. 
We also do automatic subject labeling by transferring the labels of the template after deformable registration with the subjects.  We have 93 ROIs in total. After labeling, we can get the ROI volumes of all subjects. 

\subsection*{Additional Results}

\begin{figure}[ht!]
\vspace*{0pt}
\begin{center}
\includegraphics[width=.95\textwidth,height=.56\textheight]{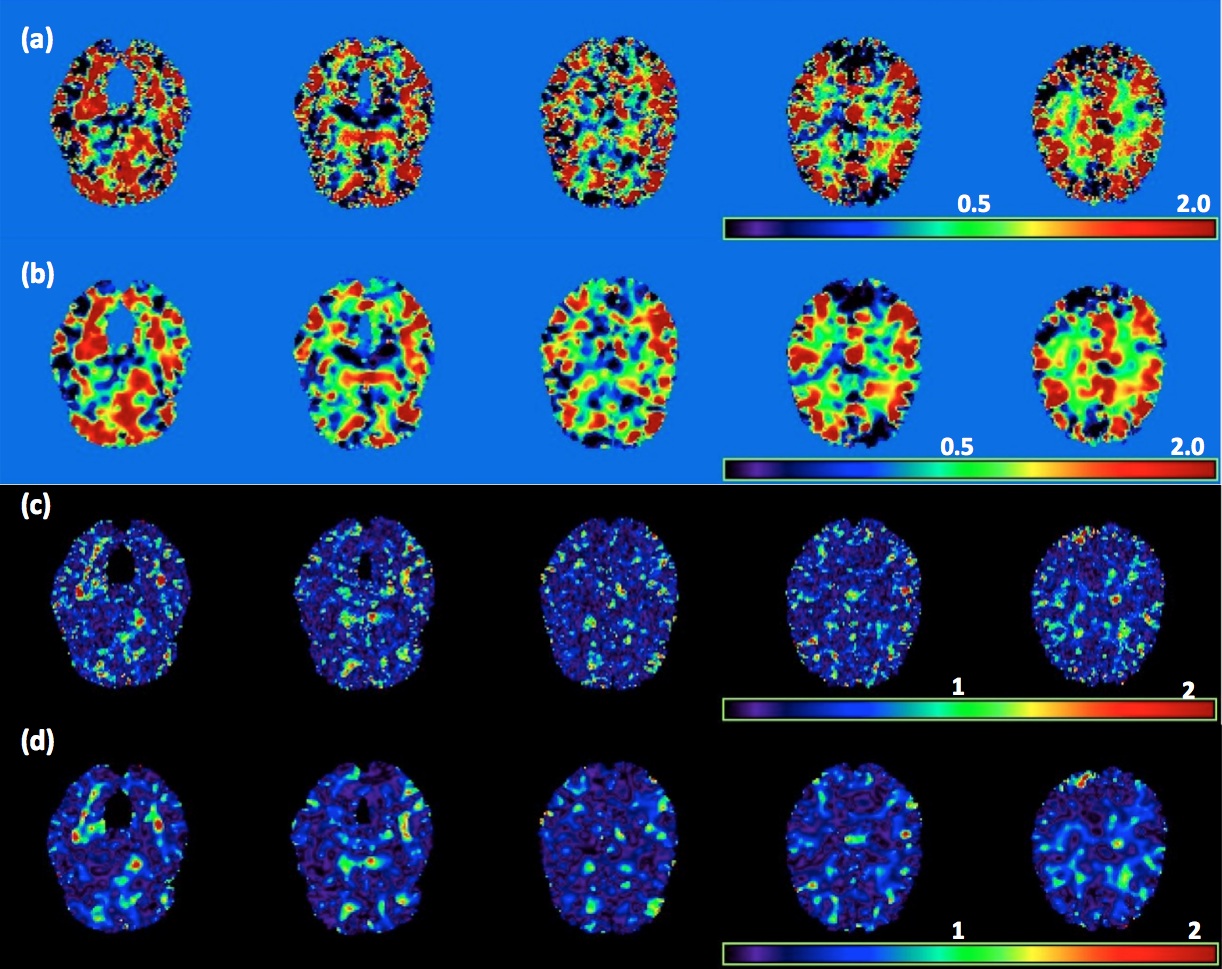}
\end{center}
\vspace*{0pt}
\caption{Results from the ADHD 200 data: five selected slices of  (a) $\hat\beta_7({\bf d}_0;  h_0)$, (b) $\hat\beta_7({\bf d}_0; h_{10})$, (c) the  $-\log_{10}(p)$ images for testing  $H_0({\bf d}_0): \beta_7({\bf d}_0)=0$  at scale $h_0$ and (d) at scale $h_{10}$. Moreover,   $\beta_7({\bf d}_0)$ is  associated with the   gender$\times$diagnosis  interaction.}
\label{svcm_s4}
\end{figure}

\begin{figure}[ht!]
\vspace*{-30pt}
\begin{center}
\includegraphics[width=.48\textwidth,height=.40\textheight]{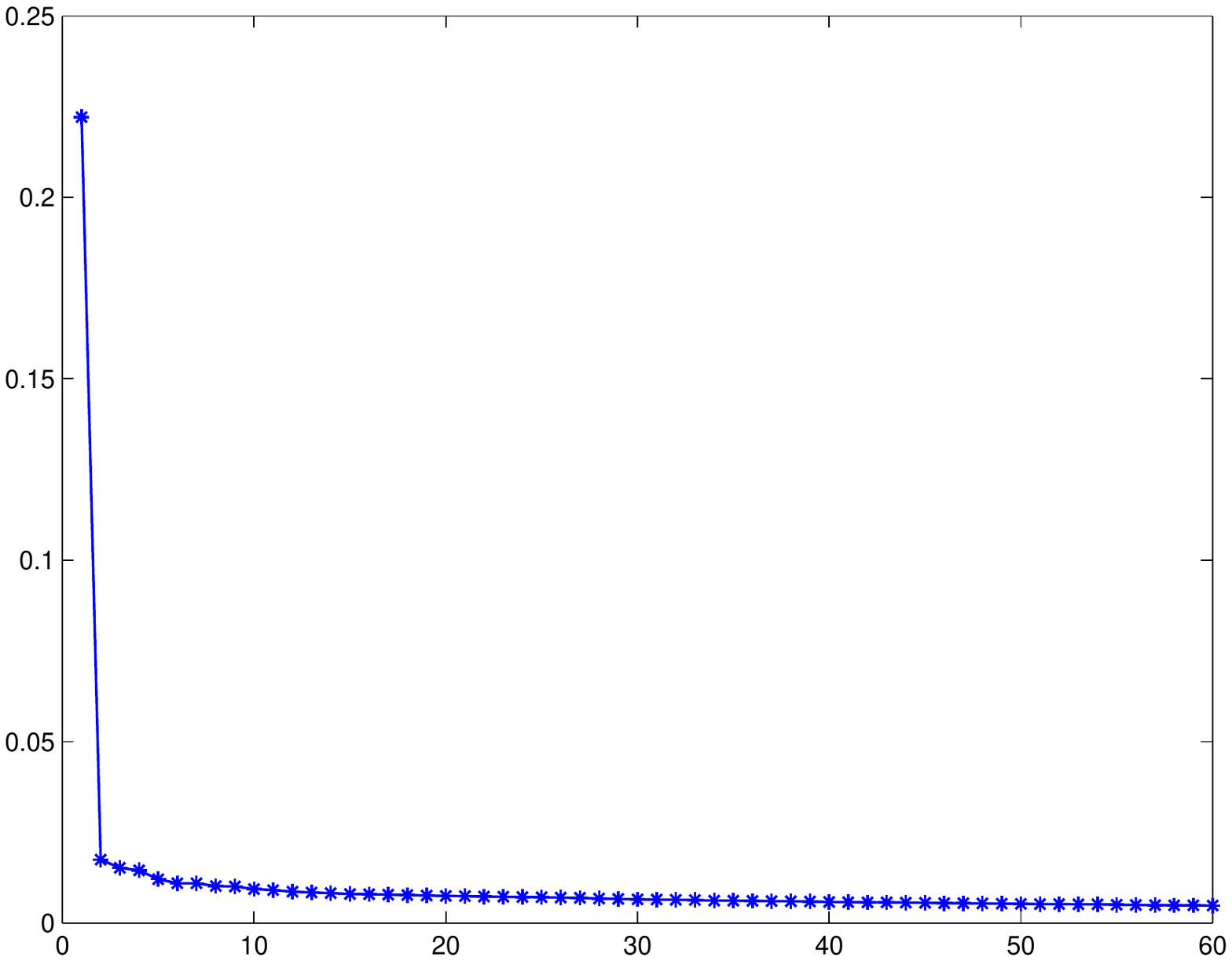}%
\includegraphics[width=.48\textwidth,height=.40\textheight]{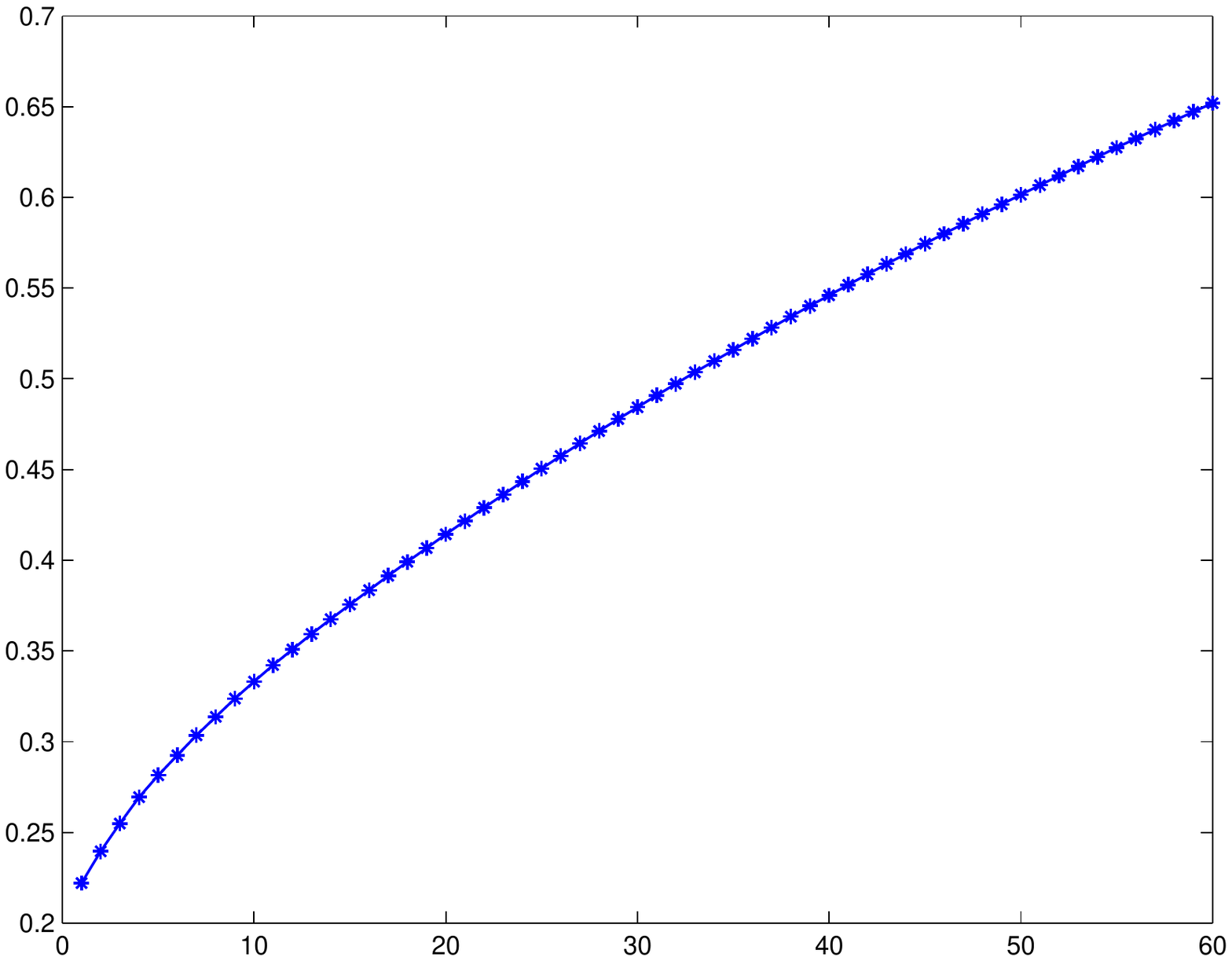}
\end{center}
\vspace*{-60pt}
\caption{Results from the ADHD 200 data: the first 60 relative eigenvalues of $\hat\Sigma_{\eta}$ (left) and their cumulative variation explained (right).}
\label{svcm_s5}
\end{figure}

We are also interested in assessing  the gender and diagnostic interaction. Specifically, we tested $H_0({\bf d}_0):~\beta_7({\bf d}_0)=0$ against $H_1({\bf d}_0): \beta_7({\bf d}_0)\not=0$
  for the gender$\times$diagnosis  interaction
  across all voxels.
As   $s$ increases from 0 to 10,  MASS shows an advantage
   in smoothing effective signals  within  relatively homogeneous ROIs, while preserving the edges of these ROIs  (Figure \ref{svcm_s4}  (a)-(b)). Inspecting  Figure ~\ref{svcm_s4} (c) and (d) reveals that  it is much easy to   identify significant ROIs in the $-\log_{10}(p)$ images at scale $h_{10}$, which   are  much smoother    than those at scale $h_0$. Thus, MASS shows a clear advantage in detecting more significant and smoothed activation regions. Furthermore,  as shown in Figure \ref{svcm_s5} ,   the largest estimated eigenvalue is much larger than the rest estimated  eigenvalues, which
  decrease very slowly to zero, and explains 22$\%$ of  variation in data after accounting for ${\bf x}_i$. This is quite common in neuroimaging data \citep{caffo2010two}. 
   
\begin{figure}[ht!]
\vspace*{0pt}
\begin{center}
\includegraphics[width=.95\textwidth,height=.28\textheight]{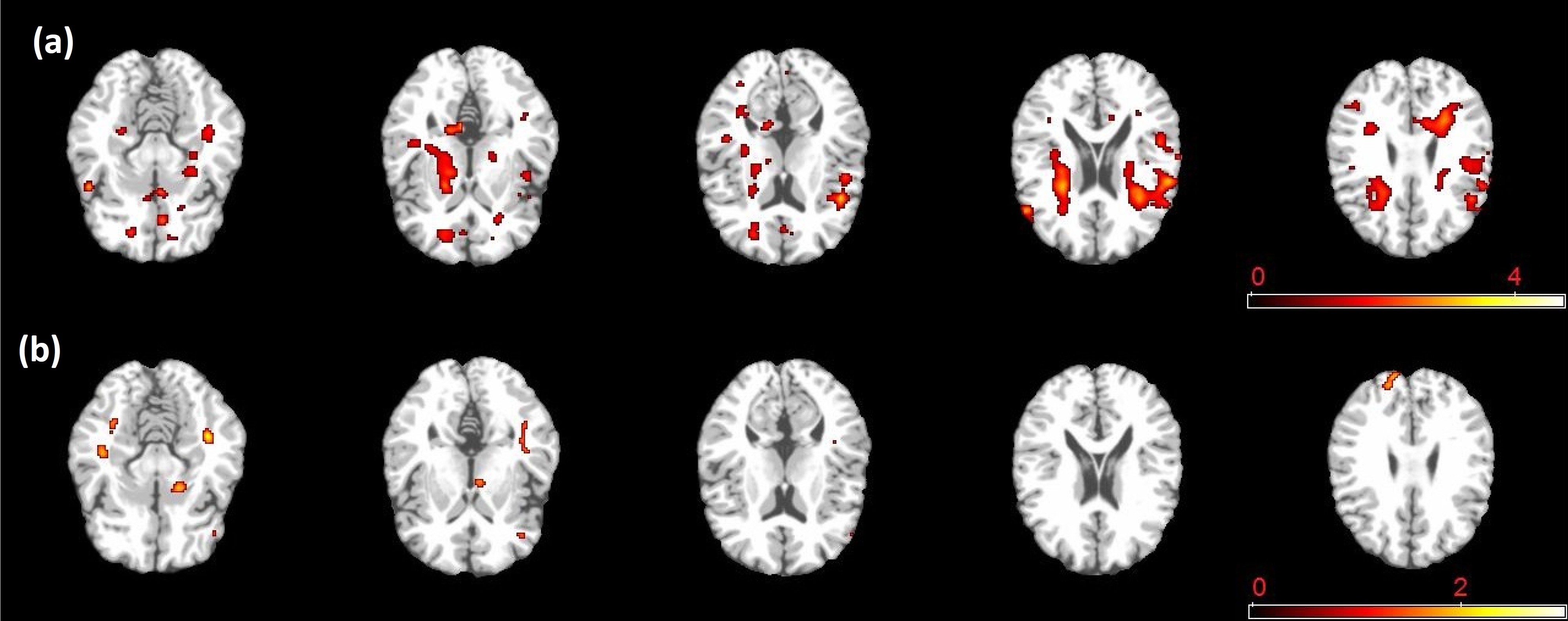}
\end{center}
\caption{Results from the ADHD 200 data:  The 26 and 10 significant blocks to test $H_0:\beta_6({\bf d})=0$ (a) and $H_0:\beta_7({\bf d})=0$ (b) overlaid with $-\log_{10}(p)$ values, respectively,  on selected slices, where $\beta_6({\bf d}_0)$ and $\beta_7({\bf d}_0)$ are, respectively, associated with the age$\times$diagnosis  and gender$\times$diagnosis  interactions.}
\label{svcm_s6}
\end{figure}

   To formally detect significant ROIs, we used
  a cluster-form of threshold of $5\%$  with a minimum voxel clustering value of 50 voxels.
     We were able to  detect  26  and 10 significant clusters for testing $H_0({\bf d}_0): \beta_6({\bf d}_0)=0$ (Figure ~\ref{svcm_s6} (a)) and $H_0({\bf d}_0): \beta_7({\bf d}_0)=0$ (Figure ~\ref{svcm_s6} (b)),  respectively, across all voxels.   Table \ref{svcm_ts1} lists the first two largest predefined regions (ROIs) within the first six largest significant blocks for testing  $H_0({\bf d}_0): \beta_6({\bf d}_0)=0$ and $H_0({\bf d}_0): \beta_7({\bf d}_0)=0$, respectively, along with their voxel sizes. Left and right frontal lobe white matter ROIs  are the largest ROIs with significant Age$\times$Diagnosis  interaction effect  while the first largest ROI  with significant  Gender$\times$Diagnosis interaction effect is temporal lobe. We can also observe that size of the significant blocks for Age$\times$Diagnosis interaction effect becomes much small starting from the fifth largest block while size of the significant blocks for Gender$\times$Diagnosis interaction effect becomes much small starting from the third largest block.

\bibliographystyle{asa}

\bibliography{HongtuNeuroimaging}
\newpage 

\newpage 

\newpage 

 \newpage 

\newpage

\newpage 

 \newpage 

\newpage

\newpage 

\newpage 

\newpage 

\begin{center}
\begin{table}
\caption{ Simulation results from LCE: Average Bias, RMS, SD, and RE  of ${\beta}_2({\bf d}_0)$ parameters in the five ROIs  at three different
bandwidths ($h_s,  h_m, h_l$), $N(0, 1)$ and $\chi^2(3)-3$ distributed data,  and 2 different sample sizes
($n=60,  80$).    BIAS denotes the bias of the mean of   estimates;   RMS denotes the
root-mean-square error; SD denotes the mean of the standard
deviation estimates; RE denotes the ratio of RMS over SD.   For each case, 200 simulated
data sets were used.
 }
\begin{center}
{\small
\begin{tabular}{|cc|cccccc|cccccc|}
\hline\hline
&&\multicolumn{6}{c}{$\chi^2(3)-3$}& \multicolumn{6}{|c|}{$N(0, 1)$}\\
&&   \multicolumn{3}{c}{$n=60$} & \multicolumn{3}{c|}{$n=80$}& \multicolumn{3}{c}{$n=60$}& \multicolumn{3}{c|}{$n=80$}\\ \hline
\multicolumn{2}{|c|}{$\beta_2({\bf d}_0)$} 		&$h_s$		&$h_m$		&$h_l$		&$h_s$		&$h_m$		&$h_l$		&$h_s$		&$h_m$		&$h_l$		&$h_s$		 &$h_m$		&$h_l$	\\ \hline
0.0	&	BIAS	&	0.01	&	0.01	&	0.03	&	0.01	&	0.01	&	0.03	&	0.01	&	0.01	&	0.03	&	0.01	&	0.01	&	0.03	\\	
	&	RMS	&	0.07	&	0.05	&	0.05	&	0.06	&	0.04	&	0.04	&	0.07	&	0.05	&	0.04	&	0.06	&	0.04	&	0.04	\\	
	&	SD	&	0.05	&	0.03	&	0.01	&	0.05	&	0.02	&	0.01	&	0.05	&	0.03	&	0.01	&	0.05	&	0.02	&	0.01	\\	
	&	RE	&	1.27	&	1.85	&	4.39	&	1.23	&	1.74	&	4.04	&	1.20	&	1.67	&	3.82	&	1.23	&	1.76	&	4.09	\\	\hline
0.2	&	BIAS	&	0.01	&	0.01	&	0.02	&	0.00	&	0.01	&	0.02	&	0.00	&	0.01	&	0.01	&	0.00	&	0.01	&	0.02	\\	
	&	RMS	&	0.07	&	0.06	&	0.05	&	0.06	&	0.05	&	0.04	&	0.07	&	0.05	&	0.04	&	0.06	&	0.05	&	0.04	\\	
	&	SD	&	0.05	&	0.03	&	0.01	&	0.05	&	0.02	&	0.01	&	0.05	&	0.03	&	0.01	&	0.05	&	0.02	&	0.01	\\	
	&	RE	&	1.35	&	2.08	&	5.28	&	1.27	&	1.89	&	4.66	&	1.25	&	1.84	&	4.51	&	1.31	&	1.99	&	5.00	\\	\hline
0.4	&	BIAS	&	-0.01	&	-0.01	&	-0.03	&	0.00	&	-0.01	&	-0.02	&	0.00	&	-0.01	&	-0.02	&	0.00	&	-0.01	&	-0.02	\\	
	&	RMS	&	0.08	&	0.06	&	0.06	&	0.06	&	0.05	&	0.04	&	0.07	&	0.05	&	0.05	&	0.06	&	0.05	&	0.04	\\	
	&	SD	&	0.06	&	0.03	&	0.01	&	0.05	&	0.02	&	0.01	&	0.06	&	0.03	&	0.01	&	0.05	&	0.02	&	0.01	\\	
	&	RE	&	1.38	&	2.15	&	5.53	&	1.28	&	1.93	&	4.80	&	1.26	&	1.88	&	4.65	&	1.33	&	2.05	&	5.20	\\	\hline
0.6	&	BIAS	&	-0.03	&	-0.06	&	-0.13	&	-0.03	&	-0.06	&	-0.13	&	-0.03	&	-0.06	&	-0.13	&	-0.03	&	-0.06	&	-0.13	\\	
	&	RMS	&	0.07	&	0.06	&	0.05	&	0.06	&	0.04	&	0.04	&	0.07	&	0.05	&	0.04	&	0.06	&	0.05	&	0.04	\\	
	&	SD	&	0.05	&	0.03	&	0.01	&	0.05	&	0.02	&	0.01	&	0.05	&	0.03	&	0.01	&	0.05	&	0.02	&	0.01	\\	
	&	RE	&	1.33	&	2.04	&	5.13	&	1.25	&	1.82	&	4.44	&	1.23	&	1.79	&	4.34	&	1.29	&	1.94	&	4.82	\\	\hline
0.8	&	BIAS	&	-0.04	&	-0.09	&	-0.20	&	-0.04	&	-0.09	&	-0.20	&	-0.04	&	-0.09	&	-0.20	&	-0.04	&	-0.09	&	-0.20	\\	
	&	RMS	&	0.08	&	0.06	&	0.06	&	0.06	&	0.05	&	0.04	&	0.07	&	0.06	&	0.05	&	0.06	&	0.05	&	0.05	\\	
	&	SD	&	0.06	&	0.03	&	0.01	&	0.05	&	0.02	&	0.01	&	0.06	&	0.03	&	0.01	&	0.05	&	0.02	&	0.01	\\	
	&	RE	&	1.40	&	2.22	&	5.73	&	1.31	&	2.00	&	5.04	&	1.28	&	1.92	&	4.82	&	1.35	&	2.11	&	5.39	\\	\hline
\hline
\end{tabular} 
}
\label{lce_t1}
\end{center}
\end{table}
\end{center}

\newpage 
 
\begin{center}
\begin{table}
\caption{ Simulation Study for $W_\beta({\bf d}_0;  h)$ from LCE:  estimates (ES) and standard errors (SE) of
rejection rates for pixels inside the five ROIs   were reported at three different
bandwidths ($h_s,  h_m, h_l$), $N(0, 1)$ and $\chi^2(3)-3$ distributed data,  and 2 different sample sizes
($n=60,  80$) at $\alpha=5\%$.  For each case, 200 simulated
data sets were used.
 }
\begin{center}
{\small 
 \begin{tabular}{|cc|cccc|cccc|}\hline\hline
   & &  \multicolumn{4}{c|}{$\chi^2(3)-3$}& \multicolumn{4}{c|}{$N(0, 1)$}\\
  & &  \multicolumn{2}{c}{$n=60$} & \multicolumn{2}{c|}{$n=80$}& \multicolumn{2}{c}{$n=60$}& \multicolumn{2}{c|}{$n=80$}\\ \hline
$\beta_{2}({\bf d}_0)$	&	h	&	ES	&	SE	&	ES	&	SE	&	ES	&	SE	&	ES	&	SE	\\	\hline
0.0	&	$h_s$	&	0.135	&	0.078	&	0.129	&	0.091	&	0.115	&	0.076	&	0.130	&	0.093	\\	
	&	$h_m$	&	0.335	&	0.206	&	0.317	&	0.215	&	0.289	&	0.211	&	0.312	&	0.217	\\	
	&	$h_l$	&	0.688	&	0.191	&	0.688	&	0.198	&	0.658	&	0.205	&	0.670	&	0.203	\\	\hline
0.2	&	$h_s$	&	0.883	&	0.070	&	0.950	&	0.045	&	0.892	&	0.071	&	0.944	&	0.047	\\	
	&	$h_m$	&	0.969	&	0.051	&	0.989	&	0.028	&	0.975	&	0.048	&	0.986	&	0.032	\\	
	&	$h_l$	&	0.992	&	0.019	&	0.998	&	0.009	&	0.994	&	0.016	&	0.996	&	0.012	\\	\hline
0.4	&	$h_s$	&	1.000	&	0.002	&	1.000	&	0.001	&	1.000	&	0.001	&	1.000	&	0.000	\\	
	&	$h_m$	&	1.000	&	0.002	&	1.000	&	0.001	&	1.000	&	0.001	&	1.000	&	0.001	\\	
	&	$h_l$	&	1.000	&	0.001	&	1.000	&	0.000	&	1.000	&	0.001	&	1.000	&	0.000	\\	\hline
0.6	&	$h_s$	&	1.000	&	0.000	&	1.000	&	0.000	&	1.000	&	0.000	&	1.000	&	0.000	\\	
	&	$h_m$	&	1.000	&	0.000	&	1.000	&	0.000	&	1.000	&	0.000	&	1.000	&	0.000	\\	
	&	$h_l$	&	1.000	&	0.000	&	1.000	&	0.000	&	1.000	&	0.000	&	1.000	&	0.000	\\	\hline
0.8	&	$h_s$	&	1.000	&	0.000	&	1.000	&	0.000	&	1.000	&	0.000	&	1.000	&	0.000	\\	
	&	$h_m$	&	1.000	&	0.000	&	1.000	&	0.000	&	1.000	&	0.000	&	1.000	&	0.000	\\	
	&	$h_l$	&	1.000	&	0.000	&	1.000	&	0.000	&	1.000	&	0.000	&	1.000	&	0.000	\\	\hline
\hline
\end{tabular}
}
\label{lce_t2}
\end{center}
\end{table}
\end{center}
 
 \newpage
 
\begin{center}
\begin{table}
\caption{ Simulation results from GKS: Average Bias, RMS, SD, and RE  of ${\beta}_2({\bf d}_0)$ parameters in the five ROIs  at three different
bandwidths ($h_s,  h_m, h_l$), $N(0, 1)$ and $\chi^2(3)-3$ distributed data,  and 2 different sample sizes
($n=60,  80$).    BIAS denotes the bias of the mean of   estimates;   RMS denotes the
root-mean-square error; SD denotes the mean of the standard
deviation estimates; RE denotes the ratio of RMS over SD.   For each case, 200 simulated
data sets were used. 
 }
\begin{center}
{\small
\begin{tabular}{|cc|cccccc|cccccc|}
\hline\hline
&&\multicolumn{6}{c}{$\chi^2(3)-3$}& \multicolumn{6}{|c|}{$N(0, 1)$}\\
&&   \multicolumn{3}{c}{$n=60$} & \multicolumn{3}{c|}{$n=80$}& \multicolumn{3}{c}{$n=60$}& \multicolumn{3}{c|}{$n=80$}\\ \hline
\multicolumn{2}{|c|}{$\beta_2({\bf d}_0)$} 		&$h_s$		&$h_m$		&$h_l$		&$h_s$		&$h_m$		&$h_l$		&$h_s$		&$h_m$		&$h_l$		&$h_s$		 &$h_m$		&$h_l$	\\ \hline
0.0	&	BIAS	&	0.01	&	0.01	&	0.03	&	0.01	&	0.01	&	0.03	&	0.01	&	0.01	&	0.03	&	0.01	&	0.01	&	0.03	\\	
	&	RMS	&	0.07	&	0.05	&	0.05	&	0.06	&	0.04	&	0.04	&	0.07	&	0.05	&	0.04	&	0.06	&	0.04	&	0.04	\\	
	&	SD	&	0.07	&	0.05	&	0.04	&	0.06	&	0.04	&	0.04	&	0.07	&	0.05	&	0.04	&	0.06	&	0.04	&	0.04	\\	
	&	RE	&	1.03	&	1.05	&	1.06	&	0.99	&	0.99	&	0.99	&	0.97	&	0.95	&	0.93	&	1.00	&	1.00	&	0.99	\\	\hline
0.2	&	BIAS	&	0.01	&	0.01	&	0.02	&	0.00	&	0.01	&	0.02	&	0.00	&	0.01	&	0.01	&	0.00	&	0.01	&	0.02	\\	
	&	RMS	&	0.07	&	0.06	&	0.05	&	0.06	&	0.05	&	0.04	&	0.07	&	0.05	&	0.04	&	0.06	&	0.05	&	0.04	\\	
	&	SD	&	0.07	&	0.06	&	0.05	&	0.06	&	0.05	&	0.04	&	0.07	&	0.06	&	0.05	&	0.06	&	0.05	&	0.04	\\	
	&	RE	&	1.05	&	1.07	&	1.09	&	0.99	&	0.98	&	0.96	&	0.97	&	0.95	&	0.93	&	1.02	&	1.03	&	1.03	\\	\hline
0.4	&	BIAS	&	-0.01	&	-0.01	&	-0.03	&	0.00	&	-0.01	&	-0.02	&	0.00	&	-0.01	&	-0.02	&	0.00	&	-0.01	&	-0.02	\\	
	&	RMS	&	0.08	&	0.06	&	0.06	&	0.06	&	0.05	&	0.04	&	0.07	&	0.05	&	0.05	&	0.06	&	0.05	&	0.04	\\	
	&	SD	&	0.07	&	0.06	&	0.05	&	0.06	&	0.05	&	0.04	&	0.07	&	0.06	&	0.05	&	0.06	&	0.05	&	0.04	\\	
	&	RE	&	1.05	&	1.08	&	1.10	&	0.98	&	0.97	&	0.96	&	0.97	&	0.95	&	0.93	&	1.02	&	1.03	&	1.04	\\	\hline
0.6	&	BIAS	&	-0.03	&	-0.06	&	-0.13	&	-0.03	&	-0.06	&	-0.13	&	-0.03	&	-0.06	&	-0.13	&	-0.03	&	-0.06	&	-0.13	\\	
	&	RMS	&	0.07	&	0.06	&	0.05	&	0.06	&	0.04	&	0.04	&	0.07	&	0.05	&	0.04	&	0.06	&	0.05	&	0.04	\\	
	&	SD	&	0.07	&	0.05	&	0.05	&	0.06	&	0.05	&	0.04	&	0.07	&	0.05	&	0.05	&	0.06	&	0.05	&	0.04	\\	
	&	RE	&	1.05	&	1.08	&	1.10	&	0.98	&	0.97	&	0.95	&	0.97	&	0.95	&	0.93	&	1.02	&	1.03	&	1.04	\\	\hline
0.8	&	BIAS	&	-0.04	&	-0.09	&	-0.20	&	-0.04	&	-0.09	&	-0.20	&	-0.04	&	-0.09	&	-0.20	&	-0.04	&	-0.09	&	-0.20	\\	
	&	RMS	&	0.08	&	0.06	&	0.06	&	0.06	&	0.05	&	0.04	&	0.07	&	0.06	&	0.05	&	0.06	&	0.05	&	0.05	\\	
	&	SD	&	0.07	&	0.06	&	0.05	&	0.06	&	0.05	&	0.05	&	0.07	&	0.06	&	0.05	&	0.06	&	0.05	&	0.05	\\	
	&	RE	&	1.05	&	1.08	&	1.09	&	0.98	&	0.97	&	0.96	&	0.96	&	0.93	&	0.92	&	1.01	&	1.02	&	1.03	\\	\hline
\hline
\end{tabular} 
}
\label{gks_t1}
\end{center}
\end{table}
\end{center}

\newpage 
 
\begin{center}
\begin{table}
\caption{ Simulation Study for $W_\beta({\bf d}_0;  h)$ from GKS:  estimates (ES) and standard errors (SE) of
rejection rates for pixels inside the five ROIs   were reported at three different
bandwidths ($h_s,  h_m, h_l$), $N(0, 1)$ and $\chi^2(3)-3$ distributed data,  and 2 different sample sizes
($n=60,  80$) at $\alpha=5\%$.  For each case, 200 simulated
data sets were used.
 }
\begin{center}
{\small 
 \begin{tabular}{|cc|cccc|cccc|}\hline\hline
   & &  \multicolumn{4}{c|}{$\chi^2(3)-3$}& \multicolumn{4}{c|}{$N(0, 1)$}\\
  & &  \multicolumn{2}{c}{$n=60$} & \multicolumn{2}{c|}{$n=80$}& \multicolumn{2}{c}{$n=60$}& \multicolumn{2}{c|}{$n=80$}\\ \hline
$\beta_{2}({\bf d}_0)$	&	h	&	ES	&	SE	&	ES	&	SE	&	ES	&	SE	&	ES	&	SE	\\	\hline
0.0	&	$h_s$	&	0.068	&	0.050	&	0.064	&	0.065	&	0.056	&	0.051	&	0.066	&	0.066	\\	
	&	$h_m$	&	0.118	&	0.194	&	0.115	&	0.218	&	0.100	&	0.203	&	0.117	&	0.216	\\	
	&	$h_l$	&	0.199	&	0.276	&	0.206	&	0.307	&	0.178	&	0.292	&	0.206	&	0.306	\\	\hline
0.2	&	$h_s$	&	0.783	&	0.110	&	0.886	&	0.085	&	0.785	&	0.117	&	0.880	&	0.085	\\	
	&	$h_m$	&	0.872	&	0.142	&	0.936	&	0.109	&	0.879	&	0.151	&	0.930	&	0.108	\\	
	&	$h_l$	&	0.897	&	0.157	&	0.941	&	0.122	&	0.901	&	0.168	&	0.937	&	0.121	\\	\hline
0.4	&	$h_s$	&	0.998	&	0.005	&	1.000	&	0.001	&	0.999	&	0.004	&	1.000	&	0.001	\\	
	&	$h_m$	&	0.998	&	0.014	&	1.000	&	0.005	&	0.999	&	0.009	&	1.000	&	0.004	\\	
	&	$h_l$	&	0.995	&	0.026	&	0.999	&	0.010	&	0.997	&	0.017	&	0.999	&	0.009	\\	\hline
0.6	&	$h_s$	&	1.000	&	0.000	&	1.000	&	0.000	&	1.000	&	0.000	&	1.000	&	0.000	\\	
	&	$h_m$	&	1.000	&	0.000	&	1.000	&	0.000	&	1.000	&	0.000	&	1.000	&	0.000	\\	
	&	$h_l$	&	1.000	&	0.000	&	1.000	&	0.000	&	1.000	&	0.000	&	1.000	&	0.000	\\	\hline
0.8	&	$h_s$	&	1.000	&	0.000	&	1.000	&	0.000	&	1.000	&	0.000	&	1.000	&	0.000	\\	
	&	$h_m$	&	1.000	&	0.000	&	1.000	&	0.000	&	1.000	&	0.000	&	1.000	&	0.000	\\	
	&	$h_l$	&	1.000	&	0.000	&	1.000	&	0.000	&	1.000	&	0.000	&	1.000	&	0.000	\\	\hline
\hline
\end{tabular}
}
\label{gks_t2}
\end{center}
\end{table}
\end{center}

 \newpage

\begin{center}
 \begin{table}
\caption{ Results from the ADHD 200 data: the first two largest significant regions of the first six largest significant blocks for hypothesis tests $H_0:\beta_6({\bf d})=0$ and $H_0:\beta_7({\bf d})=0$ with block and region voxel sizes. WM, L and R, respectively,  represent white matter, left hemisphere, and right hemisphere.  Moreover, $\beta_6({\bf d}_0)$ and $\beta_7({\bf d}_0)$ are, respectively, associated with the age$\times$diagnosis $(A\times D)$  and gender$\times$diagnosis  $(G\times D)$ interactions.
 }

{\small
 \begin{tabular}{|lll|ll|ll|}\hline\hline
   & & &  \multicolumn{2}{|c|}{1st largest predefined ROI}& \multicolumn{2}{|c|}{2nd largest predefined ROI}\\
	&	block	&	size	&	ROI label	&	size	&	ROI label	&	 size	\\	 \hline
$\mbox{A}\times\mbox{D}$ 	&	1	&	3954	&	frontal lobe WM L	&	 1567	&	 frontal lobe WM R	&	455	\\	
	&	2	&	2065	&	frontal lobe WM R	&	900	&	 anterior limb of internal capsule R	&	220	\\	
	&	3	&	1642	&	nucleus accumbens L	&	1019	&	frontal lobe WM L	&	213	 \\	
	&	4	&	1143	&	parietal lobe WM R	&	688	&	superior parietal lobule R	&	 132	\\	
	&	5	&	282	&	frontal lobe WM R	&	260	&	lateral front-orbital gyrus R	&	 22	\\	
	&	6	&	250	&	temporal lobe WM L	&	131	&	frontal lobe WM L	 &	35	\\	 \hline
 $\mbox{G}\times\mbox{D}$	&	1	&	228	&	temporal lobe WM L	&	184	 &	middle temporal gyrus L	&	22	\\	
	&	2	&	216	&	frontal lobe WM L	&	163	&	 superior frontal gyrus L	&	33	\\	
	&	3	&	95	&	temporal lobe WM R	&	66	&	lateral occipitotemporal gyrus R	 &	21	\\	
	&	4	&	94	&	medial frontal gyrus R	&	44	&	frontal lobe WM R	&	24	\\	 
	&	5	&	89	&	frontal lobe WM L	&	49	&	globus palladus L	 &	21	\\	
	&	6	&	83	&	superior occipital gyrus R	&	71	&	occipital lobe WM R	&	7	 \\	
	\hline
\hline
\end{tabular}
}
\label{svcm_ts1}
\end{table}

\end{center}